%% file: main.tex
\documentclass[preprint,12pt]{elsarticle}

\usepackage{amssymb}
\usepackage{amsmath}
\usepackage{amsthm}

\usepackage{graphicx}
\usepackage{array}
\usepackage{tabularx}
\usepackage{multirow}
\usepackage{xcolor}
\usepackage{textcomp}
\usepackage{booktabs}
\usepackage{algorithm}
\usepackage[noend]{algpseudocode}
\usepackage{algorithmicx}
\usepackage{bm}
\usepackage{enumitem}
\usepackage{subcaption}
\usepackage{xspace}

\newtheorem{theorem}{Theorem}
\newtheorem{lemma}[theorem]{Lemma}

\newtheorem{example}{Example}

\newtheorem{definition}{Definition}

\algrenewcommand\algorithmicrequire{\textbf{Input:}}
\algrenewcommand\algorithmicensure{\textbf{Output:}}

\renewcommand{\paragraph}[1]{\noindent\textbf{#1.}}

\newcommand{\prob}{\textsf{TAMICS}\xspace}
\newcommand{\cidx}{\textsf{TUC}-list\xspace}
\newcommand{\iidx}{\textsf{TIE}-tree\xspace}

\DeclareMathOperator*{\argmin}{arg\,min}

\journal{Neurocomputing}

\begin{document}

\begin{frontmatter}

\title{Topic-aware Most Influential Community Search in Social Networks}

\author[inst1]{Long Teng}
\ead{lteng@stu.ecnu.edu.cn}
\author[inst1]{Yanhao Wang\corref{cor1}}
\ead{yhwang@dase.ecnu.edu.cn}
\author[inst2]{Zhe Lin}
\ead{Zhe.Lin@xjtlu.edu.cn}
\author[inst3]{Fei Yu}
\ead{yufei@zhejianglab.com}

\affiliation[inst1]{organization={School of Data Science and Engineering, East China Normal University},
            city={Shanghai},
            postcode={200062}, 
            country={China}}
\affiliation[inst2]{organization={Department of Intelligent Operations and Marketing, International Business School Suzhou (IBSS), Xi'an Jiaotong-Liverpool University},
            city={Suzhou},
            postcode={215123},
            country={China}}
\affiliation[inst3]{organization={Zhejiang Lab},
            city={Hangzhou},
            postcode={311121},
            country={China}}
\cortext[cor1]{Corresponding author}

\input{sections-springer/0-abstract}

\end{frontmatter}

\input{sections-springer/1-introduction}
\input{sections-springer/2-related_work}
\input{sections-springer/3-definition}

\input{sections-springer/4-online_algorithm}
\input{sections-springer/5-index_algorithm}
\input{sections-springer/6-experiments}
\input{sections-springer/7-conclusion}




\bibliographystyle{elsarticle-num} 
\bibliography{ref}

\end{document}

%% file: sections-springer/0-abstract.tex
\begin{abstract}
Influential community search (ICS) finds a set of densely connected and high-impact vertices from a social network.
Although great effort has been devoted to ICS problems, most existing methods do not consider how relevant the influential community found is to specific topics.
A few attempts at topic-aware ICS problems cannot capture the stochastic nature of community formation and influence propagation in social networks.
To address these issues, we introduce a novel problem of topic-aware most influential community search (\prob) to discover a set of vertices such that for a given topic vector $\bm{q}$, they induce a $(k, l, \eta)$-core in an uncertain directed interaction graph and have the highest influence scores under the independent cascade (IC) model.
We propose an online algorithm to provide an approximate result for any \prob query with bounded errors.{}
Furthermore, we design two index structures and an index-based heuristic algorithm for efficient \prob query processing.
Finally, we experimentally evaluate the efficacy and efficiency of our proposed approaches on various real-world datasets.
{The results show that (1) the communities of \prob have higher relevance and social influence w.r.t.~the query topics as well as structural cohesiveness than those of several state-of-the-art topic-aware and influential CS methods and (2) the index-based algorithm achieves speed-ups of up to three orders of magnitude over the online algorithm with an affordable overhead for index construction.}
\end{abstract}


%% file: sections-springer/1-introduction.tex
\section{Introduction}
\label{sec_intro}

With the rapid development of online social networks, such as Facebook, X, Weibo, etc., large graphs have become widely available for analysis.
An important component of these graphs is \emph{community} \cite{girvan2002community}, typically defined as a small group of closely connected vertices within the graph.
Retrieving communities from large graphs based on query conditions, known as community search (CS) \cite{FangHQZZCL20}, is a fundamental problem in big data analytics and has attracted much attention due to its broad applications in social advertising \cite{LiLL12}, friend recommendation \cite{WangLCQW15}, and event organization \cite{SozioG10}.
{A problem closely relevant to CS is community detection (CD) {\cite{zhu2021community}}, which aims to identify all communities in a graph without specific query conditions. While CD focuses on uncovering the global community structure of a graph, CS is query-driven and retrieves local communities that satisfy specific conditions. This distinction clearly differentiates CD from CS, and the latter often involves unique challenges.}

In particular, recent efforts have been devoted to incorporating \emph{social influence} into CS problems \cite{LiQYM15, LiQYM17, LiWDYSY17, BiCLZ18, XuFWLXZ20, PengBLWY22, LuoZLGL23, ZhouFLY23}, with the aim of finding a group of vertices that are not only densely connected but also highly impactful.
We consider that influence in social networks is topic-aware and depends on the interests of vertices.
For example, a popular community of soccer fans might have little influence on vertices that are not interested in sports.
Therefore, it is essential to take into account \emph{topics} in the influential community search (ICS) problem.
In real-world scenarios, the vertices and edges of graphs contain rich semantic information about each user's topics of interest and the strengths of connections between two users across different topics.
We exploit how such semantic information can be incorporated into the ICS problem to find the most influential community on specific topics of interest.
The following are two examples where topic-aware and most influential communities can be applied.
\begin{itemize}
    \item \textbf{Academic Event Organization:} Let us consider an academic network in which authors are connected through collaborations and associated with the research topics on which they are working. Assume that we want to organize an academic event on DB+AI. We should find a group of researchers who are active in the areas of databases and artificial intelligence, as well as their integration (\emph{topic-awareness}), work closely with each other (\emph{cohesiveness}), and have a high impact on other researchers in both areas (\emph{influence}) as suitable candidates to lead the event.
    \item \textbf{Social Media Advertising:} For a social platform where users share their experience with different products with comments, blogs, videos, etc., an advertiser wants to promote a product through influencer marketing \cite{doi:10.1287/ijoc.2022.1246}.
    To identify appropriate influencers for the promotion campaign, we should select a group of users who align closely with the theme of the target product (\emph{topic-awareness}), have a large and dedicated fan base that is potentially interested in the product (\emph{influence}), and have a dense mutual connection to increase the chances of reaching target customers and affecting their purchasing behavior (\emph{cohesiveness}).
\end{itemize}

Despite extensive studies on both ICS~\cite{LiQYM15, LiQYM17, LiWDYSY17, BiCLZ18, XuFWLXZ20, PengBLWY22, LuoZLGL23, ZhouFLY23} and topic-aware CS~\cite{FangCLH16, ChenLLLZ19, ZhangHXCS19, LiuZZHXG20, LiZLWW23, XieZWY22} problems, there are only a few attempts~\cite{XieSLZL21, IslamAKSCR22} to incorporate social influence and topic awareness into the CS problem simultaneously.
And these existing studies on influential and topic-aware CS problems still have several limitations.
First, in the definition of cohesiveness, they adopt the classic $k$-core and $k$-truss models as well as their variants.
These models cannot denote the strengths of relationships between users on different topics because they only consider whether there is an edge (i.e., a connection).
Furthermore, they do not capture the uncertainty of community formation since their definitions are based on deterministic graphs.
Second, in terms of influence, most of them pre-assign a fixed influence score to each vertex.
Such schemes do not reflect that the influence of a user varies with topics and that the propagation of information in social networks is often described by stochastic diffusion models such as the Independent Cascade (IC) model \cite{KempeKT03}.
These limitations can hinder the application of existing CS methods in real-world scenarios.

\vspace{1mm}
\paragraph{Our Results}
To address the above issues, we propose a novel problem of \emph{\underline{T}opic-\underline{A}ware \underline{M}ost \underline{I}nfluential \underline{C}ommunity \underline{S}earch} (\prob) in social networks.
Specifically, for a social network $\mathcal{G}$ with $z$ topics, each \prob query is indicated by a $z$-dimensional vector $\bm{q}$ that denotes its relevance to each topic.
The social network $\mathcal{G}$ is then transformed into an uncertain directed graph $G_{\bm{q}}$ w.r.t.~$\bm{q}$ where the probability $p(e)$ associated with an edge $e = (u, v)$ signifies the strength and influence of the connection from $u$ to $v$.
We propose the $(k, l, \eta)$-core model in $G_{\bm{q}}$, which generalizes both the D-core model in directed graphs \cite{GiatsidisTV13} and the $(k, \eta)$-core model in uncertain undirected graphs \cite{bonchi2014core}, to represent the cohesiveness of subgraphs in directed and uncertain settings.
We adopt the widely used topic-aware IC model \cite{BarbieriBM13, AslayBBB14, ChenFLFTT15} to calculate the influence score of each vertex in $G_{\bm{q}}$.
Then, we consider the influence score of a subgraph as the minimum of the influences scores among all its vertices.
As such, a \prob query on the topic vector $\bm{q}$ aims to find the $(k, l, \eta)$-core subgraph with the highest influence score among all candidate $(k, l, \eta)$-cores in $G_{\bm{q}}$.

\begin{figure}[t]
    \centering
    \includegraphics[width=\linewidth]{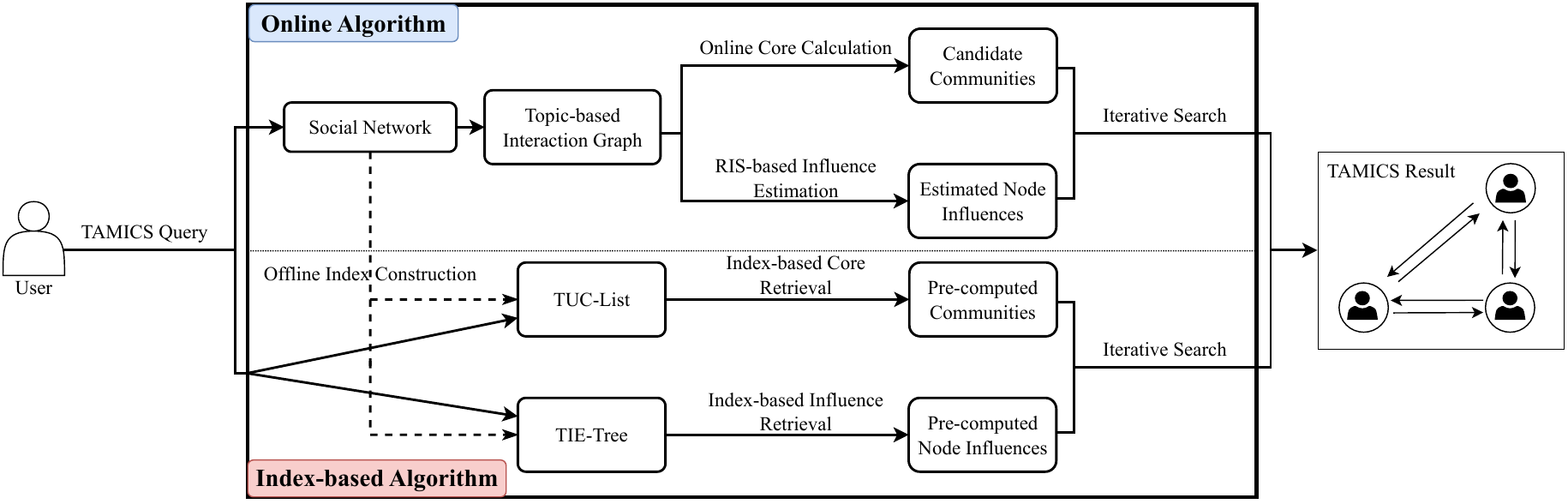}
    \caption{Workflow of our proposed algorithms for \prob query processing.}
    \label{fig_0workflow}
\end{figure}

To the best of our knowledge, no existing method can be used directly for \prob query processing.
We first propose an online algorithm for \prob.
It adopts a dynamic programming-based approach \cite{bonchi2014core} for online $(k,l,\eta)$-core computation.
Due to the \#P-hardness of computing influence scores exactly in the topic-aware IC model \cite{ChenWW10}, it utilizes the \emph{reverse influence sampling} (RIS) technique \cite{BorgsBCL14} for influence score estimation.
The online algorithm can guarantee to provide an approximate result for any \prob query with a small error bounded by that for influence score estimation.
To achieve higher efficiency, we further design an index consisting of two structures, \cidx and \iidx, for \prob query processing: (1) \cidx that maintains a list of $(k,l,\eta)$-cores in the supergraph of $G_{\bm{q}}$ for all $\bm{q}$'s to significantly reduce the search space for core computation, and (2) \iidx that utilizes a cone-tree \cite{RamG12} to store the influence scores w.r.t.~a pre-specified set of topic vectors to avoid duplicate influence calculation.
We design a more efficient algorithm based on \cidx and \iidx for \prob.
Although the index-based algorithm cannot achieve any theoretical bound on the errors for \prob, it runs much faster than the online algorithm while achieving comparable results in practice.
{The workflow of our proposed algorithms for \prob query processing is illustrated in Figure~\ref{fig_0workflow}.}

We conducted extensive experiments to demonstrate the efficacy of the \prob problem and the efficiency of our proposed algorithms for \prob on four real-world datasets.
The results show that, compared to several state-of-the-art topic-aware and influential CS methods, the communities returned by \prob have higher quality in terms of relevance and social influence w.r.t.~the query topics as well as structural cohesiveness.
Case studies on the DBLP and IMDB datasets further confirm the superior quality of communities provided by \prob.
Moreover, the index-based algorithm achieves speed-ups of up to three orders of magnitude over the online algorithm with a reasonable overhead for index construction.

The main contributions of this paper are summarized as follows.
\begin{itemize}
    \item We introduce the notion of $(k, l, \eta)$-core and formulate a novel \prob problem.
    \item We propose an online algorithm and analyze its theoretical bound for \prob.
    \item We design an index consisting of two structures, \cidx and \iidx, and devise an index-based algorithm for \prob.
    \item We demonstrate the efficacy and efficiency of our proposals through extensive experiments and case studies.
\end{itemize}

\vspace{1mm}
\paragraph{Paper Organization}
The remainder of this paper is organized as follows.
Section~\ref{sec_literature} discusses the literature related to this work.
Section~\ref{sec_define} introduces the basic notation and formally defines the \prob problem.
Section~\ref{sec_online} describes the online algorithm for \prob and analyzes it theoretically.
Section~\ref{sec_index} presents the index construction and the index-based algorithm for \prob.
Section~\ref{sec_exp} shows the setup and results of experiments and case studies.
Section~\ref{sec_conclusion} concludes the whole paper.

%% file: sections-springer/2-related_work.tex
\section{Related Work}
\label{sec_literature}

\paragraph{Influential Community Search (ICS)}
The ICS problem aims to find a cohesive subgraph that is highly influential on a graph.
Li et al.~\cite{LiQYM15, LiQYM17} first studied the ICS problem, which assigns an influence score to each vertex, defines the influence score of a community as the minimum influence value among all vertices within it, and aims to find top-$r$ subgraphs with the highest influence scores based on the $k$-core model.
Bi et al.~\cite{BiCLZ18} improved the online search method to find top-$r$ influential communities.
Peng et al.~\cite{PengBLWY22} further incorporated aggregation functions into the ICS problem.
Li et al.~\cite{LiWDYSY17} studied the most ICS problem to find a subgraph that contains at least $k$ nodes, where any two nodes can reach each other within $r$ hops, while having the maximum outer influence in the independent cascade (IC) model.
Xu et al.~\cite{XuFWLXZ20} studied the problem of personalized ICS to find the subgraphs with the largest influence on a query vertex.
Luo et al.~\cite{LuoZLGL23} proposed methods to discover all $(k, \eta)$-influential communities on an uncertain graph.
Zhou et al.~\cite{ZhouFLY23} investigated the ICS problem in heterogeneous information networks (HINs).
Zhang et al.~\cite{ZHANGHY24} investigated the ICS problem in bipartite graphs.
Zhang et al.~\cite{ZhangYLC24} studied the TopL-ICDE problem, which aims to retrieve top-$L$ seed communities with the highest influences while having high structural cohesiveness and containing user-specified query keywords.
Their approach is based on the $k$-truss model and uses a radius $r$ to constrain the size of the seed communities.
The distinction in problem formulation leads to significant differences in algorithmic design from ours.
Chang et al.~\cite{chang2024mics} proposed the Most Influenced Community Search (MICS) problem, which aims to identify a densely connected subgraph that is most significantly influenced by a given seed node set $S$ in a graph.
The aforementioned methods are primarily designed for influential community search problems and are topic-unaware.
As a result, they lack the capability to incorporate topic-specific constraints, which limits their applicability to the TAMICS problem.

\vspace{1mm}
\paragraph{Topic-Aware Community Search (TACS)}
The TACS problem aims to find a coherent subgraph relevant to given query topics.
The first line of studies on TACS focuses on keyword-based attributed graphs, where vertices or edges are associated with a set of keywords that represent the topics relevant to them.
Fang et al.~\cite{FangCLH16} first proposed the \emph{attributed community query} (ACQ) problem to find a subgraph that contains the query vertex and satisfies both structural cohesiveness (i.e., $k$-core) and topic cohesiveness (i.e., containing common query keywords).
Liu et al.~\cite{LiuZZHXG20} proposed the \emph{vertex-centric attributed community search} (VAC) problem, which also returns a subgraph containing the query vertex but defines structural cohesiveness by the $k$-truss model and topic cohesiveness by the Jaccard distance from the set of query keywords.
Li et al.~\cite{LiZLWW23} proposed the \emph{edge-attributed community query} (EACS) problem that is similar to VAC but focuses on graphs with edge attributes.
Chen et al.~\cite{ChenLLLZ19} and Zhang et al.~\cite{ZhangHXCS19} proposed to find communities related to query keywords without specifying any vertex to contain.
The second line of studies on TACS generalizes it by introducing semantic measures for topic relevance rather than keyword matching.
Al-Baghdadi and Lian~\cite{Al-BaghdadiL20} extended the TACS problem by incorporating spatial information.
Lin et al.~\cite{LinYZPLL22} studied the TACS problem focusing on semantic information and interpretability of communities.
Chowdhary et al.~\cite{ChowdharyLCZY22} investigated the problem of discovering attribute-diversified communities that maintain structural cohesion while exhibiting diversity in attributes.
Xie et al.~\cite{XieZWY22} proposed an improved TACS method considering community focusing and latent relationships.
Wang et al.~\cite{0001GXZK024} studied the TACS problem in HINs.
Since none of the above methods takes \emph{social influence} into account, they are not suitable for \prob presented in this paper.

A few attempts have been made to solve the problem of topic-aware and influential CS using models different from this work.
Xie et al.~\cite{XieSLZL21} proposed using the $(k, d)$-truss model for cohesiveness and a scoring function that combines keyword coherence and outer influence to measure community quality.
They aim to find a $(k, d)$-truss with the highest score for a query vertex and a set of query keywords.
Islam et al.~\cite{IslamAKSCR22} proposed a \emph{keyword-aware influential community query} (KICQ) to identify a maximal $k$-core that contains a set of keywords and achieves the highest score based on cohesiveness and influences.
However, both methods do not consider the stochastic nature of community formation and influence propagation in graphs and cannot work for \prob.

\vspace{1mm}
\paragraph{Topic-Aware Influence Maximization}
Another line of studies related to this work is \emph{topic-aware influence maximization}, which finds a set of vertices from a social network such that they collectively have the highest influence on the query topics.
Several works \cite{AslayBBB14, ChenLY15, ChenFLFTT15, LiZT15} proposed to generalize classic influence maximization methods \cite{KempeKT03, ChenWW10, BorgsBCL14, TangXS14} to efficiently support topic-aware influence maximization queries.
Recent work \cite{MinCYL20, TianMWP20, ChenYLYC23} also introduced deep learning methods to solve the problem.
However, these methods focus only on finding the vertices with the maximum influence on specific topics but do not consider whether these vertices can induce a coherent subgraph.
Therefore, they cannot be applied to CS problems.

%% file: sections-springer/3-definition.tex
\section{Problem Formulation}
\label{sec_define}

In this section, we first introduce the basic notation and then formally define the \prob problem.

\vspace{1mm}
\paragraph{Graph Notation}
We first define the social network considered in this paper.
\begin{definition}[Social Network]
\label{def-sn}
    A social network is denoted as a directed graph $\mathcal{G} = (\mathcal{V}, \mathcal{E}, \omega)$, where $\mathcal{V}$ is the set of vertices (i.e., users), $\mathcal{E} \subseteq \mathcal{V} \times \mathcal{V}$ is the set of directed edges (i.e., relationships), and $\omega: \mathcal{E} \mapsto \mathbb{R}^{z}_{+}$ is a mapping function that assigns each edge $e = (u, v)$ with a $z$-dimensional vector $\bm{\omega}(e) = \left(\omega_{1}(e), \dots, \omega_{z}(e)\right)$ to measure the weights of the relationship over $z$ topics.
\end{definition}
We treat an undirected graph as a special case of Definition~\ref{def-sn} with symmetric directed edges.
We assume that the topic vector associated with each edge is given as prior knowledge.
In practice, topic vectors can be obtained in many different ways, such as directly applying ground-truth topics, inferring from topic models trained on user-generated texts \cite{QiangQLYW22}, and performing low-rank approximations based on historical user behaviors \cite{WangZ13}.

In a social network, use interactions are often topic-aware and community-based, that is, users only interact with their local communities via social links on the topics they are interested in.
To model user interactions in a topic distribution denoted as a vector $\bm{q} = (q_1, \dots, q_z) \in [0, 1]^{z}$ with $\sum_{i = 1}^{z} q_i = 1$, we extract a subgraph from the social network $\mathcal{G}$, which we refer to as \emph{topic-based interaction graph}.
\begin{definition}[Topic-based Interaction Graph]
\label{def-intg}
    An interaction graph extracted from $\mathcal{G}$ w.r.t.~$\bm{q}$ is a directed uncertain graph $G_{\bm{q}} = (V_{\bm{q}}, E_{\bm{q}}, p)$, where $p: \mathcal{E} \mapsto [0, 1]$ maps each edge $e \in \mathcal{E}$ to its probability $p(e) = f(\langle \bm{\omega}(e), \bm{q} \rangle)$ under the topic distribution indicated by $\bm{q}$, $\langle \cdot, \cdot \rangle$ is the dot product of two vectors, and $f(\cdot)$ is a monotonic function to normalize any nonnegative real number to the range $[0, 1]$. Accordingly, $E_{\bm{q}} = \{e \in \mathcal{E} \,|\, p(e) > 0\}$ and $V_{\bm{q}} = \{u \,|\, (u, v) \in E_{\bm{q}}\} \cup \{v \,|\, (u, v) \in E_{\bm{q}}\}$.
\end{definition}
We adopt the possible world semantics \cite{bonchi2014core, LuoZLGL23} of uncertain graphs, wherein each possible world corresponds to an instance $G'_{\bm{q}} = (V_{\bm{q}}, E'_{\bm{q}}) \sqsubseteq G_{\bm{q}}$ with probability $\Pr[G'_{\bm{q}}] = \prod_{e \in E'_{\bm{q}}}$ $p(e) \prod_{e \in E_{\bm{q}} \setminus E'_{\bm{q}}} (1 - p(e))$.
The probability that $v \in V_{\bm{q}}$ has an in-degree (or out-degree) of at least $k \in \mathbb{Z}^{+}$ (or at least $l \in \mathbb{Z}^{+}$) is $\Pr[d^{-}_{v}(G_{\bm{q}}) \geq k] = \sum_{G'_{\bm{q}} \sqsubseteq G_{\bm{q}}} \Pr[G'_{\bm{q}}]$ (or $\Pr[d^{+}_{v}(G_{\bm{q}}) \geq l] = \sum_{G'_{\bm{q}} \sqsubseteq G_{\bm{q}}} \Pr[G'_{\bm{q}}]$), where $v$ has an in-degree of at least $k$ (or out-degree of at least $l$) in each $G'_{\bm{q}}$.
Subsequently, we define the concept of $(k, l, \eta)$-core in topic-based interaction graphs.
\begin{definition}[$(k, l, \eta)$-Core]
\label{def-core}
    Given a topic-based interaction graph $G_{\bm{q}}$ and three parameters $k, l \in \mathbb{Z}_{+}$, and $\eta \in (0, 1]$, a $(k, l, \eta)$-core is a maximal induced subgraph $C = (V_{C}, E_{C}, p)$ of $G_{\bm{q}}$ such that $\Pr[d^{-}_{v}(C) \geq k \wedge d^{+}_{v}(C) \geq l] \geq \eta$, $\forall v \in V_C$.
\end{definition}

\begin{figure}[t]
    \centering
    \begin{subfigure}[t]{0.38\textwidth}
        \centering
        \includegraphics[width=.95\linewidth]{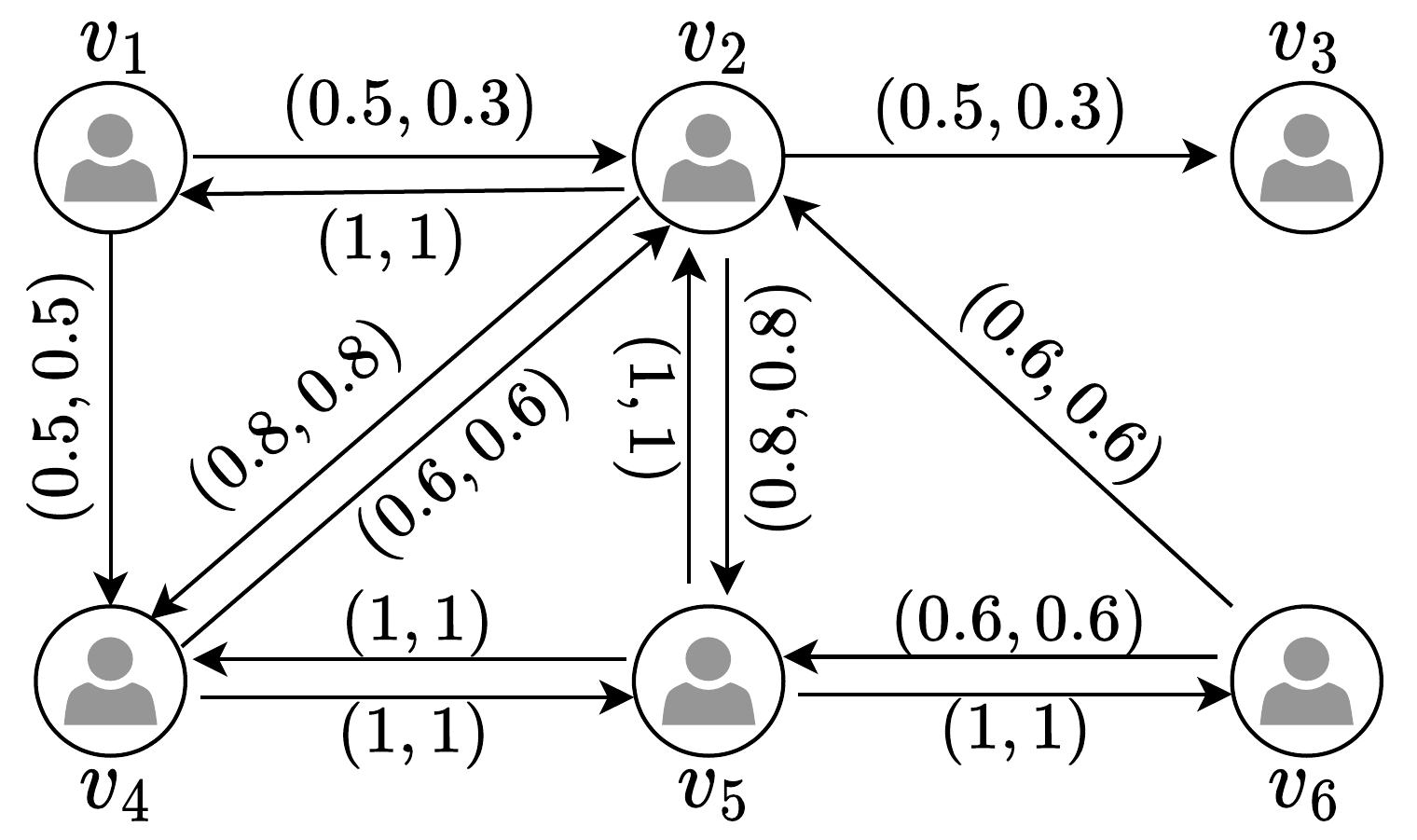}
        \caption{Social network $\mathcal{G}$}
        \label{fig_1example:left}
    \end{subfigure}
    \hfill
    \begin{subfigure}[t]{0.6\textwidth}
        \centering
        \includegraphics[width=.95\linewidth]{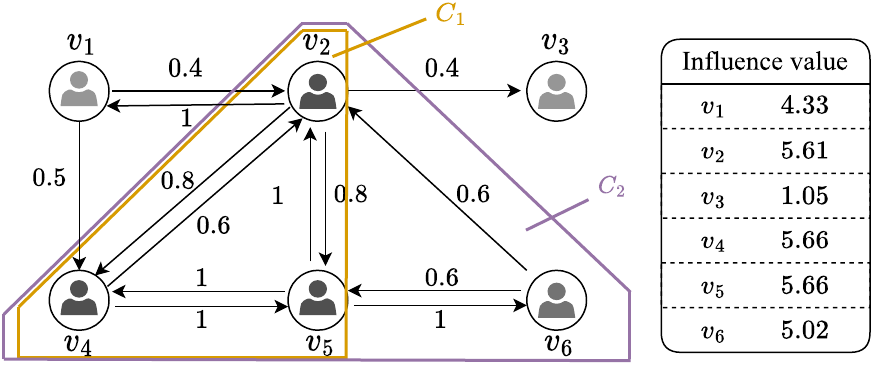}
        \caption{Topic-based interaction graph $G_{\bm{q}}$ for $\bm{q} = (0.5, 0.5)$}
        \label{fig_1example:right}
    \end{subfigure}
    \caption{Running examples of a social network with two topics ``\textsf{movie}'' and ``\textsf{music}'' and its topic-based interaction graph and influential communities for $\bm{q} = (0.5, 0.5)$.}
    \label{fig_1example}
\end{figure}

\begin{example}
    In Fig.~\ref{fig_1example}, we illustrate a social network $\mathcal{G}$ with $6$ users (i.e., vertices) and $13$ edges over two topics ``\textsf{movie}'' and ``\textsf{music}''. Each directed edge $e = (u, v)$ is associated with a two-dimensional vector to represent the strength of the relationship from $u$ to $v$ on the two topics. 
    We present an interaction graph $G_{\bm{q}}$ extracted from $\mathcal{G}$ based on a topic vector $\bm{q} = (0.5, 0.5)$.
    We show that $C_1 = \{v_2, v_4, v_5\}$ is a $(1, 2, 0.6)$-core of $G_{\bm{q}}$ since the probabilities that the in-degrees and out-degrees of $v_2$, $v_4$, and $v_5$ are greater than or equal to $1$ and $2$ are $0.64$, $0.6$, and $1.0$, respectively.
    Similarly, $C_2 = \{v_2, v_4, v_5, v_6\}$ forms a $(1, 2, 0.36)$-core and, obviously, $C_1$ is also a $(1, 2, 0.36)$-core.
\end{example}

\vspace{1mm}
\paragraph{Influence Model}
Next, we consider the topic-aware information diffusion process on interaction graphs and the computation of the influence score of each vertex.
Specifically, we employ the widely used topic-aware IC model (TIC) \cite{BarbieriBM13, AslayBBB14, ChenFLFTT15} to describe information diffusion.\footnote{Note that the topic-aware variant of any other diffusion model, such as the linear threshold (LT) and triggering models \cite{KempeKT03}, can be used in place of the TIC model.}
For an interaction graph $G_{\bm{q}}$ w.r.t.~a topic vector $\bm{q}$, the TIC model initially designates a set of vertices to be active (called ``\emph{seeds}'') and all remaining vertices to be inactive.
Then, active vertices start to activate their inactive neighbors through edges.
The activation is said to be topic-aware because its success probability depends on $\bm{q}$.
Especially, when a vertex $u$ is active, the probability that $u$ activates its neighbor $v$ through the edge $e = (u, v)$ is calculated as $pp(e) = \alpha p(e) = \alpha f(\langle \bm{\omega}(e), \bm{q} \rangle)$, where $\alpha \in (0, 1]$ is a scaling factor.
In the TIC model, each active vertex $u$ has only one chance to activate its out-neighbors.
After that, $u$ remains active and stops the activation, and each newly activated vertex will further try to activate its inactive neighbors.
The diffusion process terminates when no new vertex can be activated.
Based on the TIC model, we define the influence score $\mathbb{I}_{\bm{q}}(v)$ of a vertex $v$ w.r.t.~$\bm{q}$ as the expected number of active vertices in $G_{\bm{q}}$ when $v$ is the only seed.
It is known that computing $\mathbb{I}_{\bm{q}}(v)$ exactly is \#P-hard \cite{ChenWW10}.
In this paper, we use the reverse influence sampling (RIS) technique \cite{BorgsBCL14, TangXS14} to effectively approximate $\mathbb{I}_{\bm{q}}(v)$.
We define the influence score of an induced subgraph $C = (V_C, E_C)$ of $G_{\bm{q}}$ as the minimum of the influence scores of the vertices in $V_C$, i.e., $\mathbb{I}_{\bm{q}}(C) = \min_{v \in V_C} \mathbb{I}_{\bm{q}}(v)$.

\vspace{1mm}
\paragraph{Problem Statement}
Based on the notions of the $(k, l, \eta)$-core and its influence score, we define the $(k, l, \eta)$-influential community as follows.
\begin{definition}[$(k, l, \eta)$-Influential Community]
    Given a topic-based interaction graph $G_{\bm{q}}$ and three parameters $k, l \in \mathbb{Z}_{+}$, and $\eta \in (0,1]$, a $(k, l, \eta)$-influential community $C'$ with an influence score $\mathbb{I}_{\bm{q}}(C')$ is a subgraph in $G_{\bm{q}}$ satisfying the following conditions:
    \begin{itemize}
        \item \textbf{Connection}: $C'$ is (weakly) connected;
        \item \textbf{Cohesiveness}: Each vertex $v \in C'$ satisfies $\Pr[d^{-}_{v}(C') \geq k \wedge d^{+}_{v}(C') \geq l] \geq \eta$, i.e., $C'$ is a $(k, l, \eta)$-core in $G_{\bm{q}}$;
        \item \textbf{Maximality}: There is no other induced subgraph $C''$ containing $C'$ that satisfies the above two conditions and has an influence score $\mathbb{I}_{\bm{q}}(C'') \geq \mathbb{I}_{\bm{q}}(C')$.
    \end{itemize}
\end{definition}
Finally, we formulate the \prob problem as follows.
\begin{definition}[\textnormal{\prob}]
    Given a social network $\mathcal{G}$, a query vector $\bm{q}$, and three parameters $k$, $l$, and $\eta$, find a $(k, l, \eta)$-influential community $C^{*}_{\bm{q}}$ of the interaction graph $G_{\bm{q}}$ of $\mathcal{G}$ w.r.t.~$\bm{q}$ with the highest influence score, i.e., $C^{*}_{\bm{q}} = \arg\max_{C' \in \mathcal{C}'} \mathbb{I}_{\bm{q}}(C')$, where $\mathcal{C}'$ is the set of all $(k, l, \eta)$-influential communities in $G_{\bm{q}}$.
\end{definition}
Note that there may exist different $(k, l, \eta)$-influential communities with the same (maximum) influence score in $G_{\bm{q}}$, and any of them is considered an exact result for the \prob query on topic vector $\bm{q}$.

\begin{example}
    For the social network $\mathcal{G}$ in Fig.~\ref{fig_1example}, the result of a \prob query on $\bm{q} = (0.5, 0.5)$, $k = 1$, $l = 2$, and $\eta = 0.6$ is $C_1^* = \{v_2, v_4, v_5\}$ with an influence score $\mathbb{I}_{\bm{q}}(C^*_1) = \mathbb{I}_{\bm{q}}(v_2) \approx 5.6$.
    When $\eta = 0.36$ and $\bm{q}$, $k$, and $l$ remain unchanged, the result of the \prob query is still $C_1^*$.
    Although $C_2=\{v_2, v_4, v_5, v_6\}$ is a $(1, 2, 0.36)$-core, it has a lower influence score $\mathbb{I}_{\bm{q}}(C_2) = \mathbb{I}_{\bm{q}}(v_6) \approx 5$ than $C_1^*$ and thus is not a $(1, 2, 0.36)$-influential community.
\end{example}

Before presenting the technical details, we summarize the frequently used notations throughout this paper in Table~\ref{tab-notation}.

\input{tables/notation}

%% file: tables/notation.tex
\begin{table}[t]
    \footnotesize
    \centering
    \caption{List of frequently used symbols}
    \centering
    \label{tab-notation}
        \begin{tabular}{c p{295pt}}
            \toprule
            \textbf{Symbol} & \textbf{Description} \\
            \midrule
            $\mathcal{G}$=($\mathcal{V}$,$\mathcal{E}$,$\omega$) & Social network with a vertex set $\mathcal{V}$, an edge set $\mathcal{E}$, the edge weighting function $\omega(\cdot)$\\
            $\bm{q}$ & Topic vector in a \prob query\\
            $k$,$l$,$\eta$ & Parameters $k$, $l$, and $\eta$ in the definition of uncertain cores on directed graphs\\
            $G_{\bm{q}}$=$(V_{\bm{q}}$,$E_{\bm{q}}$,$p$) & Interaction graph w.r.t.~topic vector $\bm{q}$ with vertex set $V_{\bm{q}}$, edge set $E_{\bm{q}}$, and edge probability function $p(\cdot)$\\
            $n$,$m$,$z$ & Numbers of vertices, edges, and topics in $\mathcal{G}$\\
            $d^{-}_{v}(G)$ & In-degree of $v$ in an (uncertain) graph $G$\\
            $d^{+}_{v}(G)$ & Out-degree of $v$ in an (uncertain) graph $G$\\
            $\Delta^{-}_{v}(G)$ & Total number of in-neighbors of $v$ in a graph $G$\\
            $\Delta^{+}_{v}(G)$ & Total number of out-neighbors of $v$ in a graph $G$\\
            $C$=($V_C$,$E_C$,$p$) & Subgraph of $G_{\bm{q}}$ induced by the set $V_C \subseteq V_{\bm{q}}$ of vertices with a set of edges $E_C = \{ (u, v) \,|\, u, v \in V_C \wedge (u, v) \in E_{\bm{q}}\}$\\
            $\mathbb{I}_{\bm{q}}(v)$,$\mathbb{I}_{\bm{q}}(C)$ & Influence values of $v$ and $C$ w.r.t.~$\bm{q}$\\
            $C^*_{\bm{q}}$ & Exact result of a \prob query with topic vector $\bm{q}$\\
            \bottomrule
        \end{tabular}
\end{table}

%% file: sections-springer/4-online_algorithm.tex
\section{Online Algorithm}
\label{sec_online}

In this section, we propose an online algorithm for \prob, which proceeds in three stages: (1) computing the $(k, l, \eta)$-cores of the topic-based interaction graph $G_{\bm{q}}$; (2) evaluating the influence score $\mathbb{I}_{\bm{q}}(v)$ of each $v \in V_{\bm{q}}$; and (3) finding a $(k, l, \eta)$-core with the maximum influence score.
Next, we will present each stage in more detail.

\vspace{1mm}
\paragraph{$(k, l, \eta)$-Core Computation}
In possible world semantics \cite{bonchi2014core}, the in- and out-degrees $d^{-}_{v}(G_{\bm{q}}), d^{+}_{v}(G_{\bm{q}})$ of a vertex $v$ in $G_{\bm{q}}$ are random variables drawn from two discrete distributions in the ranges $[0, \Delta^{-}_{v}(G_{\bm{q}})]$ and $[0, \Delta^{+}_{v}(G_{\bm{q}})]$, where $\Delta^{-}_{v}(G_{\bm{q}})$ and $\Delta^{+}_{v}(G_{\bm{q}})$ are the total numbers of in- and out-neighbors of $v$ in $G_{\bm{q}}$.
Next, we show how to compute the in-degree of a vertex in $G_{\bm{q}}$.
For ease of presentation, we will drop $G_{\bm{q}}$ from the degree notation in this section when the context is clear.
Specifically, the probability that a vertex $v$ has at least $k$ in-neighbors is computed as
\begin{equation}\label{eq-indegree}
    \Pr[d^{-}_{v} \geq k] = 1 - \sum_{i=0}^{k-1} \Pr[d^{-}_{v} = i].
\end{equation}
Then, $\Pr[d^{-}_{v} = i]$ for each $i = 0, \dots, k - 1$ is computed as $\Pr[d^{-}_{v} = i] = \sum_{N \in \mathcal{N}^{-}_{v}, |N| = i}$ $\big(\prod_{u \in N}{p(e_{u, v})} \prod_{u \in \mathcal{N}^{-}_{v} \setminus N}{1 - p(e_{u, v})}\big)$, where $\mathcal{N}^{-}_{v}$ is the set of all in-neighbors of $v$, $N$ is a size-$i$ subset of $\mathcal{N}^{-}_{v}$, and $e_{u, v}$ is an edge from $u$ to $v$.
However, the number of size-$i$ subsets of $\mathcal{N}^{-}_{v}$ is $\binom{|\mathcal{N}^{-}_{v}|}{i}$, and thus computing $\Pr[d^{-}_{v} = i]$ by enumerating all subsets is prohibitive.
Therefore, we adopt a dynamic programming approach similar to that of \cite{bonchi2014core} to compute $\Pr[d^{-}_{v} \geq k]$ and update $\Pr[d^{-}_{v} \geq k]$ for the addition and removal of edges.
Formally, we specify an ordering $\pi(\mathcal{N}^{-}_{v}) = [u_1, \dots, u_{\Delta^{-}_v}]$ of $\mathcal{N}^{-}_{v}$ and use $\mathcal{N}^{-}_{v}[h]$ to denote the first $h$ vertices in $\pi(\mathcal{N}^{-}_{v})$ for any $h = 0, \dots, \Delta^{-}_{v}$.
We define a state $A(h, i)$ in the dynamic program as $\Pr[d^{-}_{v}(\mathcal{N}^{-}_{v}[h]) = i]$, i.e., the probability that exactly $i$ edges exist after checking the first $h$ vertices of $\pi(\mathcal{N}^{-}_{v})$, for any $h \in [0, \Delta^{-}_{v}]$ and $i \in [0, h]$.
Intuitively, $\Pr[d^{-}_{v} = i] = A(\Delta^{-}_v, i)$.
The state transition follows Eq.~\ref{eq-dp1}:
\begin{equation}\label{eq-dp1}
    A(h, i) = p(e_h) \cdot A(h - 1, i - 1) + (1 - p(e_h)) \cdot A(h - 1, i),
\end{equation}
where $e_h = (u_h, v)$.
The base cases include:
\begin{equation}\label{eq-dp2}
    \begin{cases}
        A(0, 0) = 1; & \\
        A(h, -1) = 0, & \forall h \in [0, \Delta^{-}_v], \\
        A(h, i) = 0,  & \forall h \in [0, \Delta^{-}_v] \text{ and } i > h.
\end{cases}
\end{equation}
Starting with $A(0, 0) = 1$, we can update the value of each state in the dynamic program progressively and obtain $\Pr[d^{-}_{v} = i]$ and $\Pr[d^{-}_{v} \geq i]$ based on Eq.~\ref{eq-indegree} for each $i \in [0, \Delta^{-}_v]$ in polynomial time.
Furthermore, when an in-edge $e$ of $v$ is removed, $\Pr[d^{-}_{v} = i]$ is updated by:
\begin{equation}\label{eq-update}
    \Pr[d^{-}_{v}(\neg e) = i] = \frac{\Pr[d^{-}_{v} = i] - p(e) \Pr[d^{-}_{v}(\neg e) = i - 1]}{1 - p(e)},
\end{equation}
for any $i \in [0, \Delta^{-}_v]$ and $p(e) < 1$.
If $p(e) = 1$, we set $\Pr[d^{-}_{v}(\neg e) = i] = \Pr[d^{-}_{v} = i + 1]$ directly for each $i \in [0, \Delta^{-}_v]$.
Similarly, the probability that $v$ has an out-degree of at least $l$ (i.e., $\Pr[d^{+}_{v} \geq l]$) for each $l \in [0, \Delta^{+}_v]$ can be computed by replacing all ``$^-$'' with ``$^+$''.
Since the existence of each edge is assumed to be independent, $\Pr[d^{-}_{v} \geq k \wedge d^{+}_v \geq l]$ is exactly equal to $\Pr[d^{-}_{v} \geq k] \cdot \Pr[d^{+}_{v} \geq l]$.

Given the above results, we present the procedure for $(k, l, \eta)$-core computation in Algorithm~\ref{alg_core}.
First, it calculates $\Pr[d^{-}_{v}(G_{\bm{q}}) \geq k]$ and $\Pr[d^{+}_{v}(G_{\bm{q}}) \geq l]$ of each $v \in V_{\bm{q}}$ using the dynamic program described in the previous paragraph.
Then, it iteratively finds and removes each vertex that does not satisfy the degree constraints of $(k, l, \eta)$-cores in $G_{\bm{q}}$.
To remove a vertex $v$, it deletes each incoming and outgoing edge of $v$ one by one and updates $\Pr[d^{+}_{u}(G_{\bm{q}}) \geq l]$ and $\Pr[d^{-}_{u}(G_{\bm{q}}) \geq k]$ for each in- and out-neighbor $u$ of $v$.
After removing all vertices that do not meet the degree constraints, each weakly connected component in $G_{\bm{q}}$ forms a maximal $(k, l, \eta)$-core of $G_{\bm{q}}$ and is added to $\mathcal{C}$ accordingly.

\begin{algorithm}[t]
    \small
    \caption{$(k,l,\eta)$-Core Computation}
    \label{alg_core}
    \begin{algorithmic}[1]
        \Require Topic-based interaction graph $G_{\bm{q}}$, parameters $k$, $l$, and $\eta$
        \Ensure Set $\mathcal{C}$ of all maximal $(k, l, \eta)$-cores of $G_{\bm{q}}$
        \State Initialize the maximal $(k, l, \eta)$-core set $\mathcal{C} \gets \emptyset$;
        \State Compute $\Pr[d^{-}_{v}(G_{\bm{q}}) \geq k]$ and $\Pr[d^{+}_{v}(G_{\bm{q}}) \geq l]$ for each $v \in V_{\bm{q}}$ using the dynamic program in Eqs.~\ref{eq-dp1} and~\ref{eq-dp2};
        \While{$\exists v \in V_{\bm{q}}$ s.t.~$\Pr[d^{-}_{v}(G_{\bm{q}}) \geq k] \cdot \Pr[d^{+}_{v}(G_{\bm{q}}) \geq l] < \eta$}
            \ForAll{$u \in \mathcal{N}^{-}_{v}(G_{\bm{q}})$}
                \State Update $\Pr[d^{+}_{u}(G_{\bm{q}}) \geq l]$ for deleting $e = (u, v)$ as Eq.~\ref{eq-update};
            \EndFor
            \ForAll{$u \in \mathcal{N}^{+}_{v}(G_{\bm{q}})$}
                \State Update $\Pr[d^{-}_{u}(G_{\bm{q}}) \geq k]$ for deleting $e = (v, u)$ as Eq.~\ref{eq-update};
            \EndFor
            \State Removing vertex $v$ and all its connected edges from $G_{\bm{q}}$;
        \EndWhile
        \ForAll{weakly connected subgraph $C$ in $G_{\bm{q}}$}
            \State $\mathcal{C} \gets \mathcal{C} \cup \{C\}$;
        \EndFor
        \State \Return{$\mathcal{C}$};
    \end{algorithmic}
\end{algorithm}

\vspace{1mm}
\paragraph{Influence Score Calculation}
We next present how to calculate the influence score $\mathbb{I}_{\bm{q}}(v)$ of each vertex $v$ w.r.t.~a topic vector $\bm{q}$.
Since computing $\mathbb{I}_{\bm{q}}(v)$ exactly is \#P-hard \cite{ChenWW10}, we adopt the widely used \emph{reverse influence sampling} (RIS) technique \cite{BorgsBCL14} to effectively approximate $\mathbb{I}_{\bm{q}}(v)$.
In particular, the \emph{reverse reachable} (RR) sets \cite{BorgsBCL14} obtained through RIS are defined as follows.
\begin{definition}[RR Set]\label{def-rrs}
    Given a topic-based interaction graph $G_{\bm{q}} = (V_{\bm{q}}, E_{\bm{q}}, p)$ and a scaling factor $\alpha \in (0, 1]$, $G'$ is a subgraph of $G_{\bm{q}}$ obtained by randomly removing each edge $e \in E_{\bm{q}}$ with probability $1 - pp(e)$, where $pp(e) = \alpha p(e)$.
    An RR set $RR(v, G')$ of a vertex $v$ is the set of all vertices that can reach $v$ in $G'$.
\end{definition}
As indicated in \cite{BorgsBCL14}, the probability that $u$ influences $v$ in the IC model is equal to the probability that $u$ appears in the RR set of $v$.
Based on this, we can estimate $\mathbb{I}_{\bm{q}}(u)$ for each $u \in V_{\bm{q}}$ as follows:
\begin{itemize}
    \item Sample $\theta$ subgraphs $G'_1, \dots, G'_{\theta}$ of $G_{\bm{q}}$ and compute $\theta$ RR sets $RR(u, G'_1),$\\$ \dots,$ $RR(u, G'_{\theta})$ for each vertex $u \in V_{\bm{q}}$;
    \item For each vertex $u \in V_{\bm{q}}$, obtain the average number of RR sets containing $u$ in each subgraph as its influence score, i.e.,
    \begin{equation}
    \label{eq-ris-inf}
        \widetilde{\mathbb{I}}_{\bm{q}}(u) = \frac{1}{\theta} \sum_{i = 1}^{\theta} \left| \{ RR(v, G'_i) \,|\, v \in V_{\bm{q}} \wedge u \in RR(v, G'_i) \} \right|.
    \end{equation}
\end{itemize}
As shown in \cite{BorgsBCL14,TangXS14}, the RIS technique produces a near-optimal solution to the influence maximization problem when $\theta$ is large enough.
However, their results cannot be used for \prob because it requires a bounded error in the estimation of $\mathbb{I}_{\bm{q}}(u)$ for each $u \in V_{\bm{q}}$, rather than for the seed set with the highest influence.
In the case of \prob, we have the following lemma to indicate the number $\theta$ of random subgraphs required for influence estimation.
\begin{lemma}\label{lm-inf-est}
    Let $\theta = O(\frac{1}{\epsilon^2} \log{\frac{n}{\delta}})$ and $G'_1, \dots, G'_{\theta}$ be a set of subgraphs sampled from $G_{\bm{q}}$ as Definition~\ref{def-rrs}. For each vertex $v \in V_{\bm{q}}$, we have $\widetilde{\mathbb{I}}_{\bm{q}}(v) = \mathbb{I}_{\bm{q}}(v) \pm \epsilon n$ with probability at least $1 - \delta$, where $\widetilde{\mathbb{I}}_{\bm{q}}(v)$ is the estimated influence score of $v$ obtained from $G'_1, \dots, G'_{\theta}$.
\end{lemma}
\begin{proof}
    For any vertex $v \in V_{\bm{q}}$, the number $\mathbb{I}_{G'_{i}}(v)$ of RR sets that are generated from $G'_{i}$ and contain $v$ is an unbiased estimator of $\mathbb{I}_{\bm{q}}$ for each $i \in [\theta]$ and $\mathbb{I}_{G'_{i}}(v) \in [0, n]$.
    Since $\widetilde{\mathbb{I}}_{\bm{q}}(v) = \frac{1}{\theta} \sum_{i = 1}^{\theta} \mathbb{I}_{G'_{i}}(v)$, according to Hoeffding's inequality \cite{Hoeffding1963}, we have
    \begin{equation}\label{eq-hoff}
        \Pr[|\widetilde{\mathbb{I}}_{\bm{q}}(v) - \mathbb{I}_{\bm{q}}(v)| > \epsilon n] \leq 2 \exp(-2 \epsilon^2 \theta).
    \end{equation}
    According to Eq.~\ref{eq-hoff}, when $\theta = O(\frac{1}{\epsilon^2}\log{\frac{1}{\delta}})$, $|\widetilde{\mathbb{I}}_{\bm{q}}(v) - \mathbb{I}_{\bm{q}}(v)| \leq \epsilon n$ holds with probability at least $1 - \delta$ for a vertex $v \in V_{\bm{q}}$.
    To ensure that the above inequality holds for all vertices in $V_{\bm{q}}$, we need to set $\theta = O(\frac{1}{\epsilon^2} \log{\frac{n}{\delta}})$ according to the union bound.
\end{proof}

\begin{algorithm}[t]
    \small
    \caption{Online Algorithm for \prob}
    \label{alg_basic}
    \begin{algorithmic}[1]
        \Require Social network $\mathcal{G}$, topic vector $\bm{q}$, parameters $k, l, \eta$
        \Ensure Result $C'_{\bm{q}}$ for the \prob query
        \State Obtain $G_{\bm{q}} = (V_{\bm{q}}, E_{\bm{q}},p)$ from $\mathcal{G}$ w.r.t.~$\bm{q}$ as Definition~\ref{def-intg};
        \State Compute the set $\mathcal{C}$ of maximal $(k, l, \eta)$-cores of $G_{\bm{q}}$ using Algorithm~\ref{alg_core};
        \State Sample $\theta = O(\frac{1}{\epsilon^2} \log{\frac{n}{\delta}})$ subgraphs $G'_1, \dots, G'_{\theta}$ of $G_{\bm{q}}$ as Definition~\ref{def-rrs};
        \State Calculate $\widetilde{\mathbb{I}}_{\bm{q}}(v)$ of each vertex $v \in V_{\bm{q}}$ based on $G'_1, \dots, G'_{\theta}$ as Eq.~\ref{eq-ris-inf};
        \State Initialize $C'_{\bm{q}} \gets \emptyset$ and $\mathbb{I}_{max} \gets 0$;\label{ln-sub-s}
        \State Create a max-heap $\mathcal{H}$ with all $C \in \mathcal{C}$ ordered by $\widetilde{\mathbb{I}}_{\bm{q}}(C)$;
        \While{$\mathcal{H}$ is not empty}
            \State $C \gets \mathcal{H}.\mathtt{pop}()$;
            \State \textbf{if} $\widetilde{\mathbb{I}}_{\bm{q}}(C) > \mathbb{I}_{max}$ \textbf{then} $C'_{\bm{q}} \gets C$ and $\mathbb{I}_{max} \gets \widetilde{\mathbb{I}}_{\bm{q}}(C)$;
            \State Find the vertex $v^* = \argmin_{v \in C} \widetilde{\mathbb{I}}_{\bm{q}}(v)$;
            \State \textsc{Delete}$(v^*, C)$;
            \ForAll{weakly connected subgraph $C'$ in $C$}
                \State $\mathcal{H}.\mathtt{insert}(C')$;
            \EndFor
        \EndWhile
        \State \Return{$C'_{\bm{q}}$};\label{ln-sub-t}
        \Statex
        \Procedure{Delete}{$v^*, C$}
            \ForAll{$u \in \mathcal{N}^{-}_{v^*}(C)$ \label{ln-del-s}}
                \State Update $\Pr[d^{+}_{u}(C) \geq l]$ for deleting $e = (u, v^*)$ as Eq.~\ref{eq-update};
            \EndFor
            \ForAll{$u \in \mathcal{N}^{+}_{v^*}(C)$}
                \State Update $\Pr[d^{-}_{u}(C) \geq k]$ for deleting $e = (v^*, u)$ as Eq.~\ref{eq-update};
            \EndFor
            \State Remove $v^*$ and all its connected edges from $C$; \label{ln-del-t}
            \If{$\exists u \in C$ s.t.~$\Pr[d^{-}_{u}(C) \geq k] \cdot \Pr[d^{+}_{u}(C) \geq l] < \eta$}
                \State \textsc{Delete}($u, C$);
            \EndIf
        \EndProcedure
    \end{algorithmic}
\end{algorithm}

\paragraph{Online Algorithm for \prob}
Finally, we present the online search algorithm to obtain the result of a \prob query for the topic vector $\bm{q}$ and parameters $k, l, \eta$ in Algorithm~\ref{alg_basic}.
The preparation steps before the search procedure are \emph{(i)} extracting $G_{\bm{q}}$ from the social network $\mathcal{G}$ w.r.t.~$\bm{q}$, \emph{(ii)} finding all maximal $(k, l, \eta)$-cores of $G_{\bm{q}}$ using Algorithm~\ref{alg_core}, and \emph{(iii)} estimating the influence score $\widetilde{\mathbb{I}}_{\bm{q}}(v)$ of each vertex $v$ using RIS.
The search procedure involves enumerating all $(k, l, \eta)$-cores and maintaining the one with the highest (estimated) influence score among them.
Specifically, it begins with the maximal $(k, l, \eta)$-cores in $G_{\bm{q}}$, which are kept in a max-heap $\mathcal{H}$ in descending order of their influence scores.
Then, it proceeds iteratively in the following three steps: \emph{(i)} retrieve the $(k, l, \eta)$-core $C$ with the highest influence score in $\mathcal{H}$ and compare its influence score with the most influential community so far; if the influence score of $C$ is greater, set $C$ as the most influential community so far; (2) find the vertex $v^*$ with the lowest influence score in $C$, remove $v^*$ from $C$, and identify the weakly connected subgraphs of $C$ that are still $(k,l,\eta)$-cores after removing $v^*$; (3) add these $(k,l,\eta)$-cores to $\mathcal{H}$ for further checking.
The above steps are repeated until $\mathcal{H}$ is empty.
After that, the $(k, l, \eta)$-core with the highest influence score among all checked ones is returned as the result of the \prob query.

\begin{figure}[t]
    \centering
    \begin{subfigure}{0.95\textwidth}
        \includegraphics[width=\linewidth]{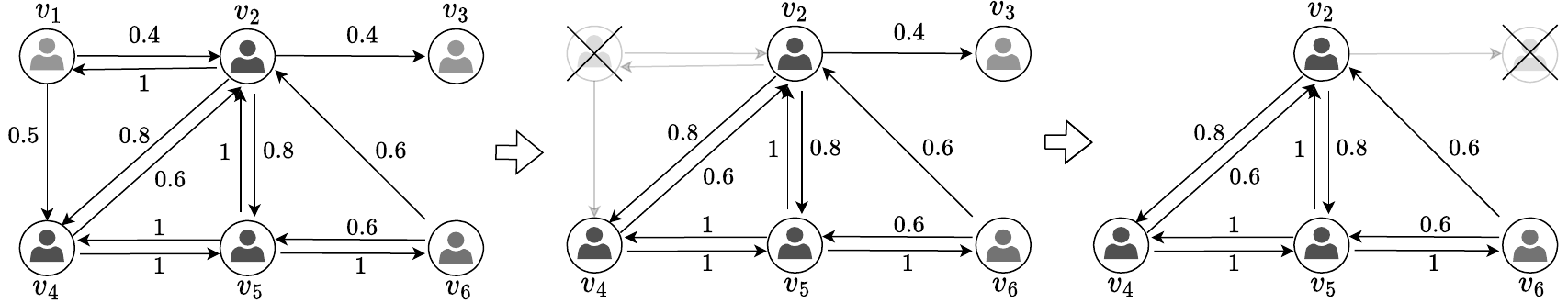}
        \caption{Compute the maximal $(1,2,0.36)$-core set $\mathcal{C}$ on $G_{\bm{q}}$ w.r.t.~$\bm{q} = (0.5, 0.5)$}
        \label{fig_2example:left}
    \end{subfigure}
    \begin{subfigure}{0.95\textwidth}
        \includegraphics[width=\linewidth]{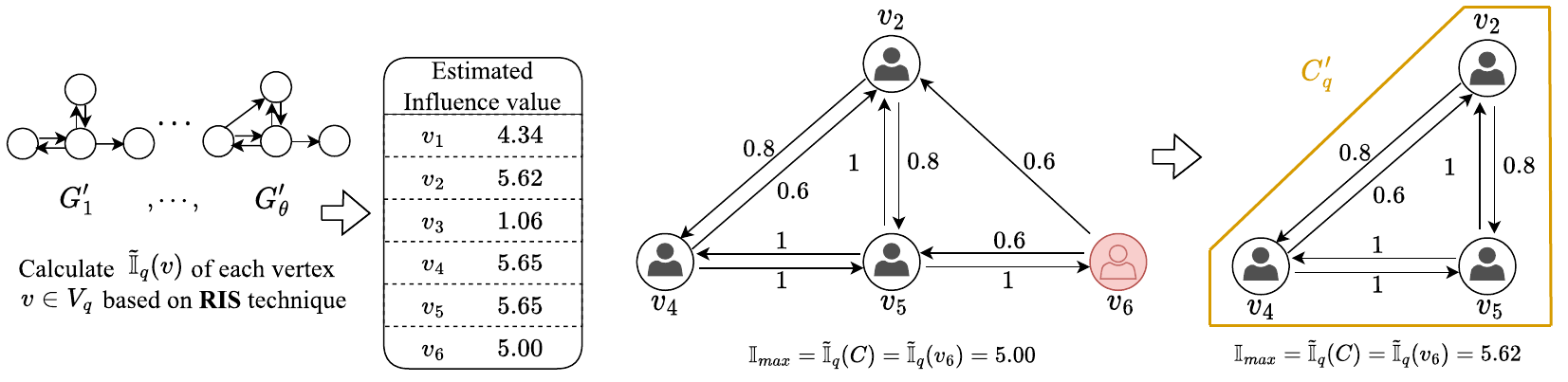}
        \caption{Search $C'_{\bm{q}}$ based on the maximal $(1,2,0.36)$-core set $\mathcal{C}$ and influence scores}
        \label{fig_2example:right}
    \end{subfigure}
    \caption{Running examples of the online algorithm for \prob with $\bm{q} = (0.5, 0.5)$, $k = 1$, $l = 2$, and $\eta = 0.6$.}
    \label{fig_2example}
\end{figure}

\begin{example}
    For $G_{\bm{q}}$ with $\bm{q} = (0.5, 0.5)$ in Fig.~\ref{fig_1example:right}, we show how to compute its maximal $(1, 2, 0.36)$-core set $\mathcal{C}$ using Algorithm~\ref{alg_core} in Fig.~\ref{fig_2example:left}.
    In the first iteration, since $\Pr[d^{-}_{v_1}(G_{\bm{q}}) \geq 1] \cdot \Pr[d^{+}_{v_1}(G_{\bm{q}}) \geq 2] = 0.2 \leq \eta = 0.36$, $v_1$ and all its connected edges are removed from $G_{\bm{q}}$.
    Then, in the second iteration, $v_3$ and its connected edges are removed.
    After that, the remaining vertices do not meet the deletion criteria, thus constituting the maximal $(1, 2, 0.36)$-core set $\mathcal{C} = \{C_1 = \{v_2,v_4,v_5,v_6\}\}$.

    Then, we show how to find the result of \prob for $\bm{q} = (0.5, 0.5)$ in Fig.~\ref{fig_2example:right}.
    First, we use the RIS technique to calculate the estimated influence score of each vertex in $G_{\bm{q}}$.
    Next, we add $C_1$ to $\mathcal{H}$, as $C_1$ is the only weakly connected component in $\mathcal{C}$.
    We retrieve $C_1$ from $\mathcal{H}$. Since $\mathbb{I}_{max}$ is 0, we update $C'_{\bm{q}} = C_1$, $\mathbb{I}_{max} \approx 5$, and $v^* = v_6$.
    Subsequently, we remove $v_6$ and add $C_2 = \{v_2,v_4,v_5\}$ to $\mathcal{H}$. $C_2$ is removed from $\mathcal{H}$ immediately and serves as $C'_{\bm{q}}$.
    We have $\mathbb{I}_{max} \approx 5.6$ and $v^* = v_2$ accordingly.
    After removing $v_2$, $\mathcal{H}$ is empty.
    Therefore, $C'_{\bm{q}} = \{v_2, v_4, v_5\}$ is returned as the result of \prob for $\bm{q} = (0.5, 0.5)$.
\end{example}

\vspace{1mm}
\paragraph{Theoretical Analysis}
We present the theoretical bound and time complexity of the online algorithm in Algorithm~\ref{alg_basic} for \prob in the following theorem.
\begin{theorem}\label{thm-basic}
    Algorithm~\ref{alg_basic} returns a $(k, l, \eta)$-core $C'_{\bm{q}}$ of $G_{\bm{q}}$ such that $\mathbb{I}_{\bm{q}}(C'_{\bm{q}}) \geq \mathbb{I}_{\bm{q}}(C^{*}_{\bm{q}}) - 2 \epsilon n$ with probability at least $1 - 2 \delta$ in $O(\frac{m}{\epsilon^2} \log{\frac{n}{\delta}})$ time.
\end{theorem}
\begin{proof}
    From Lemma~\ref{lm-inf-est}, we have $|\widetilde{\mathbb{I}}_{\bm{q}}(C) - \mathbb{I}_{\bm{q}}(C)| \leq \epsilon n$ with probability $1 - \frac{\delta |C|}{n}$ for any subgraph $C$ in $G_{\bm{q}}$.
    Accordingly, with probabilities $1 - \frac{\delta |C^{*}_{\bm{q}}|}{n}$ and $1 - \frac{\delta |C'_{\bm{q}}|}{n}$, we have $|\widetilde{\mathbb{I}}_{\bm{q}}(C^{*}_{\bm{q}}) - \mathbb{I}_{\bm{q}}(C^{*}_{\bm{q}})| \leq \epsilon n$ and $|\widetilde{\mathbb{I}}_{\bm{q}}(C'_{\bm{q}}) - \mathbb{I}_{\bm{q}}(C'_{\bm{q}})| \leq \epsilon n$, respectively.
    Since $C^{*}_{\bm{q}}$ must be a $(k, l, \eta)$-core of $G_{\bm{q}}$, we have $\widetilde{\mathbb{I}}_{\bm{q}}(C^{*}_{\bm{q}}) \leq \widetilde{\mathbb{I}}_{\bm{q}}(C'_{\bm{q}})$ based on the procedure of Algorithm~\ref{alg_basic}.
    Therefore, we have $\mathbb{I}_{\bm{q}}(C'_{\bm{q}}) \geq \mathbb{I}_{\bm{q}}(C^{*}_{\bm{q}}) - 2 \epsilon n$ with probability $1 - \frac{\delta (|C^*_{\bm{q}}| + |C'_{\bm{q}}|)}{n} \geq 1 - 2\delta$ since $|C^*_{\bm{q}}|, |C'_{\bm{q}}| \leq n$.
    
    The total time to compute $\Pr[d^{-}_{v}(G_{\bm{q}}) \geq k]$ and $\Pr[d^{+}_{v}(G_{\bm{q}}) \geq l]$ for each vertex $v \in V_{\bm{q}}$ using dynamic programming is $O( \sum_{v \in V_{\bm{q}}} (k \cdot$ $\Delta^{-}_{v}(G_{\bm{q}}) + l \cdot \Delta^{+}_{v}(G_{\bm{q}})))$.
    When a vertex $v$ is deleted, it takes $O(k \cdot \Delta^{-}_{v}(G_{\bm{q}}) + l \cdot \Delta^{+}_{v}(G_{\bm{q}}))$ time to update the probabilities of its neighbors. Subsequently, to obtain weakly connected components, we simply run the BFS algorithm in $O(n + m)$ time.
    In summary, the time complexity of Algorithm ~\ref{alg_core} is $O((k+l) m)$.
    Algorithm~\ref{alg_basic} takes $O(mz)$ time to obtain $G_{\bm{q}}$ and $m \theta = O(\frac{m}{\epsilon^2} \log{\frac{n}{\delta}})$ time to estimate the influence score of each vertex using RIS.
    Finally, the iterative online search procedure takes the same $O((k+l) m)$ time as Algorithm~\ref{alg_core}.
    Therefore, the time complexity of Algorithm~\ref{alg_basic} is
    \begin{equation*}
        O(mz) + O((k+l) m) + O(\frac{m}{\epsilon^2} \log{\frac{n}{\delta}}) = O\Big(m (k + l + z + \frac{1}{\epsilon^2}\log{\frac{n}{\delta}}) \Big).
    \end{equation*}
    Since $k$, $l$, and $z$ are typically much smaller than $\frac{1}{\epsilon^2} \log{\frac{n}{\delta}}$, the time complexity of Algorithm~\ref{alg_basic} can be simplified as $O(\frac{m}{\epsilon^2} \log{\frac{n}{\delta}})$.
\end{proof}

%% file: sections-springer/5-index_algorithm.tex
\section{Index-based Algorithm}
\label{sec_index}

In this section, we propose an index-based heuristic algorithm to improve the efficiency of the online algorithm for \prob queries.
In particular, we build an index consisting of (1) the \emph{topic-aware uncertain core list} (\cidx) to efficiently generate candidate communities for any \prob query and (2) the \emph{topic-aware influence estimation tree} (\iidx) to estimate the influence scores of candidate communities, which jointly obtain the query results.
Next, we will describe the construction procedures of \cidx and \iidx in Sections~\ref{subsec-ticu} and~\ref{sub_TRIS}, respectively, and the index-based query algorithm in Section~\ref{subsec-query}.

\subsection{TUC-List Construction}
\label{subsec-ticu}

Since there exists an infinite number of topic vectors, each of which corresponds to a distinct interaction graph, it is impossible to precompute and maintain all $(k, l, \eta)$-cores in the interaction graph for any possible \prob query.
To address this issue, we propose to build a \cidx by computing a supergraph of any possible interaction graph and maintaining the $(k, l, \eta)$-cores on the supergraph for different combinations of $k$, $l$, and $\eta$.
In this way, upon receiving any \prob query with the topic vector $\bm{q}$ and parameters $k, l, \eta$, we quickly find the supergraphs of the $(k, l, \eta)$-cores in $G_{\bm{q}}$ from the \cidx and then perform an online search procedure only on these candidates rather than on the entire $G_{\bm{q}}$ to efficiently obtain the result of the \prob query.

To construct the \cidx, we first find a graph $G^{\perp} = (V^{\perp}, E^{\perp}, p^{\perp})$ based on $\mathcal{G}$ that is a supergraph of an interaction graph $G_{\bm{q}}$ w.r.t.~any topic vector $\bm{q}$.
First, we show how to construct such a supergraph $G^\perp$ based on the social network $\mathcal{G} = (\mathcal{V}, \mathcal{E}, \omega)$.
For a given $\mathcal{G}$, we build an uncertain graph $G^{\perp} = (V^{\perp}, E^{\perp}, p^{\perp})$ as follows: \emph{(i)} set $V^{\perp} = \mathcal{V}$ and $E^{\perp} = \mathcal{E}$; \emph{(ii)} for each edge $e \in \mathcal{E}$, calculate $p^{\perp}(e) = f(\omega_{max}(e))$, where $\omega_{max}(e) = \max_{i \in [z]} \omega_{i}(e)$.
For any $\bm{q} = (q_1, \dots, q_z) \in [0, 1]^{z}$ with $\sum_{i = 1}^{z} q_i = 1$, we have $\langle \bm{\omega}(e), \bm{q} \rangle \leq \omega_{max}(e)$.
Then, since $f(\cdot)$ is monotonic, we have $p^{\perp}(e) \geq p(e)$ for all $e \in E_{\bm{q}}$.
Therefore, $G^{\perp}$ must be a supergraph of $G_{\bm{q}}$ w.r.t.~any $\bm{q}$.
Furthermore, due to Definition~\ref{def-core}, for any $(k, l, \eta)$-core $C$ in $G_{\bm{q}}$, a subgraph induced by the same set of vertices $V_C$ must also be a $(k, l, \eta)$-core $C^{\perp}$ in $G^\perp$.

After obtaining the graph $G^{\perp}$, the next step of \cidx construction is to maintain all its $(k, l, \eta)$-cores for different combinations of $k$, $l$, and $\eta$.
In particular, the \cidx is organized as a two-dimensional array $\mathcal{L}$, where each cell $\mathcal{L}[k, l]$ keeps all $(k,l,\eta)$-cores for a fixed pair of $k, l$ and all different values of $\eta \in (0, 1)$.
The width and height of $\mathcal{L}$ are set to the maximum values $k_{\text{max}}, l_{\text{max}}$ of $k, l$ such that a $(k, 0, 0)$-core and a $(0, l, 0)$-core exist in $G^\perp$, respectively.
The values of $k_{\text{max}}$ and $l_{\text{max}}$ can be obtained by performing any D-core decomposition method on $G^{\perp}$ \cite{GiatsidisTV13,LiaoLJHXC22}.
Then, to identify each value of $\eta$ corresponding to a distinct $(k,l,\eta)$-core in $G^\perp$, we generalize the notion of \emph{$\eta$-threshold} \cite{YangWQZCL19} for each vertex $v \in V^{\perp}$ to directed graphs:
\begin{definition}[$\eta$-Threshold]\label{def-threshold}
    For an uncertain graph $G^{\perp}$ and two integers $k, l$, the $\eta$-threshold $\eta_{k,l}(v)$ for a vertex $v \in V^{\perp}$ is the maximum value of $\eta$ such that there is a $(k, l, \eta)$-core containing $v$ in $G^{\perp}$.
\end{definition}
As analyzed in \cite{YangWQZCL19}, since a $(k, l, \eta)$-core is also a $(k, l, \eta')$-core for any $\eta' \leq \eta$, it suffices to enumerate all $\eta$-thresholds of vertices to identify all distinct $(k, l, \eta)$-cores of a graph for fixed $k, l$.
The method for $\eta$-threshold computation is similar to that for $(k, l, \eta)$-core computation in Algorithm~\ref{alg_core}.
Specifically, the computation procedure is as follows:
\emph{(i)} Initialize $\eta_{cur} = 0$ and $\eta_{k,l}(v) = 0$ for each $v \in V^{\perp}$, remove each $v$ with $\Pr[d^{-}_{v}(G^\perp) \geq k \wedge d^{+}_{v}(G^\perp) \geq l] = 0$ and its adjacent edges from $G^\perp$;
\emph{(ii)} Find $v^* = \argmin_{v \in V^{\perp}} \Pr[d^{-}_{v}(G^\perp) \geq k \wedge d^{+}_{v}(G^\perp) \geq l]$, update $\eta_{cur} = \max(\eta_{cur}, \Pr[d^{-}_{v^*}(G^\perp$\\$) \geq k \wedge d^{+}_{v^*}(G^\perp) \geq l])$, and set $\eta_{k,l}(v^*) = \eta_{cur}$;
\emph{(iii)} Remove $v^*$ and its adjacent edges and update $\Pr[d^{-}_{u}(G^\perp) \geq k \wedge d^{+}_{u}(G^\perp) \geq l]$ for each $u \in \mathcal{N}^{-}_{v^*}(G^{\perp}) \cup \mathcal{N}^{+}_{v^*}(G^{\perp})$.
Steps \emph{(ii)}--\emph{(iii)} are repeated until $G^\perp$ is empty.

\begin{figure}[t]
    \centering
    \begin{subfigure}[t]{0.45\textwidth}
        \includegraphics[width=\linewidth]{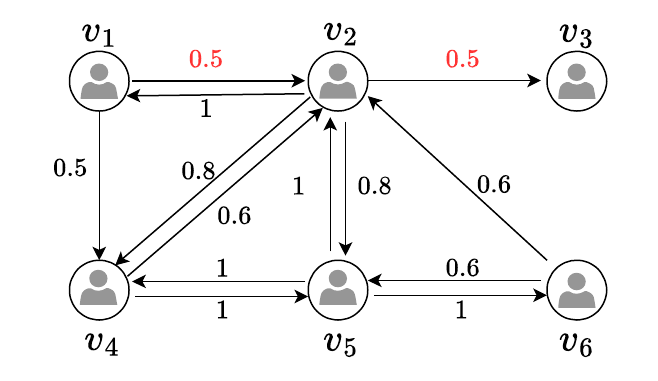}
        \caption{Supergraph $G^{\perp}$ of $\mathcal{G}$}
        \label{fig-eta_threshold:left}
    \end{subfigure}
    \begin{subfigure}[t]{0.45\textwidth}
        \includegraphics[width=\linewidth]{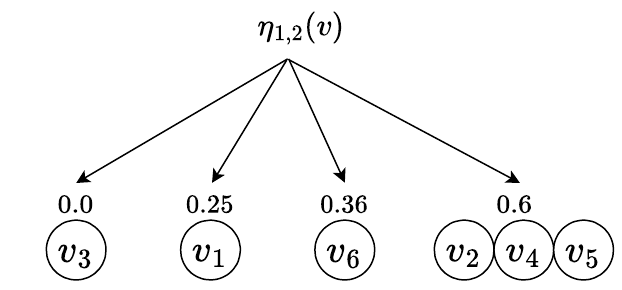}
        \caption{$\eta$-thresholds of vertices in $G^{\perp}$}
        \label{fig-eta_threshold:right}
    \end{subfigure}
    \caption{Examples of a supergraph and the $\eta$-thresholds of vertices in the social network $\mathcal{G}$ of Fig.~\ref{fig_1example:left} when $k = 1$ and $l = 2$.}
    \label{fig-eta_threshold}
\end{figure}

\begin{example}
    Fig.~\ref{fig-eta_threshold:left} illustrates the supergraph $G^\perp$ of the social network $\mathcal{G}$ in Fig.~\ref{fig_1example:left}.
    Then, we present the $\eta$-thresholds of vertices in $G^\perp$.
    Since $v_6$ is not in any $(1, 2)$-core, we first remove $v_6$ and its adjacent edges from $G^\perp$ and set $\eta_{1,2}(v_6) = 0$.
    Then, in the first iteration, $v^* = v_1$ with $\Pr[d^{-}_{v_1}(G^\perp) \geq 1 \wedge d^{+}_{v_1}(G^\perp) \geq 2] = 0.25$ is found.
    We update $\eta_{\text{cur}} = 0.25$, set $\eta_{1,2}(v_1) = 0.25$, and remove $v_1$ and its adjacent edges from $G^\perp$.
    In the second iteration, we have $v^* = v_6$ and $\eta_{\text{cur}} = 0.36$. We set $\eta_{1,2}(v_6) = 0.36$ and remove $v_6$ and its adjacent edges from $G^\perp$.
    In the third iteration, we have $v^* = v_4$, $\eta_{\text{cur}} = 0.6$, and $\eta_{1,2}(v_4) = 0.6$.
    In the last two iterations, we have $v^* = v_2$ and $v_5$, respectively, and find that they cannot form any $(1, 2)$-core.
    Therefore, we set $\eta_{1,2}(v_2) = \eta_{1,2}(v_5) =\eta_{\text{cur}}=0.6$.
    Finally, since $G^\perp$ is empty, the computation procedure is finished.
\end{example}

When the $\eta$-thresholds of all vertices for fixed $k, l$ are computed, the final step of \cidx construction is to store these thresholds and their corresponding vertices within a cell $\mathcal{L}[k, l]$ as a sub-list $\mathcal{I}$.
First, the vertices with $\eta$-thresholds equal to $0$ are removed because they are not in any $(k,l,\eta)$-core with $\eta > 0$.
Then, the remaining $\eta$-thresholds are sorted ascendingly.
Here, we use all distinct $\eta$-threshold values as keys in the sub-list $\mathcal{I}$, and assign each vertex $v$ to the key $\mathcal{I}[j]$ with $\eta_{k,l}(v) = \mathcal{I}[j]$ as a tuple.

\begin{algorithm}[t]
    \small
    \caption{\cidx Construction}
    \label{alg_TUC_create}
    \begin{algorithmic}[1]
        \Require Social network $\mathcal{G}$
        \Ensure \cidx $\mathcal{L}$
        \State Construct the supergraph $G^{\perp}$ from $\mathcal{G}$;
        \State Run any D-core decomposition algorithm \cite{LiaoLJHXC22} to compute $k_{max}$ and $l_{max}$;
        \State Initialize a two-dimensional array $\mathcal{L}$ of size $k_{max} \times l_{max}$;
        \For{$k = 1$ \textbf{to} $k_{max}$}
            \For{$l = 1$ \textbf{to} $l_{max}$}
                \State Run \textsc{$\eta$-Threshold}($k, l, G^{\perp}$) and sort all $\eta$-thresholds ascendingly (with duplicates and zeros removed) as the keys of a sub-list $\mathcal{I}$;
                \ForAll{$v \in V^{\perp}$}
                    \State Add $v$ as a tuple w.r.t.~the key $\mathcal{I}[j]$ if $\eta_{k,l}(v) = \mathcal{I}[j]$;
                \EndFor
                \State Add $\mathcal{I}$ to $\mathcal{L}[k, l]$;
            \EndFor
        \EndFor
        \State \Return{$\mathcal{L}$}
        \Statex
        \Procedure{$\eta$-Threshold}{$k, l, G^{\perp}$}
            \State Initialize $\eta_{cur} \gets 0$ and $\eta_{k,l}(v) \gets 0$ for all $ v \in V^{\perp}$;
            \ForAll{$v \in V^{\perp}$ with $\Pr[d^{-}_{v}(G^{\perp}) \geq k \wedge d^{+}_{v}(G^{\perp}) \geq l] = 0$}
                \State Remove $v$ and its adjacent edges from $G^{\perp}$;
            \EndFor
             \While{$V^{\perp} \neq \emptyset$}
                \State Find $v^* = \arg\min_{v \in V^{\perp}} \Pr[d^{-}_{v}(G^{\perp}) \geq k \wedge d^{+}_{v}(G^{\perp}) \geq l]$;     
                \State Set $\eta_{cur} \gets \max(\eta_{cur}, \Pr[d^{-}_{v^*}(G^{\perp}) \geq k \wedge d^{+}_{v^*}(G^{\perp}) \geq l])$ and $\eta_{k,l}(v^*) \gets \eta_{cur}$;
                \State Invoke \textsc{Delete}$(v^*, G^{\perp})$ (Lines~\ref{ln-del-s}--\ref{ln-del-t}) of Algorithm \ref{alg_basic};
             \EndWhile
        \EndProcedure
    \end{algorithmic}
\end{algorithm}

The pseudocode of \cidx construction is presented in Algorithm~\ref{alg_TUC_create}.
First, it obtains $G^\perp$ in $O(mz)$ time.
Then, the D-core decomposition algorithm in \cite{LiaoLJHXC22} runs in $O(m \cdot \Delta_{max}(G^{\perp}))$ time, where $\Delta_{max}(G^{\perp})$ is the maximum (in-/out-)degree of any vertex in $G^{\perp}$.
Next, computing and sorting the $\eta$-thresholds of all vertices for fixed $k, l$ takes $O(n \log{n} + m (k + l))$ and $O(n \log{n})$ time, respectively.
Subsequently, it also requires $O(n \log{n})$ time to assign all vertices to a sub-list.
Therefore, the overall time complexity of Algorithm~\ref{alg_TUC_create} is $O\left(n k_{max} l_{max} \log{n}  + m \Delta_{max}(G^{\perp})\right)$.
Additionally, the space complexity of \cidx is $O(n k_{max} l_{max})$, because each cell $\mathcal{L}[k, l]$ keeps up to $n$ $\eta$-thresholds and vertices.

\begin{figure}[t]
    \centering
    \begin{subfigure}[t]{0.6\textwidth}
        \includegraphics[width=\linewidth]{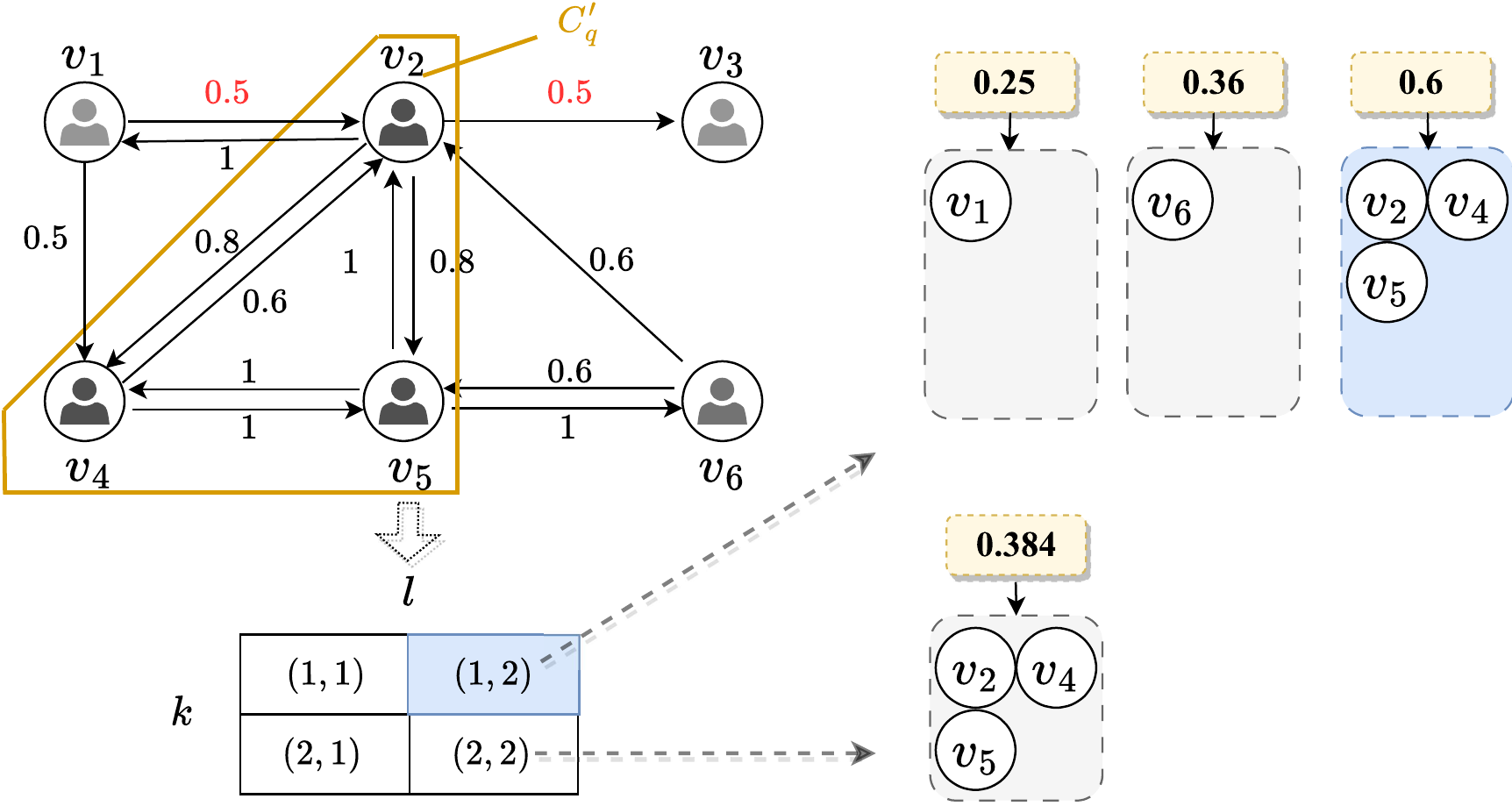}
        \caption{\cidx}
        \label{fig:TUC-list}
    \end{subfigure}
    \hfill
    \begin{subfigure}[t]{0.35\textwidth}
        \includegraphics[width=\linewidth]{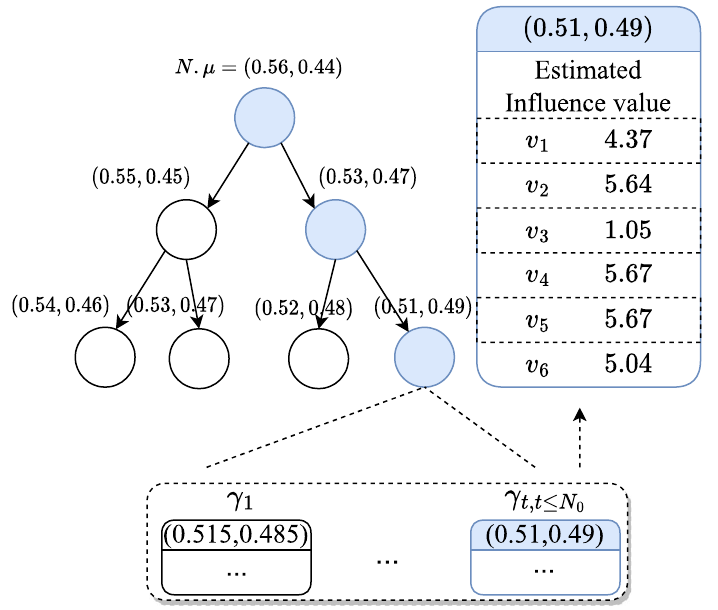}
        \caption{\iidx}
        \label{fig:TIE-tree}
    \end{subfigure}
    \caption{Illustration of the \cidx and \iidx.}
    \label{fig:Index}
\end{figure}

\subsection{TIE-Tree Construction}
\label{sub_TRIS}

In this subsection, we show how to construct the \iidx to efficiently estimate the influence score of a vertex or community for any \prob query.
The main idea behind the \iidx is to utilize the RIS technique \cite{BorgsBCL14,TangXS14} to precompute and maintain the influence scores w.r.t.~a specific set of topic vectors.
Then, for each \prob query with a topic vector $\bm{q}$, we find a pre-specified topic vector that is the closest to $\bm{q}$ and directly use its corresponding influence scores to estimate those for $\bm{q}$.
Here, we need to measure the ``similarity'' between two topic vectors in terms of social influence.
Following the existing literature on topic-aware influence maximization \cite{AslayBBB14,ChenFLFTT15}, we adopt the angular distance of two topic vectors $\bm{q}, \bm{q}'$, that is, $sim(\bm{q},\bm{q}') = \frac{\langle \bm{q}, \bm{q}' \rangle}{\lVert{\bm{q}}\rVert \cdot \lVert{\bm{q}'}\rVert}$, as the similarity measure.
When $sim(\bm{q},\bm{q}')$ is higher, the influence score of any vertex $v$ w.r.t.~$\bm{q}'$ tends to be closer to that w.r.t.~$\bm{q}$.

\begin{algorithm}[t]
    \small
    \caption{\iidx Construction}
    \label{alg_TIE_Construct}
    \begin{algorithmic}[1]
        \Require Social network $\mathcal{G}$, set $\Gamma = \{\bm{\gamma}_1, \dots, \bm{\gamma}_h\}$ of topic vectors
        \Ensure \iidx $\mathcal{T}$
        \State Initialize an empty cone tree $\mathcal{T}$;
        \State Call \textsc{Split}($\mathcal{T}.root$, $\Gamma$);
        \State \Return{$\mathcal{T}$};
        \Statex
        \Procedure{Split}{Node $N$, Set $S$}
            \State $N.S \gets S, N.\bm{\mu} \gets \frac{1}{|S|} \sum_{\bm{\gamma} \in S} \bm{\gamma}, N.\theta \gets \max_{\bm{\gamma} \in S} \theta(N.\bm{\mu}, \bm{\gamma})$;
            \If{$|S| \leq N_0$}
                \ForAll{$\bm{\gamma} \in S$}
                    \State Obtain $G_{\bm{\gamma}}$ w.r.t.~$\bm{\gamma}$ from $\mathcal{G}$;
                    \State Build a list to store $\widetilde{\mathbb{I}}_{\bm{\gamma}}(v)$ for each $v \in V_{\bm{\gamma}}$ computed by RIS \cite{BorgsBCL14,TangXS14};
                \EndFor
            \Else
                \State Pick a vector $\bm{\gamma}_x$ randomly from $S$ and find $\bm{\gamma}_y \gets \max_{\bm{\gamma} \in S} \theta(\bm{\gamma}_x, \bm{\gamma})$;
                \State Divide $S$ into $S_x \gets \{\bm{\gamma} \in S : \theta(\bm{\gamma}_x, \bm{\gamma}) \leq \theta(\bm{\gamma}_y, \bm{\gamma})\}$ and $S_y \gets S \setminus S_x$;
                \State Create the two child nodes $N.lc$ and $N.rc$ of $N$;
                \State Call \textsc{Split}$(N.lc, S_x)$ and \textsc{Split}$(N.rc, S_y)$;
            \EndIf
        \EndProcedure
    \end{algorithmic}
\end{algorithm}

The first step in the construction of \iidx is to select a set of $h$ topic vectors, denoted as $\Gamma = \{\bm{\gamma}_1, \dots, \bm{\gamma}_h\}$, which can best cover all possible topic vectors of the \prob queries.
To this end, we randomly sample a set of ``representative'' topic vectors (e.g., based on historical query logs) and run the k-means++ \cite{ArthurV07} clustering algorithm on the vectors sampled to identify the $h$ cluster centers as $\Gamma$.
Then, we precompute the influence scores of the vertices in $\mathcal{G}$ w.r.t.~each topic vector in $\Gamma$ and store them as a list.
To efficiently find the vector $\bm{\gamma}^* \in \Gamma$ that is most similar to the topic vector $\bm{q}$, we build a cone tree \cite{RamG12}, a binary data-partitioning tree specific to high-dimensional similarity search w.r.t.~the angular distance, for the set of vectors $\Gamma$.
The construction of a cone tree begins by assigning all vectors in $\Gamma$ to the root node.
Then, it recursively splits a node into two child nodes by picking two vectors as pivots and assigning the remaining vectors to a pivot closer to them in terms of angular distance.
Each node in the cone tree is associated with its assigned vectors and an open cone denoted by a vector corresponding to its axis and an angle corresponding to the maximum
angle made by any of its assigned vectors with the axis at the origin.
When a node contains at most $N_0$ vectors, the node will no longer be split, and the estimated influence scores of vertices w.r.t.~each vector within the node are maintained as a table in this node. The \iidx is effective for social networks of different types with varying topic distributions because the partitioning of the tree nodes is based on the extent
to which the $h$ topic vectors best cover all possible topic vectors
of the \prob queries and thus is adaptive to different topic distributions. This will be confirmed by experimental results in Section \ref{sec_index}.
The pseudocode of \iidx construction is shown in Algorithm~\ref{alg_TIE_Construct}.
Calculating the influence scores of vertices w.r.t.~the $h$ topic vectors in $\Gamma$ takes $O(\frac{mh}{\varepsilon^2} \log{\frac{n}{\delta}})$ time.
Then, it takes $O(hz \log h)$ time to split each node, and thus the time complexity of building a cone tree is $O(hz \log^2{h})$.
The \iidx uses $O(n h)$ space to store all influence scores and $O(hz \log h)$ space to maintain the cone tree.
Thus, the space complexity of the \iidx is $O(h (n + z \log h))$.

\subsection{Index-based TAMICS Query Processing}
\label{subsec-query}

Next, we show how to efficiently answer a \prob query using the \cidx and \iidx.
Specifically, it first performs a top-down traversal on the \iidx to find a topic vector $\bm{\gamma}^* \in \Gamma$ that is similar to the query vector $\bm{q}$.
The influence scores of vertices w.r.t.~$\bm{\gamma}^*$ are used to estimate those w.r.t.~$\bm{q}$ in the remaining steps.
Then, it obtains the set $\mathcal{C}'$ of candidate $(k, l, \eta)$-cores from $\mathcal{L}$.
This involves collecting the set $V'$ of vertices from each $\mathcal{L}[k, l].\mathcal{I}[j']$ such that $\mathcal{L}[k, l].\mathcal{I}[j'] \geq \eta$, that is, all the vertices contained by any $(k, l, \eta)$-core in $G^{\perp}$ and thus $G_{\bm{q}}$.
Next, it induces a subgraph $G'_{\bm{q}}$ from $V'$ and computes its edge probabilities as Definition~\ref{def-intg}.
After obtaining $G'_{\bm{q}}$, it performs the same search procedure as Algorithm~\ref{alg_core} to find the set $\mathcal{C}'$ of candidate $(k, l, \eta)$-cores.
Finally, the final result $C'_{\bm{q}}$ for the \prob query is returned by running Lines~\ref{ln-sub-s}--\ref{ln-sub-t} of Algorithm~\ref{alg_basic} on $\mathcal{C}'$.

\begin{algorithm}[t]
    \small
    \caption{Index-based Algorithm for \prob}
    \label{alg_Index_query}
    \begin{algorithmic}[1]
        \Require Social network $\mathcal{G}$, topic vector $\bm{q}$, parameters $k, l, \eta$, \cidx $\mathcal{L}$, \iidx $\mathcal{T}$
        \Ensure Result $C'_{\bm{q}}$ for \prob query
        \State Initialize the node $cur$ to visit as $\mathcal{T}.root$;
        \While{$cur$ is a non-leaf node}
            \State \textbf{if} $\theta(cur.lc.\bm{\mu}, \bm{q}) \leq \theta(cur.rc.\bm{\mu}, \bm{q})$ \textbf{then} $cur \gets cur.lc$ \textbf{else} $cur \gets cur.rc$;
        \EndWhile
        \State $\bm{\gamma}^* \gets \argmin_{\bm{\gamma} \in cur.S} \theta(\bm{\gamma}, \bm{q})$;
        \State Set $\widetilde{\mathbb{I}}_{\bm{q}}(v) \gets \widetilde{\mathbb{I}}_{\bm{\gamma}^*}(v)$ for each vertex $v \in \mathcal{V}$;
        \State Find $j^* \gets \argmin_{1 \leq j \leq |\mathcal{L}[k, l].\mathcal{I}|} \mathcal{L}[k, l].\mathcal{I}[j] \geq \eta$;
        \State Obtain the set $V'$ of vertices included in all sub-lists $\mathcal{L}[k, l].\mathcal{I}[j']$ with $j' \geq j^*$;
        \State Set $E' \gets \{e \,|\, e = (u, v) \in \mathcal{E}: u \in V' \wedge v \in V'\}$ with $p(e)$ as Definition~\ref{def-intg};
        \State Compute the set $\mathcal{C}'$ of maximal $(k, l, \eta)$-cores of $G'_{\bm{q}} = (V', E', p)$ using Algorithm~\ref{alg_core};
        \State Run Lines~\ref{ln-sub-s}--\ref{ln-sub-t} of Algorithm~\ref{alg_basic} on $\mathcal{C}'$ to compute $C'_{\bm{q}}$;
        \State \Return{$C'_{\bm{q}}$};
    \end{algorithmic}
\end{algorithm}

The pseudocode of the index-based algorithm for \prob queries is shown in Algorithm~\ref{alg_Index_query}.
It first takes $O(z (\log h + N_0))$ time to find $\bm{\gamma}^*$ and its corresponding influence scores from the \iidx.
Using a binary search to find the index $j^*$ requires $O(\log n)$ time, and collecting the vertices in $G'_{\bm{q}}$ from the \cidx requires $O(n')$ time, where $n'$ is the number of vertices in $G'_{\bm{q}}$.
The remaining steps are similar to Algorithm~\ref{alg_basic} and take $O(m' (z + k + l))$ time in total, where $m'$ is the number of edges in $G'_{\bm{q}}$.
In summary, the time complexity of Algorithm~\ref{alg_Index_query} is also $O(m' (k + l + z))$.
Compared to Algorithm~\ref{alg_basic}, Algorithm~\ref{alg_Index_query} has a much lower time complexity because (1) it does not need to perform the time-consuming RIS for every query vector and (2) we typically have $m' \ll m$ in practice.
Since influence estimates based on angular similarity cannot provide any approximation bound, the result returned by $C'_{\bm{q}}$ might be arbitrarily bad for \prob queries.
However, we will demonstrate that Algorithm~\ref{alg_Index_query} provides high-quality results for \prob queries in almost all cases through experiments.

\begin{example}
    Fig.~\ref{fig:Index} illustrates how the index-based algorithm is used to find the result of the \prob query with $\bm{q} = (0.5, 0.5)$, $k = 1$, $l = 2$, and $\eta = 0.6$. First, we find the topic vector closest to $\bm{q}$ from the \iidx in Fig.~\ref{fig:TIE-tree}. We traverse \iidx from the root to a leaf node and identify the topic vector with the highest similarity, i.e., $\gamma^* = (0.51, 0.49)$, from the leaf node.
    Then, we directly obtain the estimated influence score for each vertex from the list associated with $\gamma^*$.
    Subsequently, as shown in Fig.~\ref{fig:TUC-list}, we use $\mathcal{L}[1, 2]$ in the \cidx.
    We get $j^* = 2$ w.r.t.~$\eta = 0.6$ and return $C' = \{v_2, v_4, v_5\}$) as the candidate subgraph to search.
    Finally, we perform an online search procedure on $C'$ to obtain the final result $C'_{\bm{q}} = C'$.
\end{example}

%% file: sections-springer/6-experiments.tex
\section{Experiments}
\label{sec_exp}

In this section, we perform extensive experiments and case studies on real-world data sets to evaluate the efficacy and efficiency of the \prob problem and algorithms.

\subsection{Experimental Setup}

\noindent\textbf{Data Sets.}
The following five publicly available real-world data sets are used in our experiments.
\begin{itemize}
    \item \textbf{Epinion} is a trust network between users on a product review website, where each directed edge $e = (u, v)$ means that user $v$ trusts user $u$.
    We perform a nonnegative matrix factorization (NMF) \cite{WangZ13} on the user-item rating matrix to generate a topic vector $\bm{\omega}(u)$ for each vertex (user) $u$ that denotes the user's preference for different items.
    The weight vector $\bm{\omega}(e)$ of an edge $e = (u, v)$ is calculated as $\bm{\omega}(e) = \langle \bm{\omega}(u), \bm{\omega}(v) \rangle$.
    \item \textbf{IMDB} is a collaboration network of actresses and actors in movies, where each (undirected) edge indicates that two vertices collaborated in a movie.
    We use the $20$ movie genres as ground-truth topics and calculate the weight $\omega_i(e)$ of an edge $e = (u, v)$ on the $i$-th topic (genre) based on the number of movies labeled with the $i$-th genre in which $u$ and $v$ collaborated.
    \item \textbf{DBLP} is a co-author network of researchers in the field of computer science, where each (undirected) edge indicates that two researchers co-authored at least one paper.
    We adopt a similar method to that used for IMDB to generate the topic vectors: we select $10$ subcategories in Engineering \& Computer Science from Google Scholar Metrics as ground-truth topics and calculate the weight $\omega_i(e)$ of an edge $e = (u, v)$ on the $i$-th topic (subcategory) based on the number of papers co-authored by $u$ and $v$ and published in any venue of subcategory $i$.
    \item \textbf{Reddit} is an interaction network built from posts and comments on Reddit, where each edge $e = (u, v)$ means that $v$ comments on any of $u$'s posts. We use a pre-trained model to infer a $19$-dimensional topic vector for each subreddit. Then, we compute a topic vector for each edge $e = (u, v)$ by linearly combining the topic vectors of subreddits where $u$ and $v$ both posted or commented.
    \item \textbf{Wiki-Topcats} is a directed network of Wikipedia hyperlinks, where each directed edge $e = (u, v)$ indicates that a Wikipedia page $u$ hyperlinks to page $v$. Each vertex (page) $u$ is associated with a $5$-dimensional topic vector $\bm{\omega}(u)$, generated using a pre-trained model to follow a Dirichlet distribution, representing the page's topic distribution. The weight vector $\bm{\omega}(e)$ of an edge $e = (u, v)$ is computed as $\bm{\omega}(e) = \langle \bm{\omega}(u), \bm{\omega}(v) \rangle$.
\end{itemize}
The statistics of the data sets are shown in Table~\ref{tab_data set}, where $n$ is the number of vertices, $m$ is the number of edges, $\Delta^{-}_{max}$ and $\Delta^{+}_{max}$ are the maximal in-degree and out-degree among all vertices, and $z$ is the number of topics by default.

\input{tables/dataset}

\vspace{1mm}
\paragraph{Algorithms}
We compare \prob with the following competitors for the evaluation of community quality.
\begin{itemize}
    \item \textsf{TIM} \cite{ChenFLFTT15}: Select a set of vertices (seeds) with the largest influence spread for a topic vector under the TIC model.
    \item \textsf{UICS} \cite{LuoZLGL23}: Return a $(k, \eta)$-core subgraph with the highest influence score.
    \item \textsf{KICQ} \cite{IslamAKSCR22}: Find a $k$-core subgraph whose score in terms of cohesiveness, relevance to query keywords, and influence is the highest.
    \item \textsf{VAC} \cite{LiuZZHXG20}: Return a $k$-truss subgraph where (1) the query vertex is contained and (2) all vertices are associated with at least one query keyword.
    \item \textsf{EACS} \cite{LiZLWW23}: Return a $k$-truss subgraph that contains the query vertex. Unlike \cite{LiuZZHXG20}, EACS requires that all edges of the subgraph are associated with at least one query keyword.
    \item \textsf{MICS} \cite{chang2024mics}: Return a $(k, l)$-core subgraph where a given set of seeds has the maximum influence by expectation.
    \item \textsf{TOPL-ICDE} \cite{ZhangYLC24}: Return $L$ $k$-truss subgraphs with the highest scores in terms of cohesiveness, relevance to query keywords, and influence.
\end{itemize}
Note that none of the above methods can be used directly for \prob.
Thus, we adapt them to the same setting as \prob as follows.
For \textsf{TIM}, we use the topic-based interaction graph w.r.t.~a topic vector as input and set the number of seeds to the size of the community returned by \prob.
For \textsf{UICS}, which is topic-unaware and specific to undirected graphs, we convert the topic-based interaction graph to undirected as input.
{For \textsf{MICS}, We use the nodes picked by \textsf{TIM} as the seed set and search for the $(k,l)$-core subgraph where the seeds have the maximum influence accordingly.}
For keyword-based methods (\textsf{VAC}, \textsf{KICQ}, {\textsf{TOPL-ICDE}}, and \textsf{EACS}), we assign a set of keywords to each vertex.
In the Epinion data set, the keywords of a vertex are the items that the vertex has rated; in the IMDB, DBLP, and Reddit data sets, the keywords of a vertex are extracted from the description of movies in which the vertex has participated, titles of papers that the vertex has authored, and comments that the vertex has posted, respectively.
Then, we extract a set of frequent keywords w.r.t.~the topic vector $\bm{q}$ as the query keywords for them.
For \textsf{EACS}, we set the keywords of an edge as the common keywords in its two connected vertices.
{For \textsf{TOPL-ICDE}, we only need to return the top-$1$ subgraph with the highest influence.}
Similarly to \textsf{UICS}, directed graphs are also converted to undirected for \textsf{VAC}, \textsf{KICQ}, and \textsf{EACS}.
We then evaluate the efficiency of Algorithms~\ref{alg_basic} and~\ref{alg_Index_query} in terms of query time and index overhead.

\vspace{1mm}
\paragraph{Query Formulation}
For each data set, we generate \prob queries on $100$ topic vectors.
These topic vectors consist of (1) $z$ one-hot vectors corresponding to all pre-specified topics and (2) $100 - z$ vectors inferred from random items, movies, papers, posts, and pages in the Epinion, IMDB, DBLP, Reddit, and Wiki-Topcats data sets.
In each suite of experiments, we run a method for all $100$ queries and use the average measures for evaluation.

\vspace{1mm}
\paragraph{Parameter and Implementation}
In the experiments, we tested different methods in various parameter settings.
The values of $k$, $l$, and $\eta$ in the definition of $(k, l, \eta)$-cores vary over $[1, 2, \dots, 5]$, $[2, 3, \dots, 6]$, and $[0.1, 0.2, \dots, 0.5]$, respectively, with default values $k = 2$, $l = 5$, and $\eta = 0.2$. We use these default values of $k$, $l$,
and $\eta$ because they can ensure the existence of results for most queries.
Higher values of $k$, $l$, and $\eta$ often make it difficult to find a valid community.
For the Epinion data set, we generate user and item vectors with dimensionality $z$ ranging from $10$ to $50$ to evaluate the effect of $z$.
For \iidx construction, we always select $h = 1, 000$ vectors and use $N_0 = 5$ to build the cone tree. We have tried different values of $h$ and $N_0$ and choose
$1, 000$ and $5$ by default because they achieve the best trade-off
between construction and search costs and query accuracy.
All algorithms were implemented in C++11 and compiled with the ``-O3'' flag.
All experiments were conducted on a Linux server with an Intel\textsuperscript{\textregistered} Core\textsuperscript{\texttrademark} Xeon Processor CPU @3.0GHz and 64GB RAM. 

\subsection{Experimental Results}

\paragraph{Exp-1: Community Quality Evaluation}
{We compare the quality of communities retrieved by \prob, \textsf{TIM}, \textsf{UICS}, \textsf{KICQ}, \textsf{VAC}, \textsf{EACS}, \textsf{TOPL-ICDE}, and \textsf{MICS}.}
Suppose that $C = (V_C, E_C)$ is a community returned by one of the above methods for a query.
We evaluate the quality of $C$ using the following three metrics:
\emph{(i)} \textbf{edge density} $\rho(C) = \frac{|E_C|}{|V_C| \cdot (|V_C| - 1)}$;
\emph{(ii)} \textbf{topic similarity} $\mathrm{sim}(C) = \frac{1}{|V_C|} \sum_{u \in V_C} f(\langle\bm{\omega}(u), \bm{q}\rangle)$;
and \emph{(iii)} \textbf{influence} $\mathbb{I}_{\bm{q}}(C)$, which is estimated by running the Monte Carlo simulation in $10,000$ rounds.
For each measure, higher values mean better community quality.
We also report the size $|V_C|$ of the community $C$ for comparison.

\begin{figure}[t]
    \centering
    \includegraphics[width=0.8\textwidth]{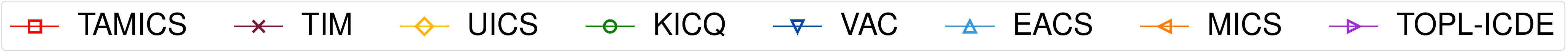}
    \\
    \includegraphics[width=0.192\textwidth]{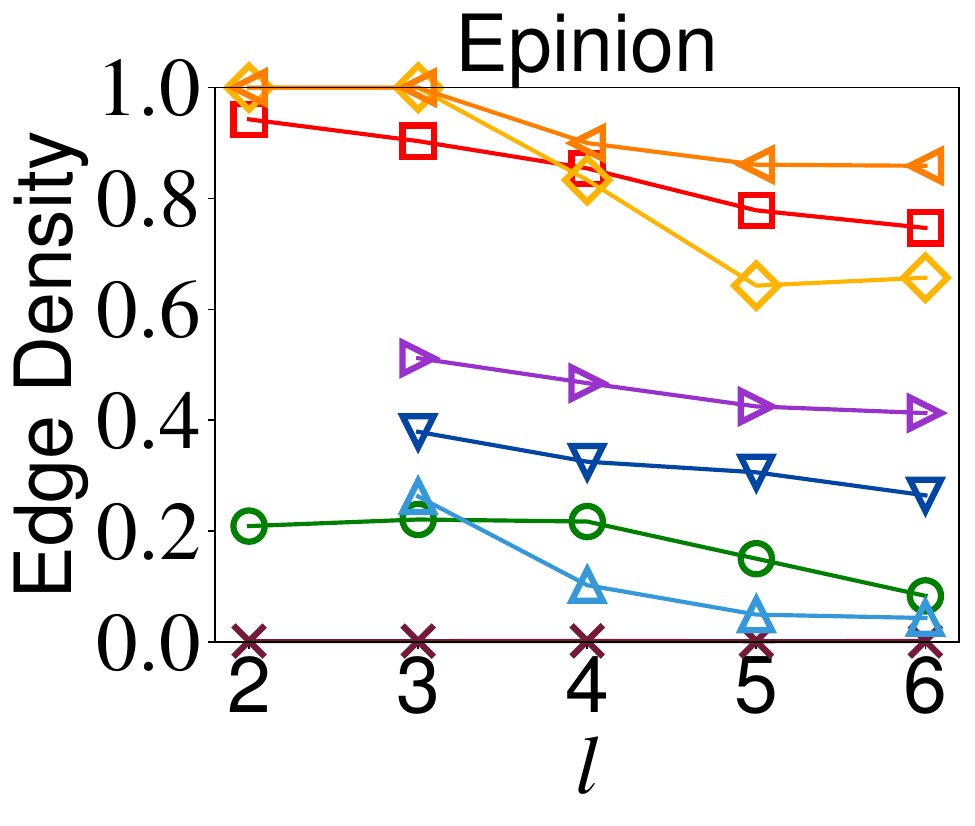}
    \includegraphics[width=0.192\textwidth]{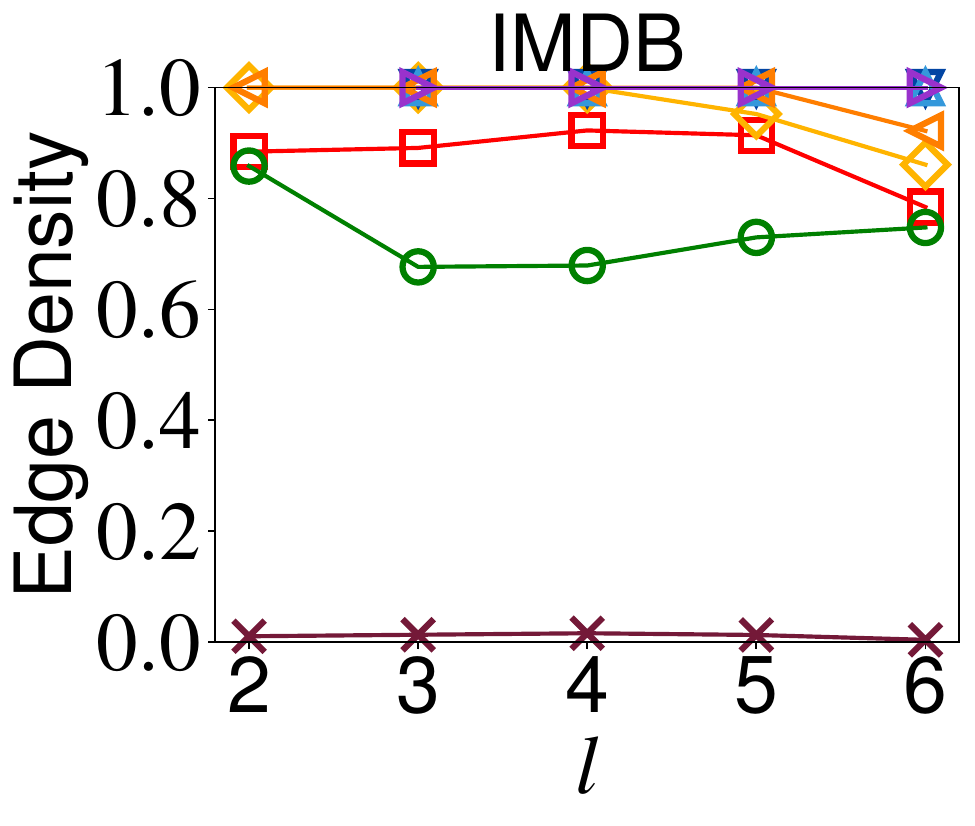}
    \includegraphics[width=0.192\textwidth]{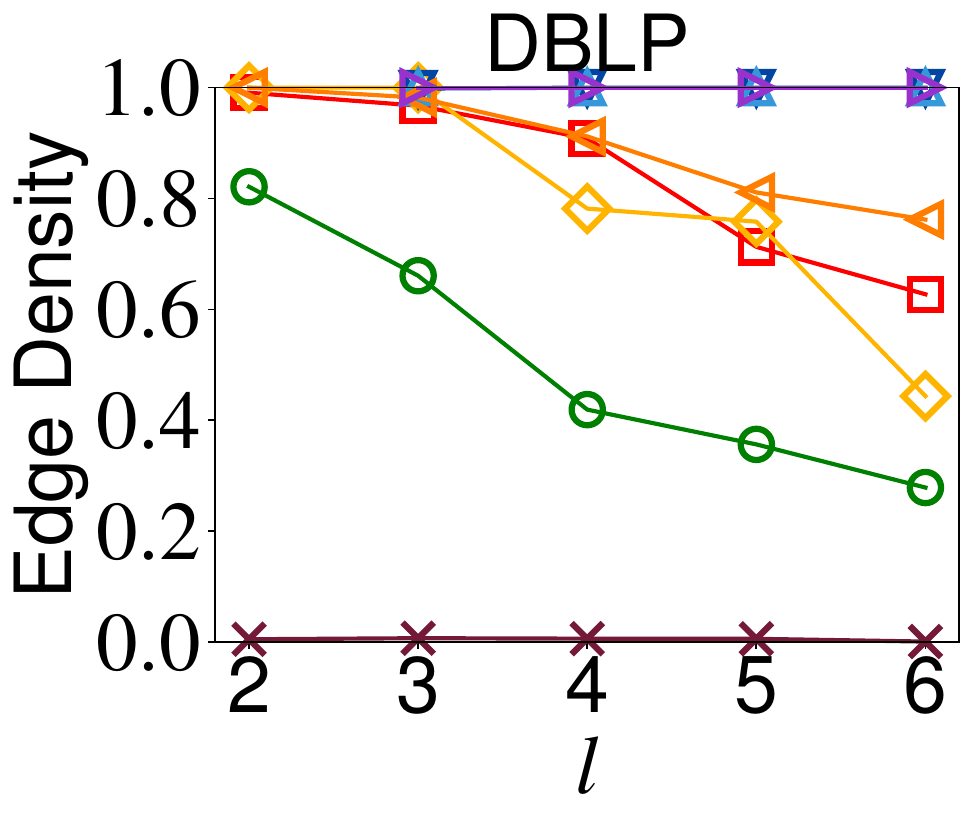}
    \includegraphics[width=0.192\textwidth]{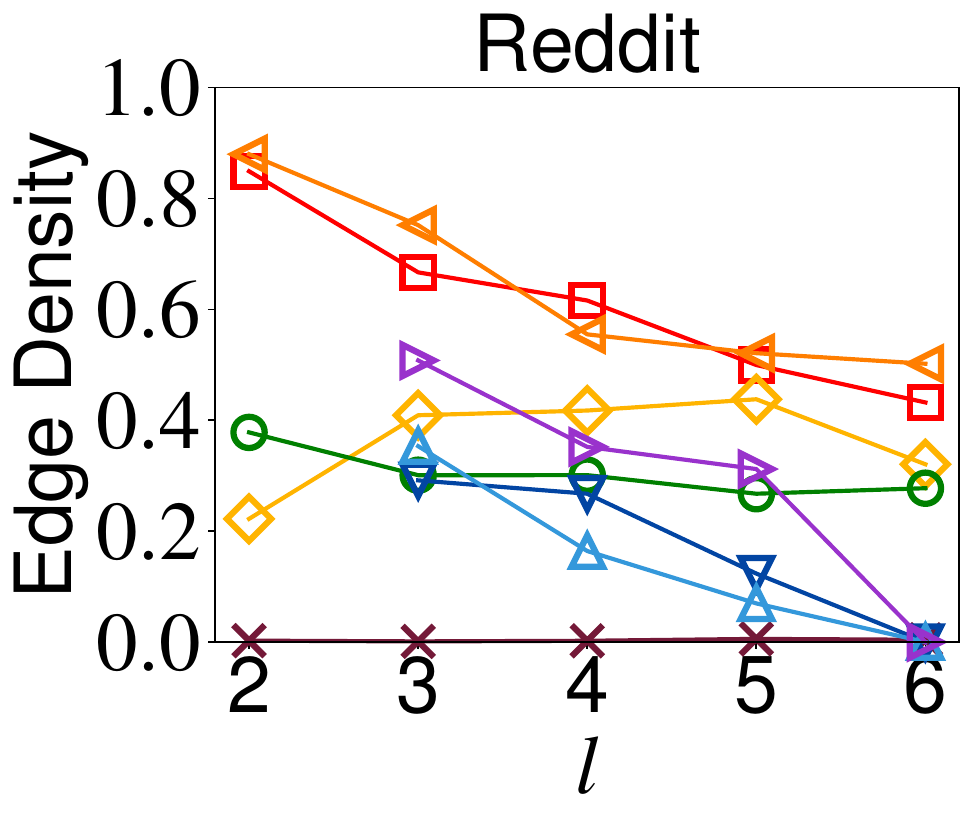}
    \includegraphics[width=0.192\textwidth]{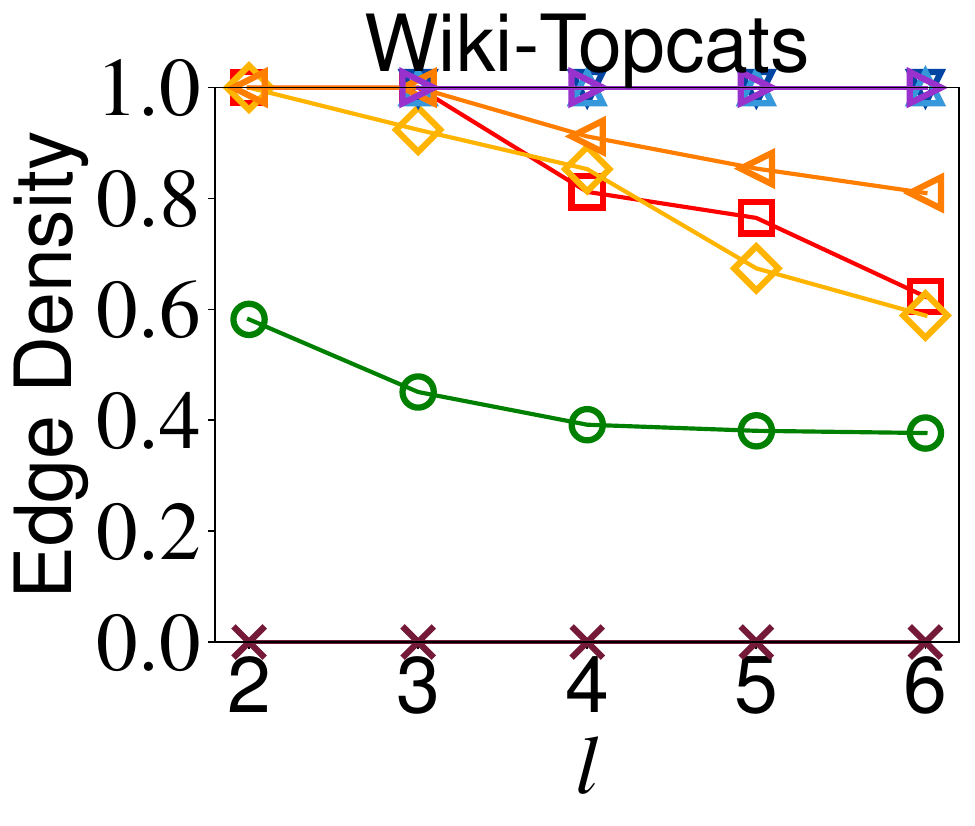}
    \\
    \vspace{1mm}
    \includegraphics[width=0.192\textwidth]{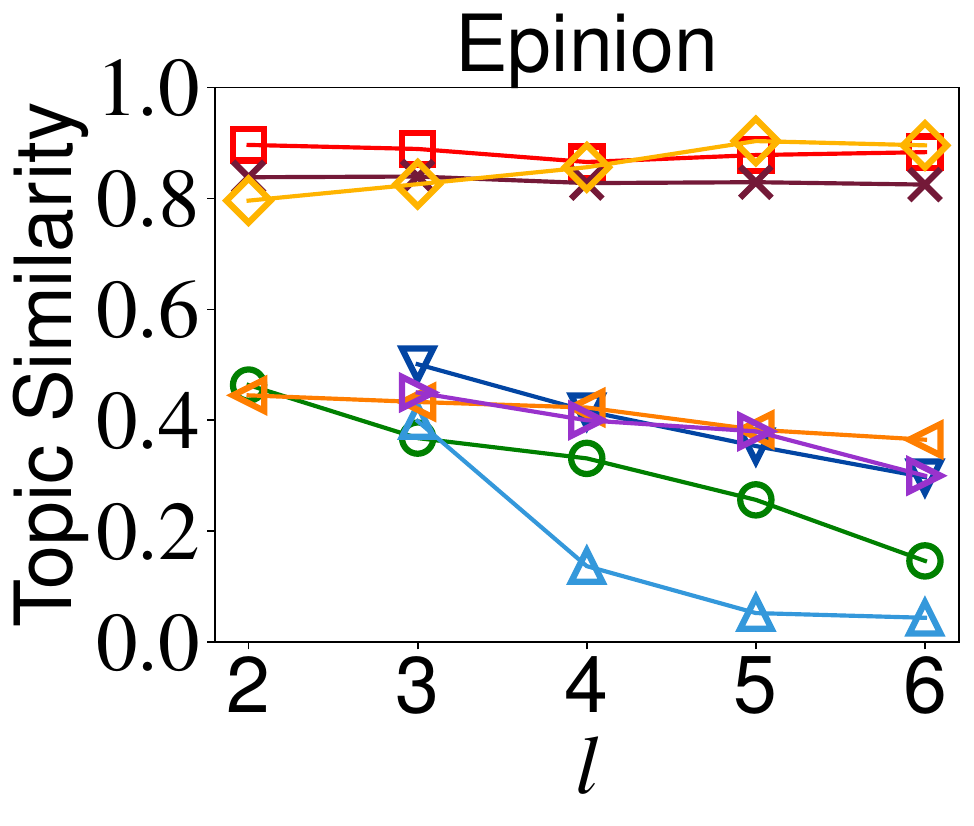}
    \includegraphics[width=0.192\textwidth]{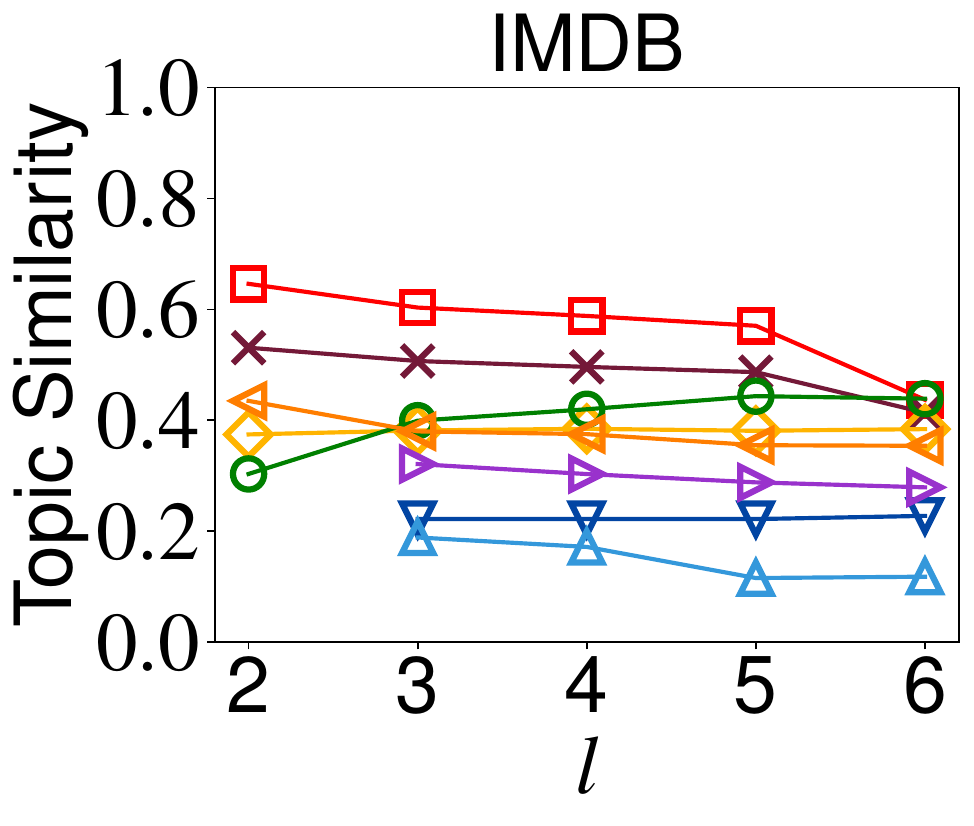}
    \includegraphics[width=0.192\textwidth]{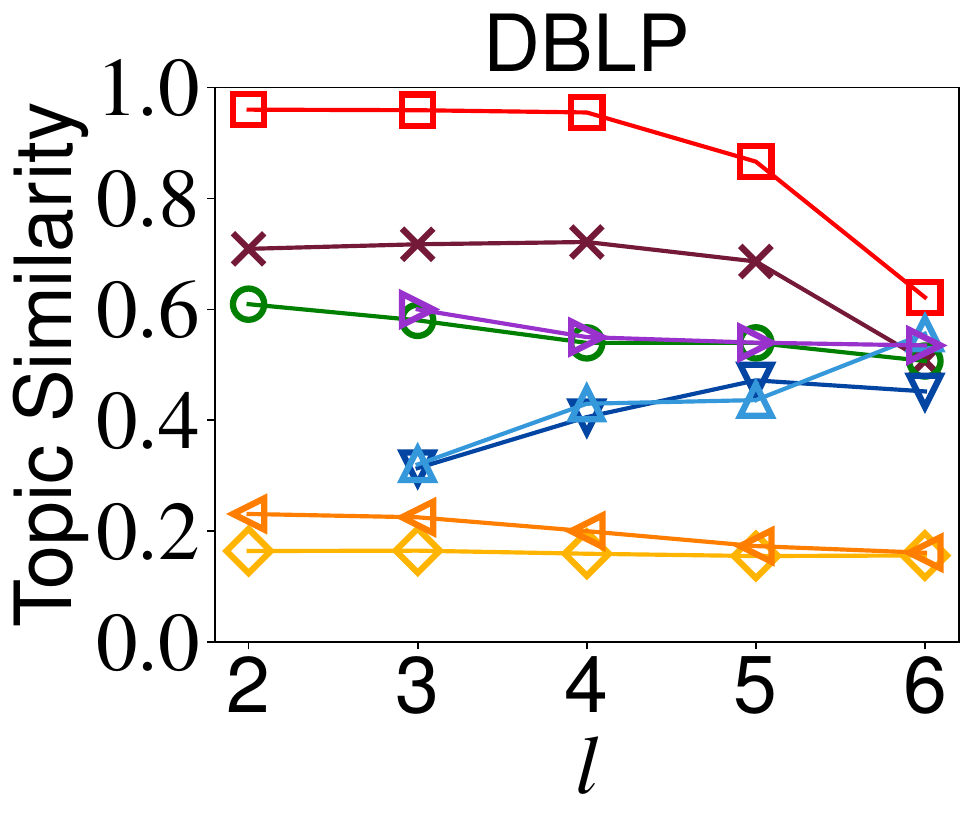}
    \includegraphics[width=0.192\textwidth]{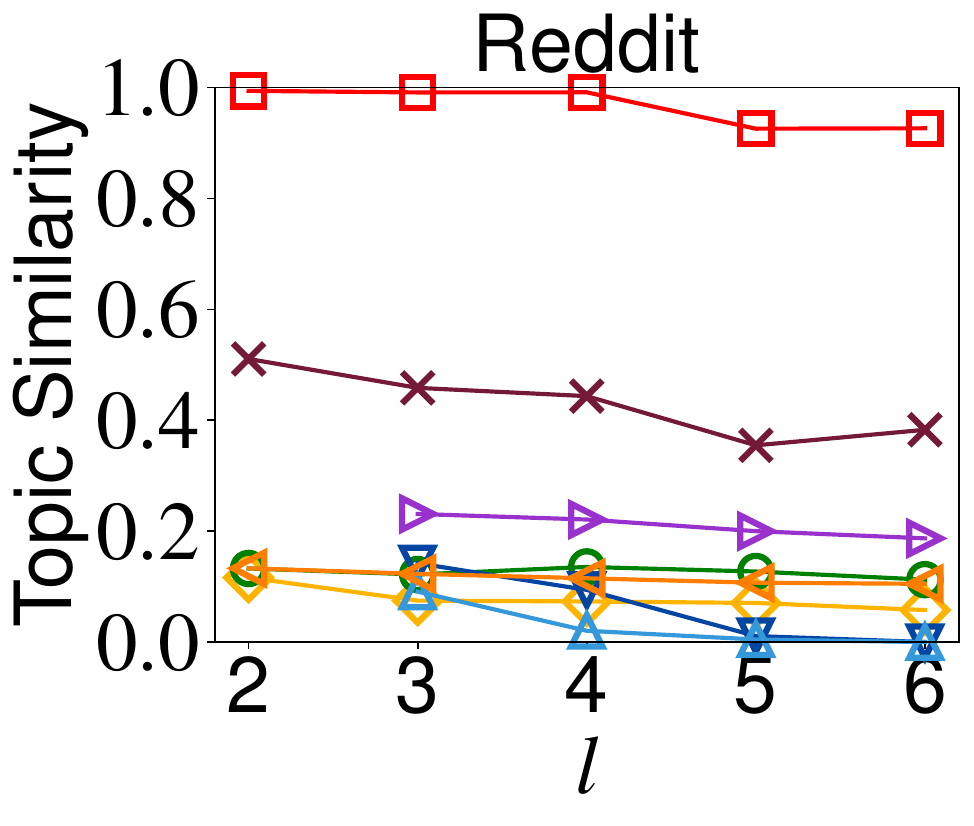}
    \includegraphics[width=0.192\textwidth]{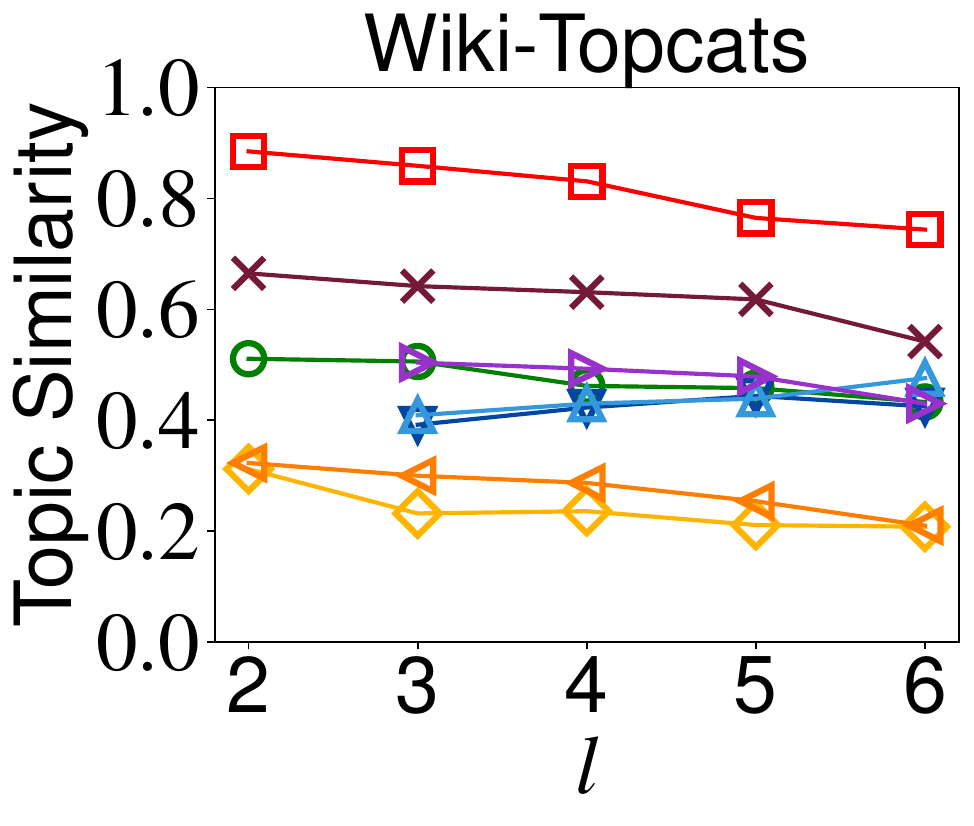}
    \\
    \vspace{1mm}
    \includegraphics[width=0.192\textwidth]{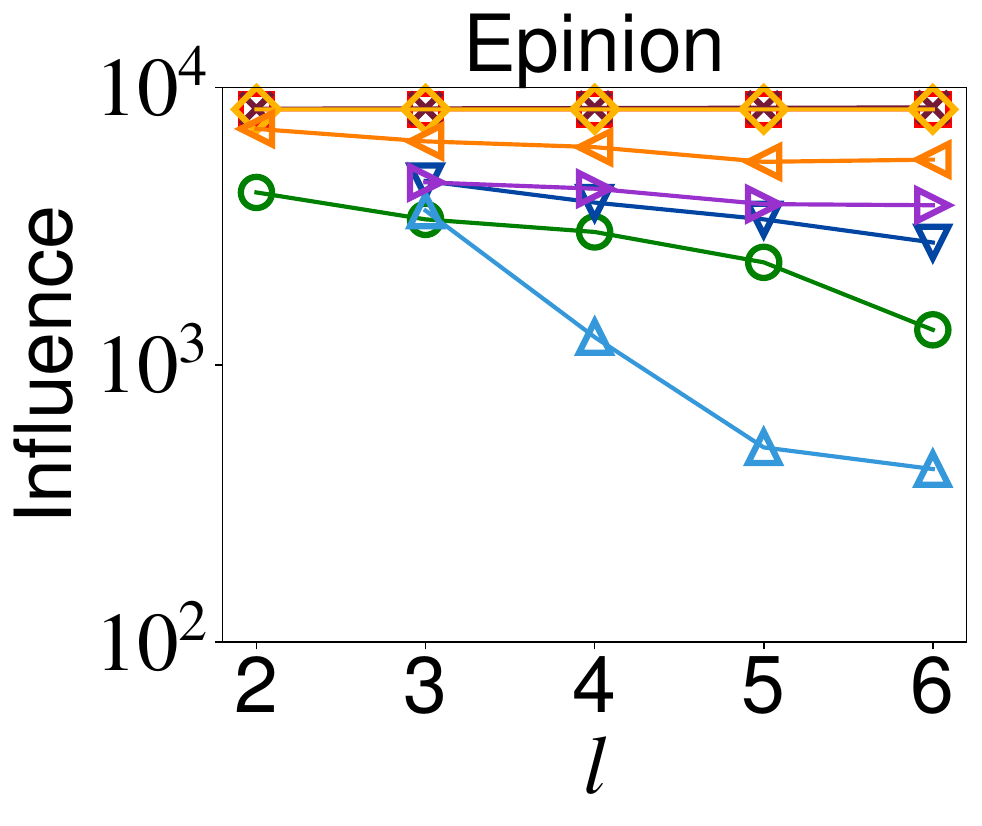}
    \includegraphics[width=0.192\textwidth]{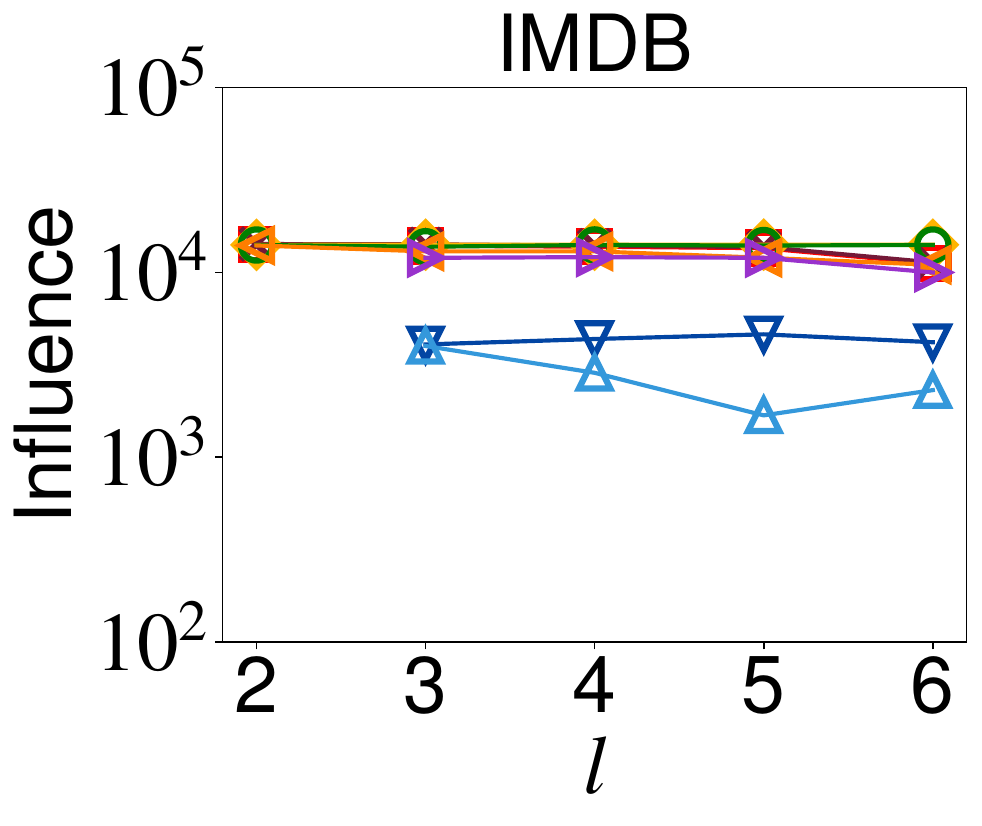}
    \includegraphics[width=0.192\textwidth]{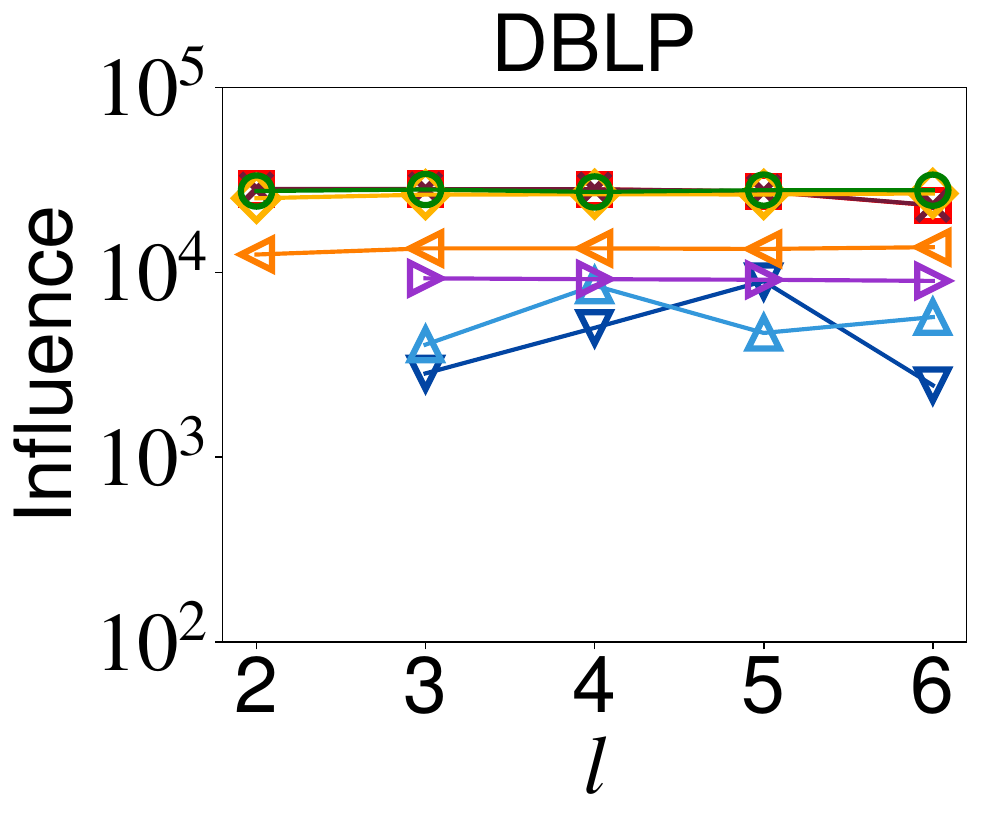}
    \includegraphics[width=0.192\textwidth]{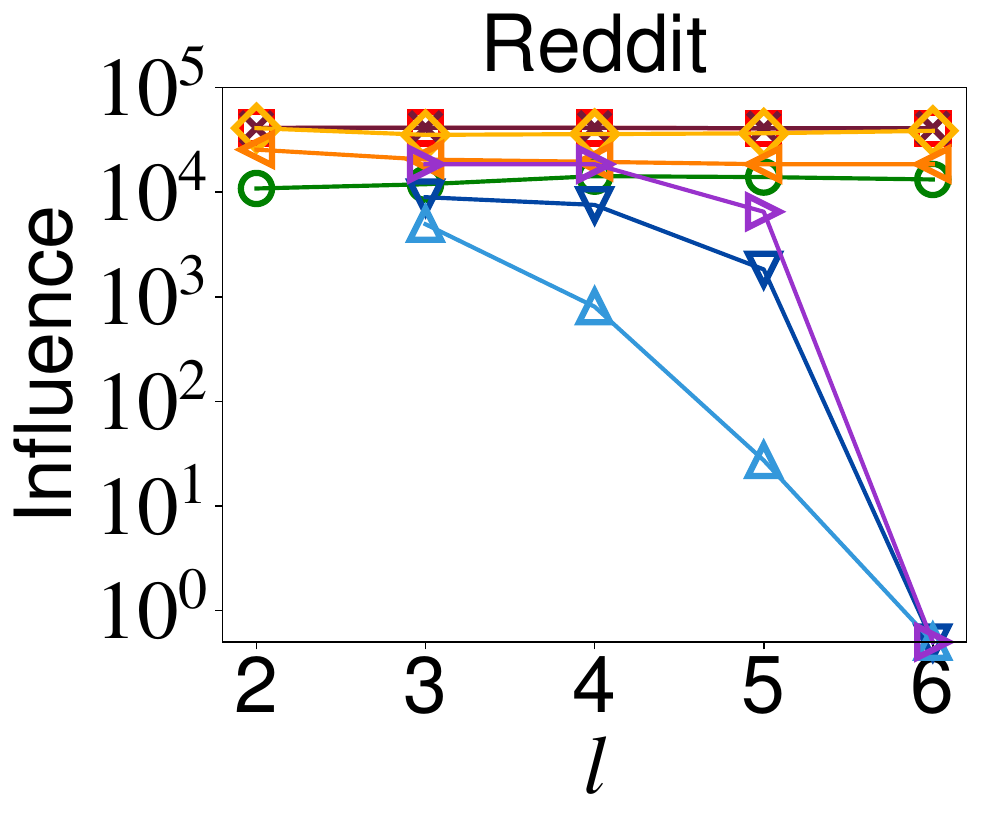}
    \includegraphics[width=0.192\textwidth]{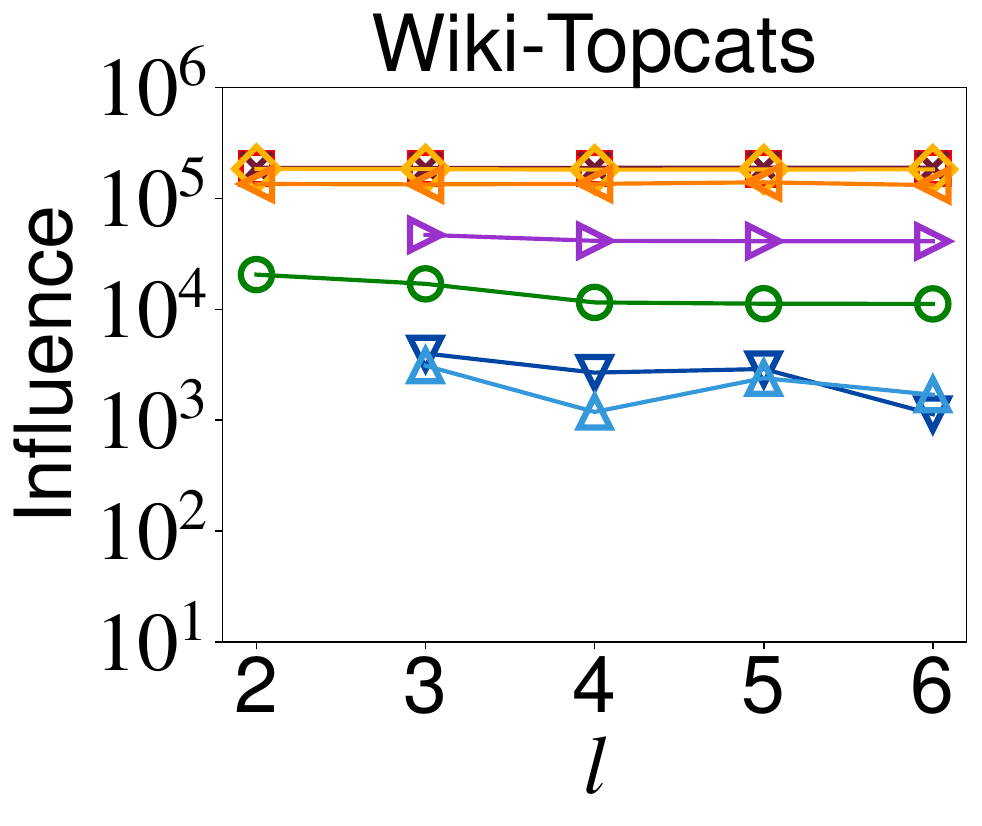}
    \\
    \vspace{1mm}
    \includegraphics[width=0.192\textwidth]{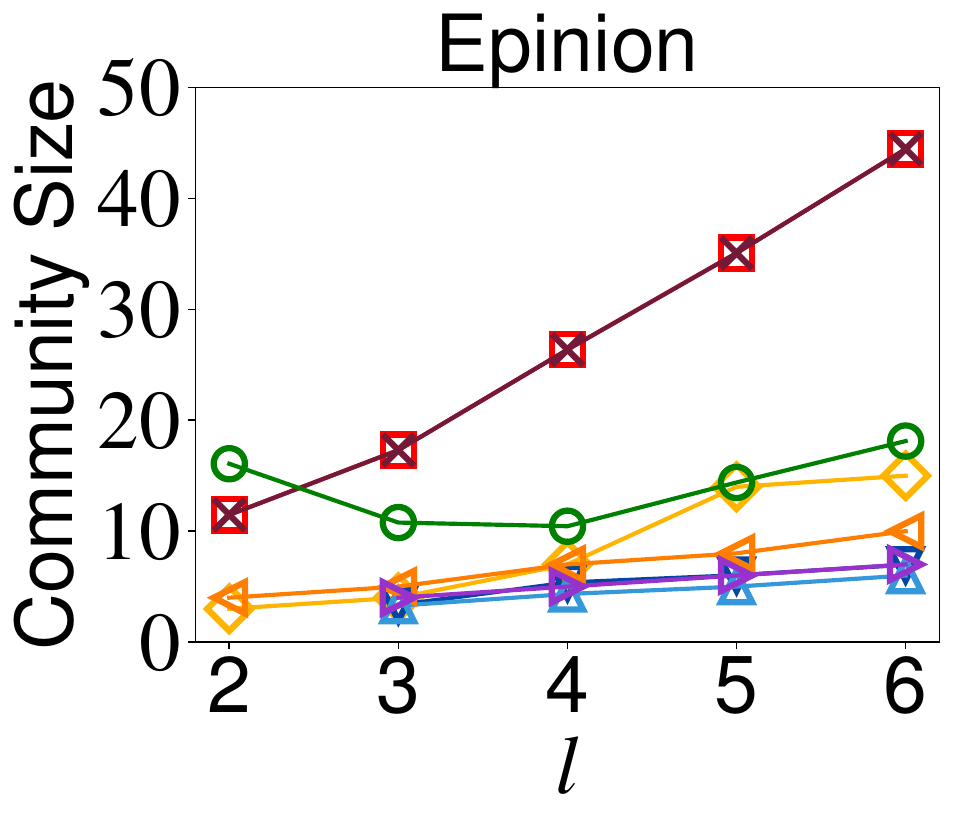}
    \includegraphics[width=0.192\textwidth]{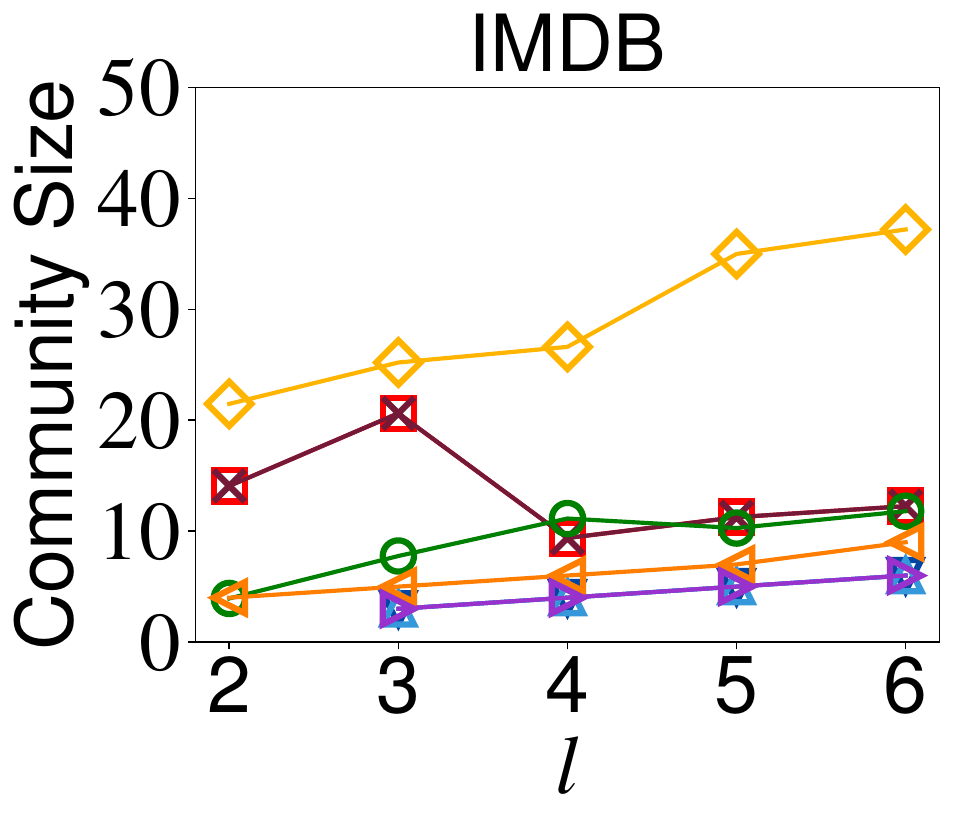}
    \includegraphics[width=0.192\textwidth]{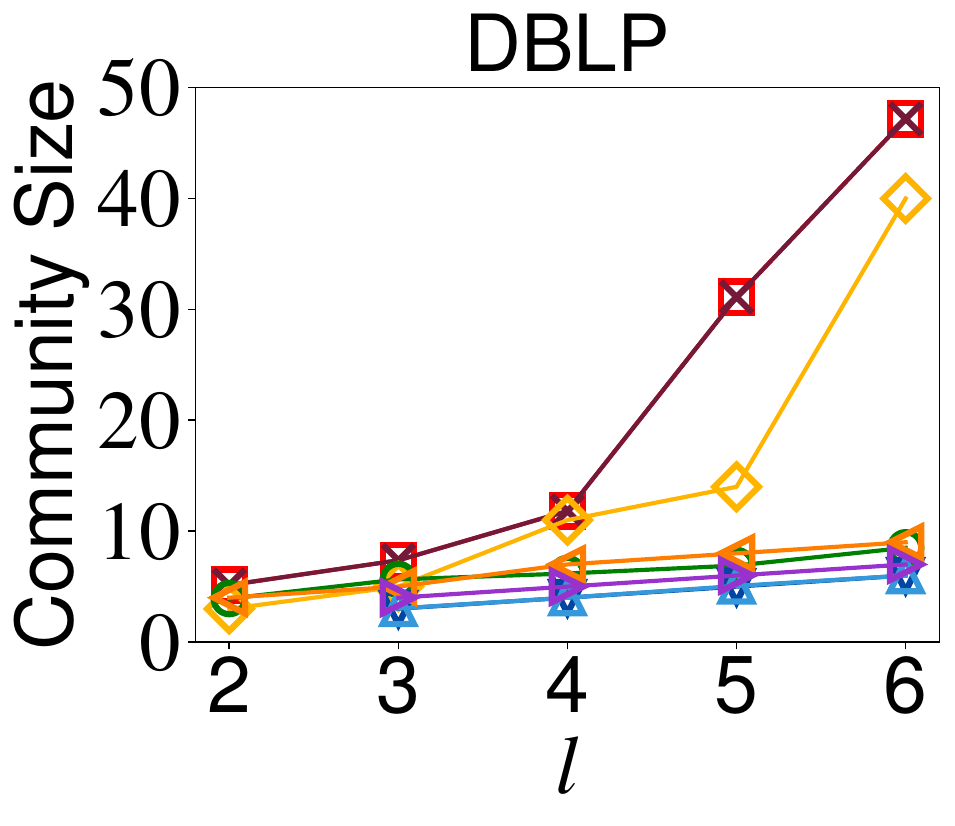}
    \includegraphics[width=0.192\textwidth]{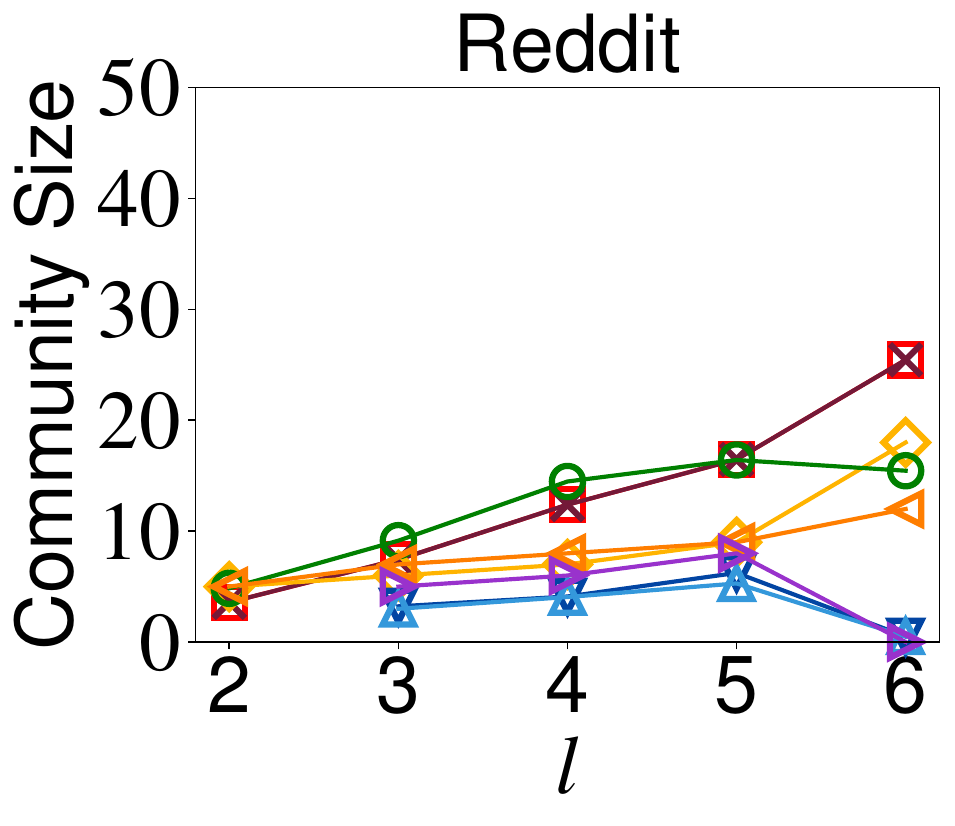}
    \includegraphics[width=0.192\textwidth]{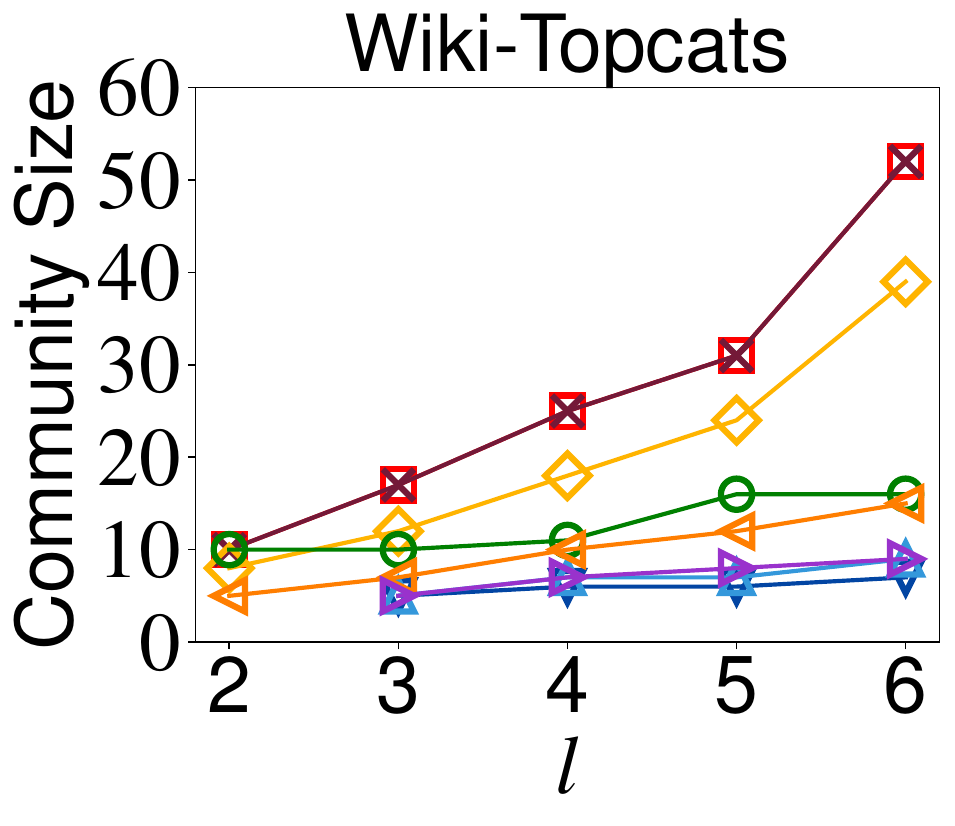}
    \\
    \caption{{Quality of the communities retrieved by each method in terms of cohesiveness, topic similarity, influence, and size when $k = 2$, $l \in \{2, \dots, 6\}$, and $\eta = 0.2$.}}
    \label{fig_exp1}
\end{figure}

The results of different CS methods for the four measures are shown in Fig.~\ref{fig_exp1}.
Generally, we observe that \prob shows exceptionally good performance across all datasets for each measure, while each baseline method excels for only one or two measures.
In terms of \emph{edge density}, \textsf{TIM} is always close to $0$ because it does not take into account the cohesiveness of the returned vertices.
{\textsf{MICS} is based on directed $(k, l)$-cores and, since it only considers deterministic graphs, its edge density is better than \textsf{TAMICS}.}
\textsf{UICS} is based on uncertain $(k, \eta)$-cores and performs similarly to \prob.
\textsf{KICQ} is based on $k$-cores and provides looser communities than \prob.
\textsf{VAC}, \textsf{EACS}, {and \textsf{TOPL-ICDE}}, all based on the $k$-truss model, provide more coherent communities than \prob on the IMDB, DBLP, and Wiki-Topcats datasets.
This is because the $k$-truss model is stricter than the $k$-core model for cohesiveness.
As a result, \textsf{VAC}, \textsf{EACS}, {and \textsf{TOPL-ICDE}} return communities much smaller than \prob.
Furthermore, on sparser datasets (Epinion and Reddit), they often fail to find any results for some queries.
In terms of \emph{topic similarity}, \prob consistently achieves the best performance across all datasets.
Keyword-based CS methods (\textsf{KICQ}, \textsf{VAC}, {\textsf{EACS}, and \textsf{TOPL-ICDE})} are inferior to \prob because topical information cannot be fully represented by keywords.
\textsf{TIM} ranks second in terms of topic similarity because it adopts the same topic model as \prob but often picks less relevant vertices as seeds for coverage.
{The reason for the poor performance of \textsf{TOPL-ICDE} is that it does not consider the variations in propagation influence across different topics.}
In terms of \emph{influence}, \textsf{TIM} is always the highest because its main goal is to maximize the influence.
\prob, {\textsf{MICS}}, and \textsf{UICS} are very close to \textsf{TIM} because they also consider influence as an important factor.
\textsf{VAC} and \textsf{EACS} cannot achieve high influence scores because they do not take into account influence in their models.
It is worth noting that our algorithm ensures high influence within identified communities by defining the community's influence score as the minimum score among its vertices, guaranteeing strong influence across all members.
In terms of \emph{community size}, core-based CS methods typically return larger communities than truss-based ones, as has been analyzed.

\begin{figure}
    \centering
    \includegraphics[height=3.6mm]{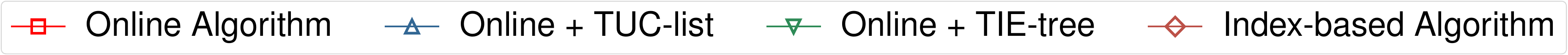}
    \\
    \vspace{1mm}
    \includegraphics[width=0.192\textwidth]{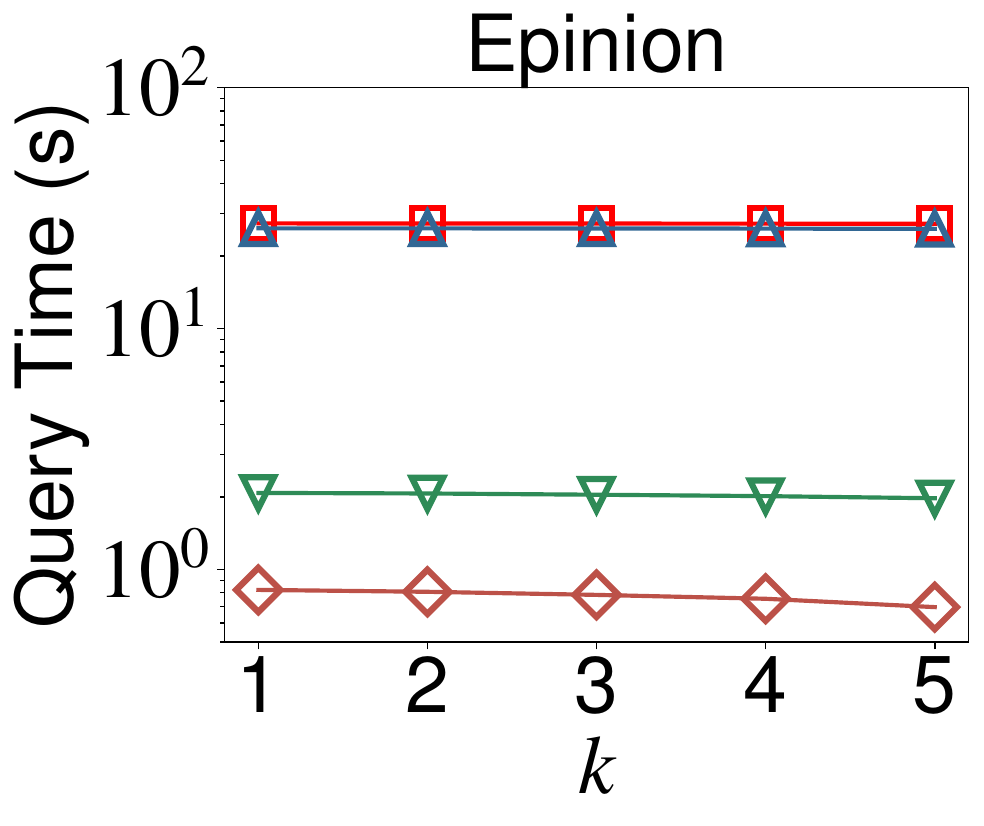}
    \includegraphics[width=0.192\textwidth]{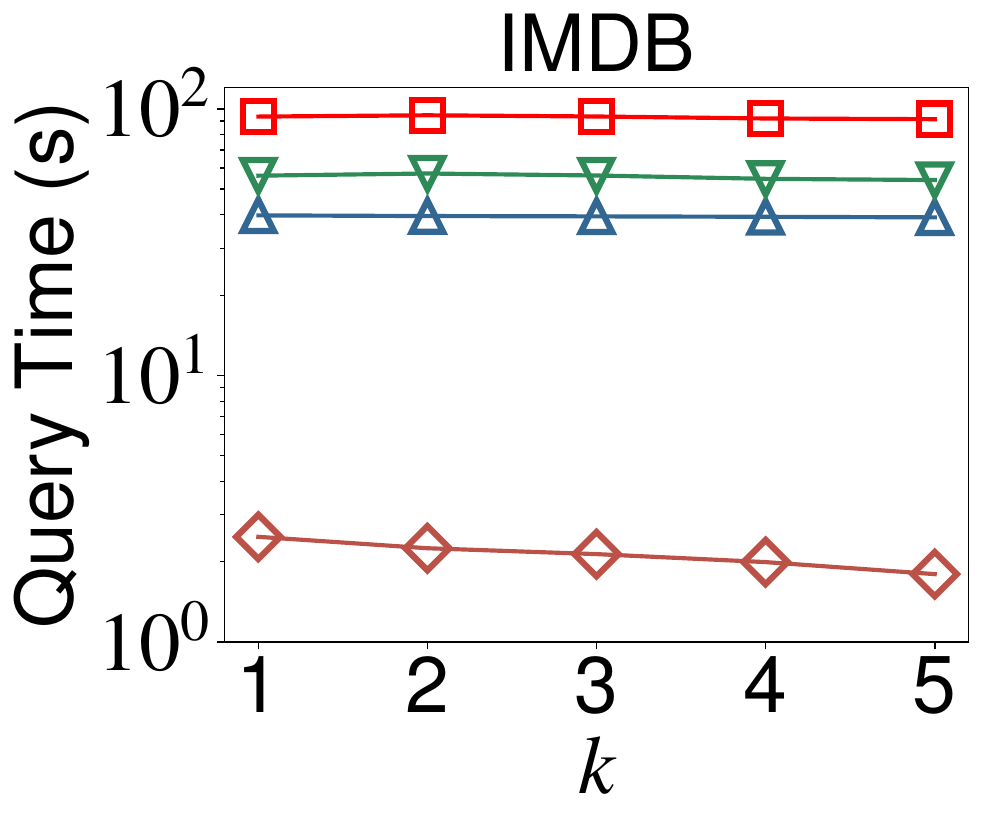}
    \includegraphics[width=0.192\textwidth]{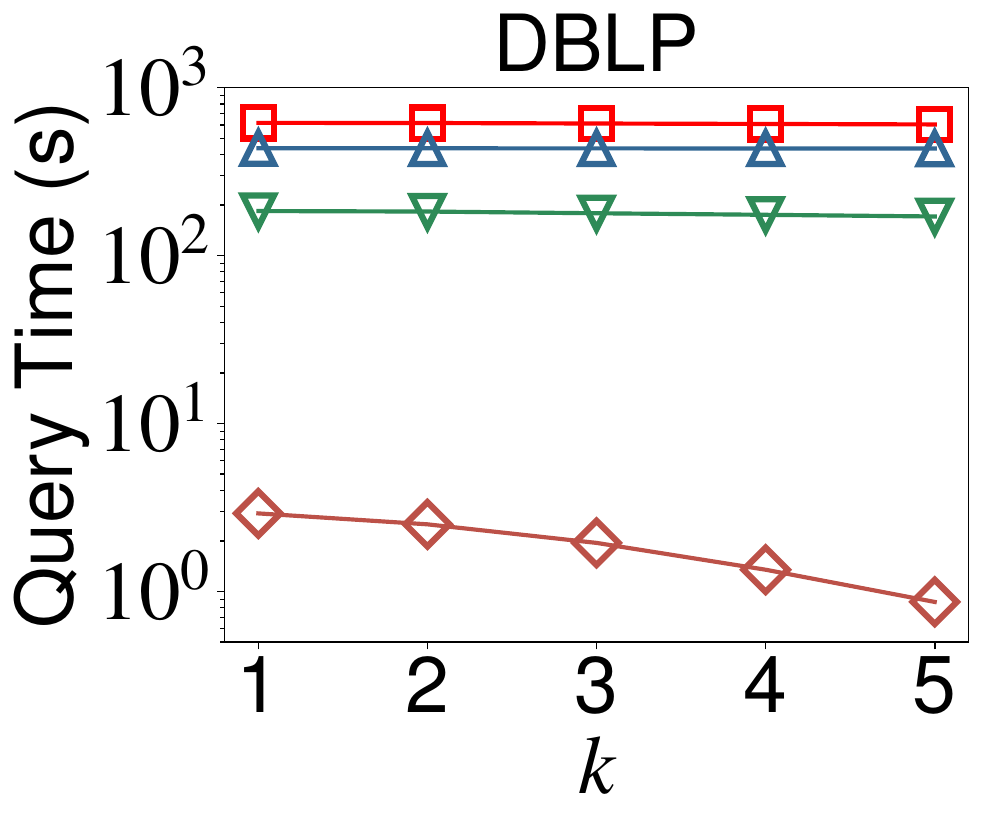}
    \includegraphics[width=0.192\textwidth]{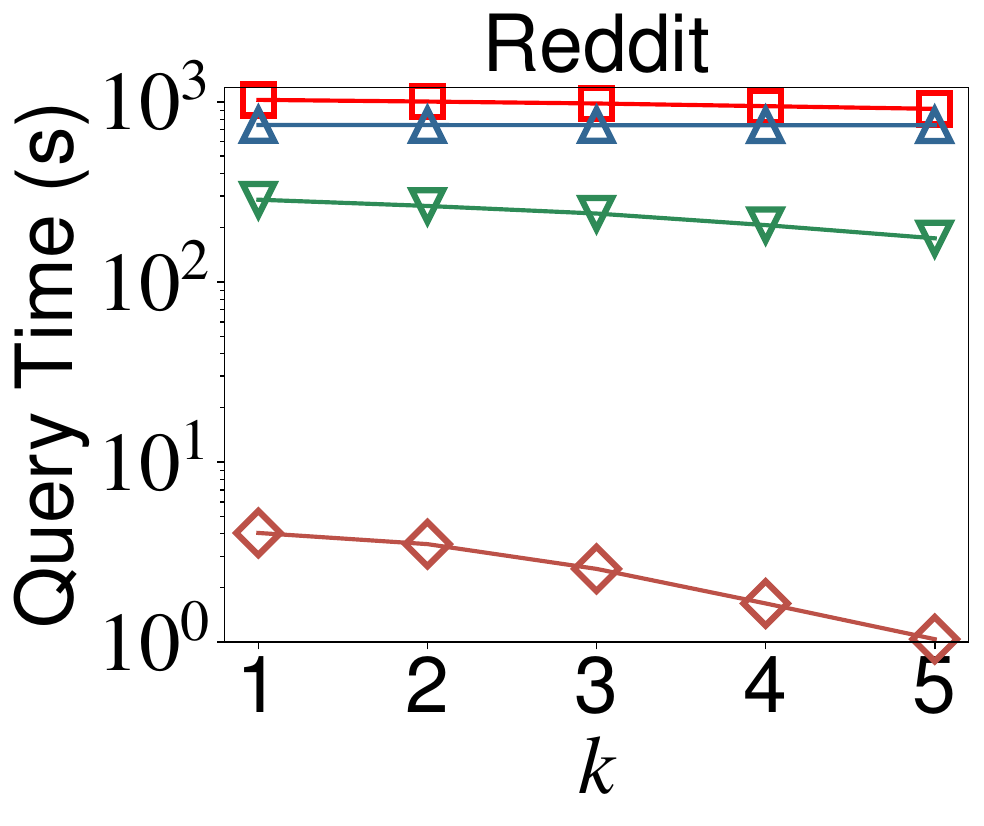}
    \includegraphics[width=0.192\textwidth]{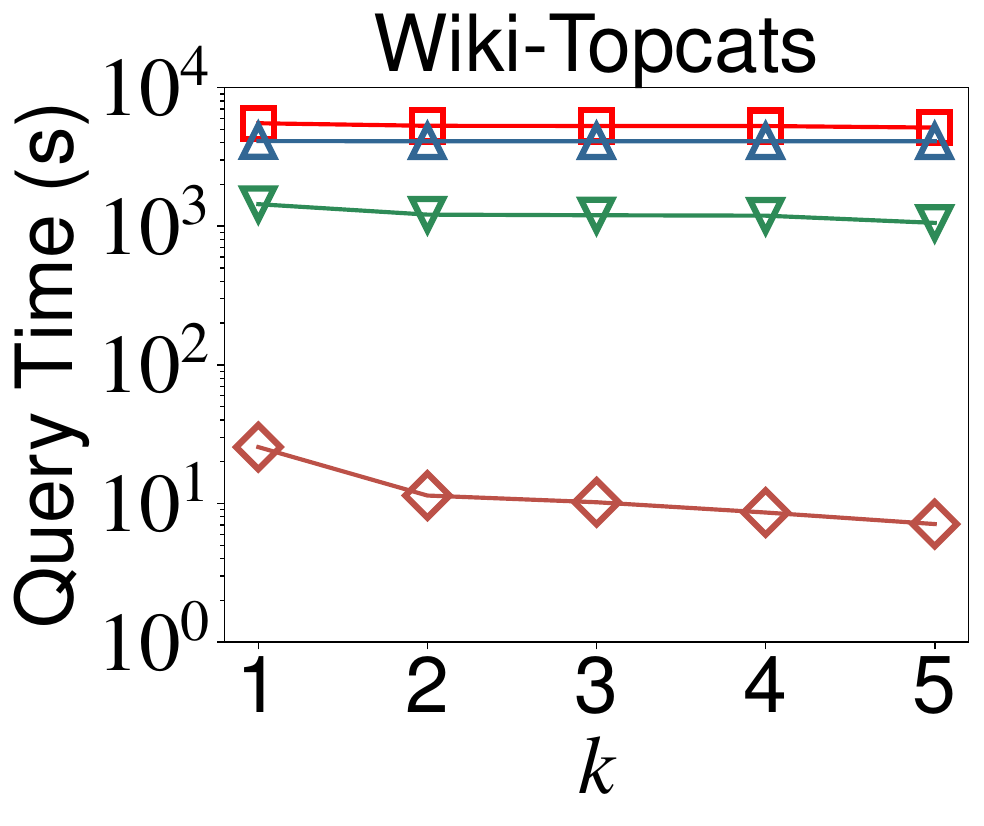}
    \\
    \vspace{1mm}
    \includegraphics[width=0.192\textwidth]{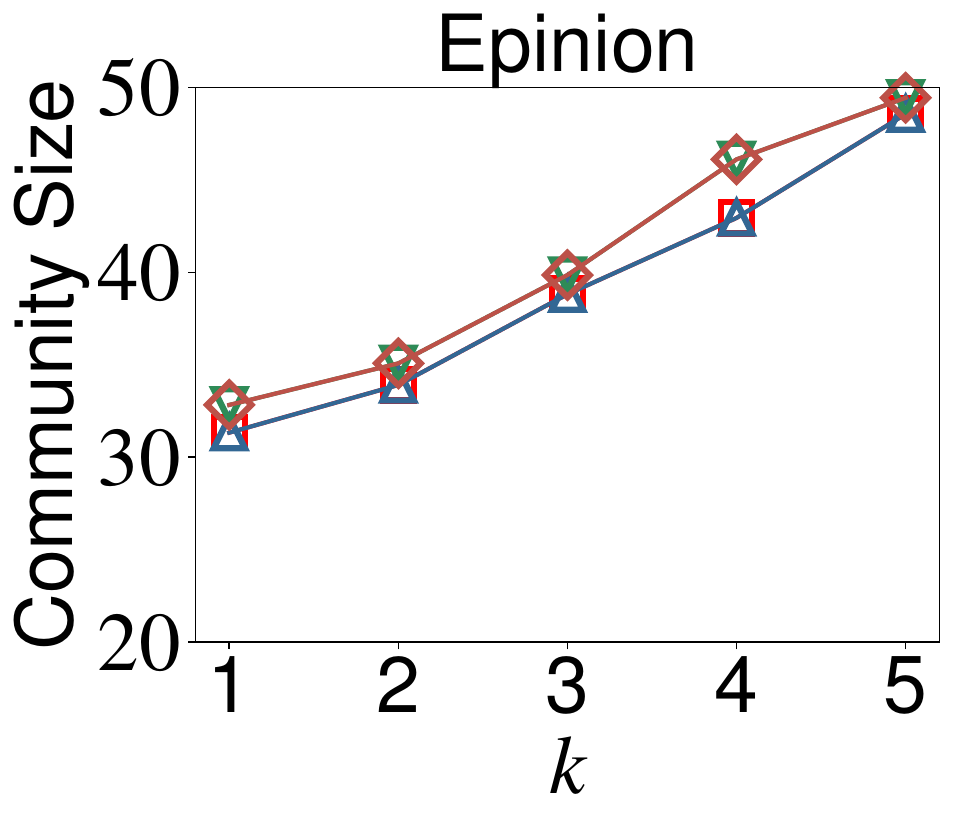}
    \includegraphics[width=0.192\textwidth]{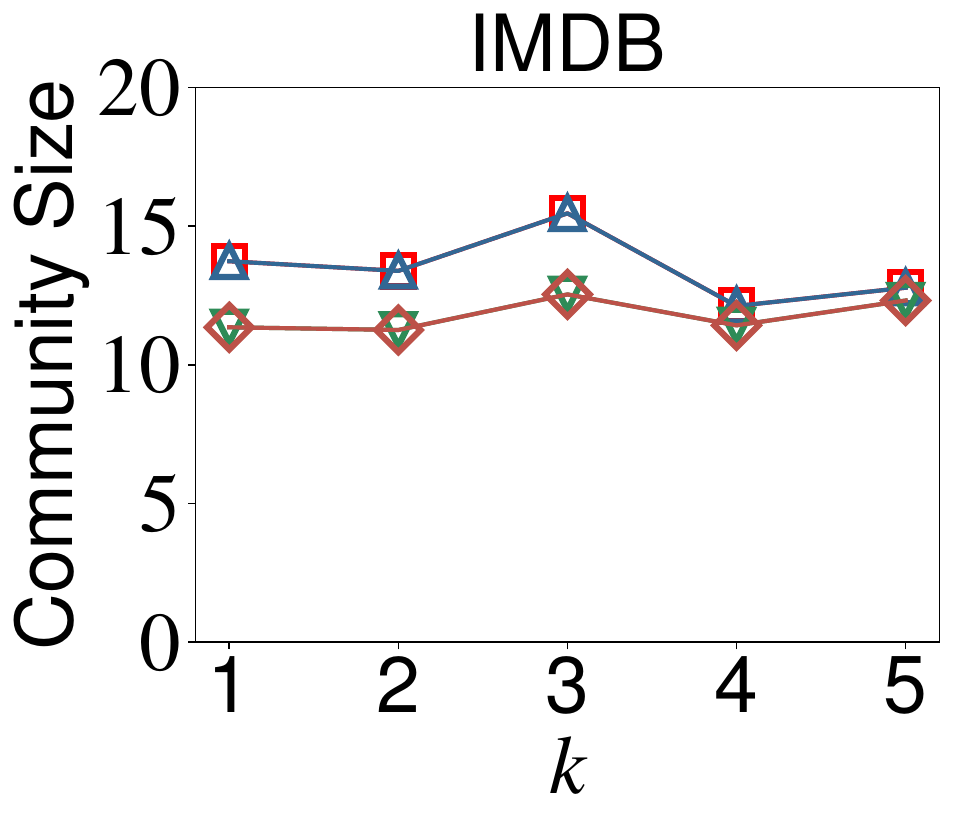}
    \includegraphics[width=0.192\textwidth]{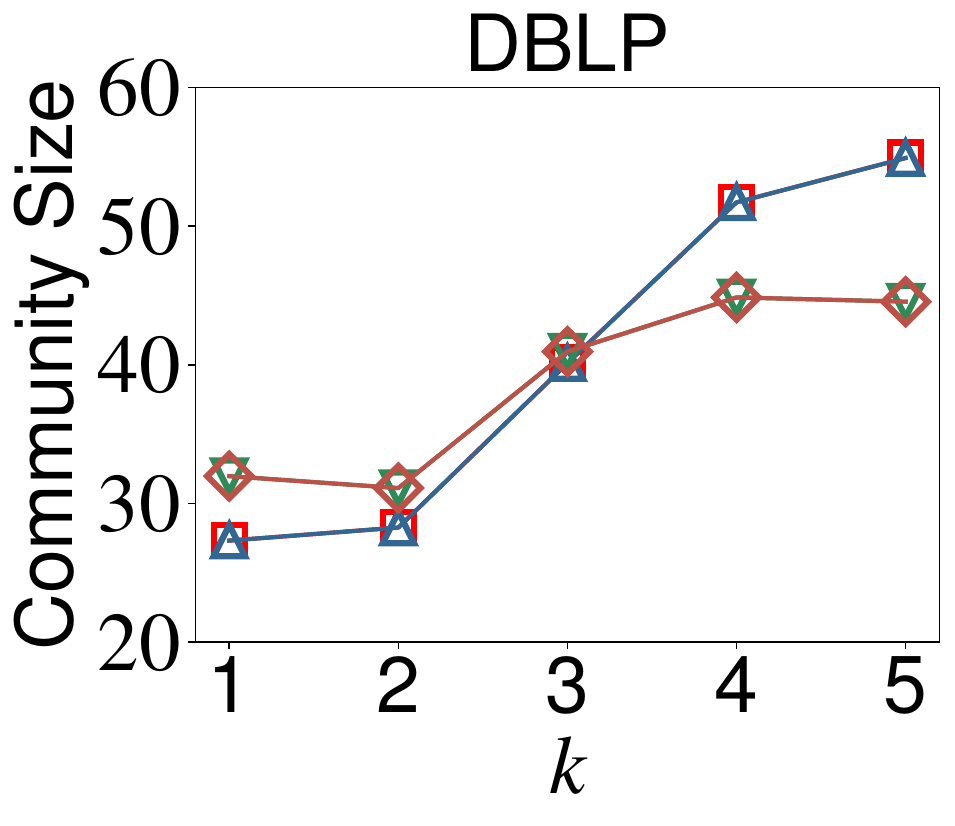}
    \includegraphics[width=0.192\textwidth]{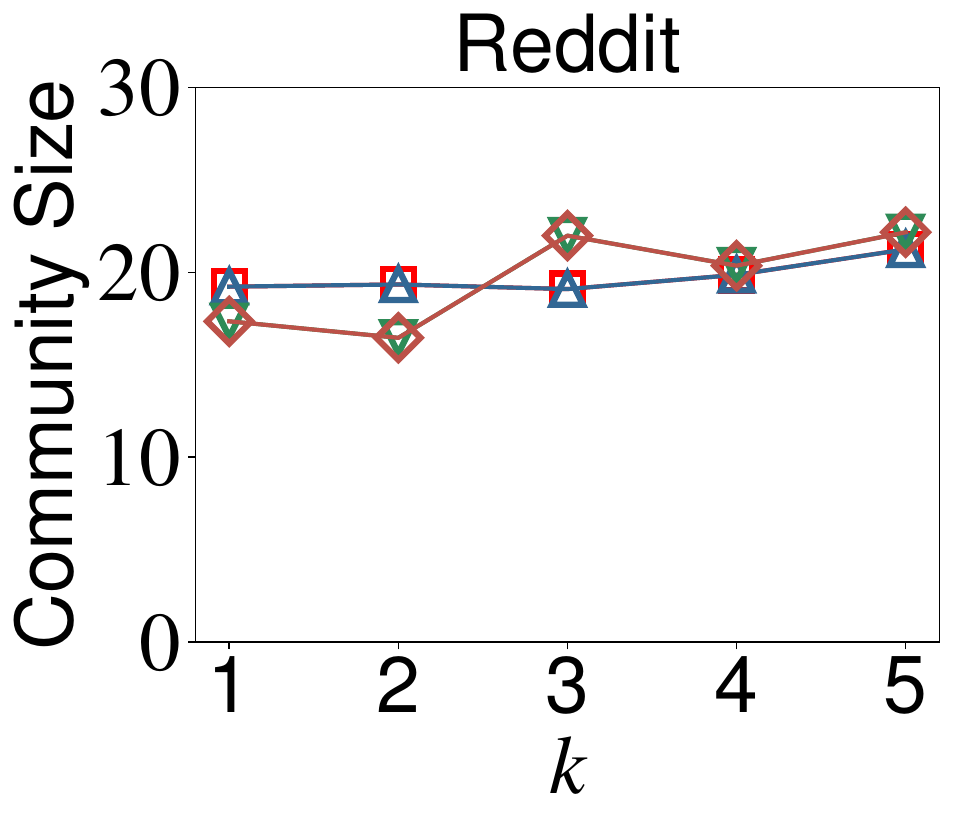}
    \includegraphics[width=0.192\textwidth]{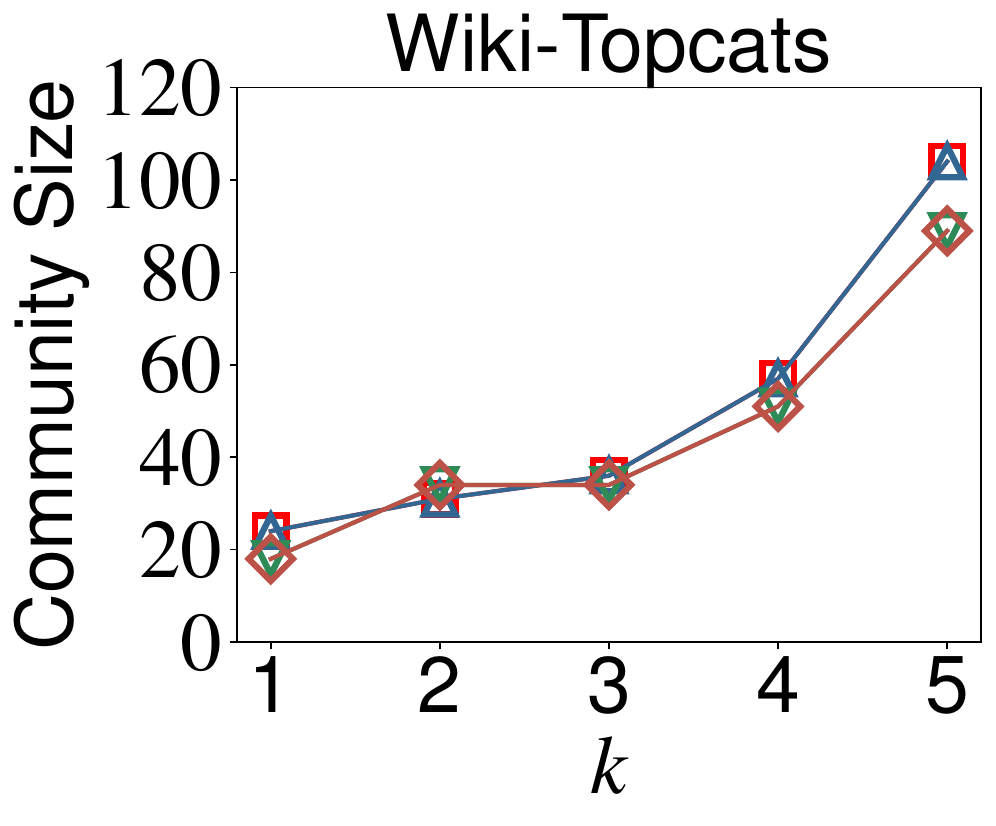}
    \\
    \vspace{1mm}
    \includegraphics[width=0.192\textwidth]{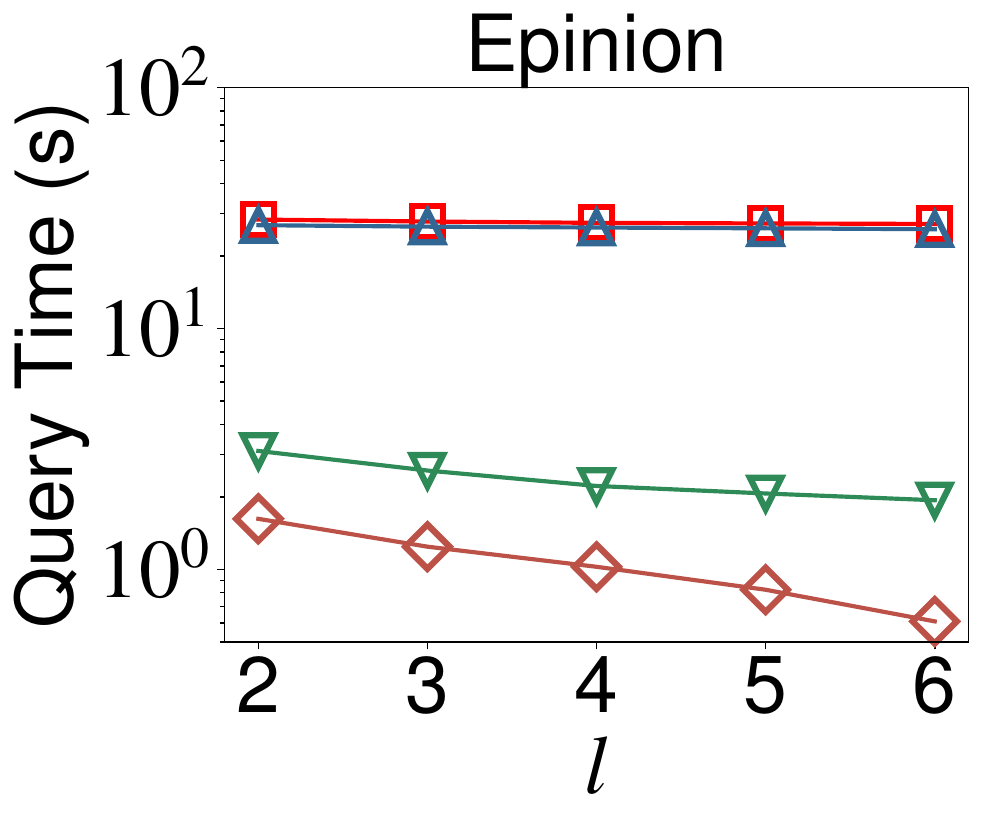}
    \includegraphics[width=0.192\textwidth]{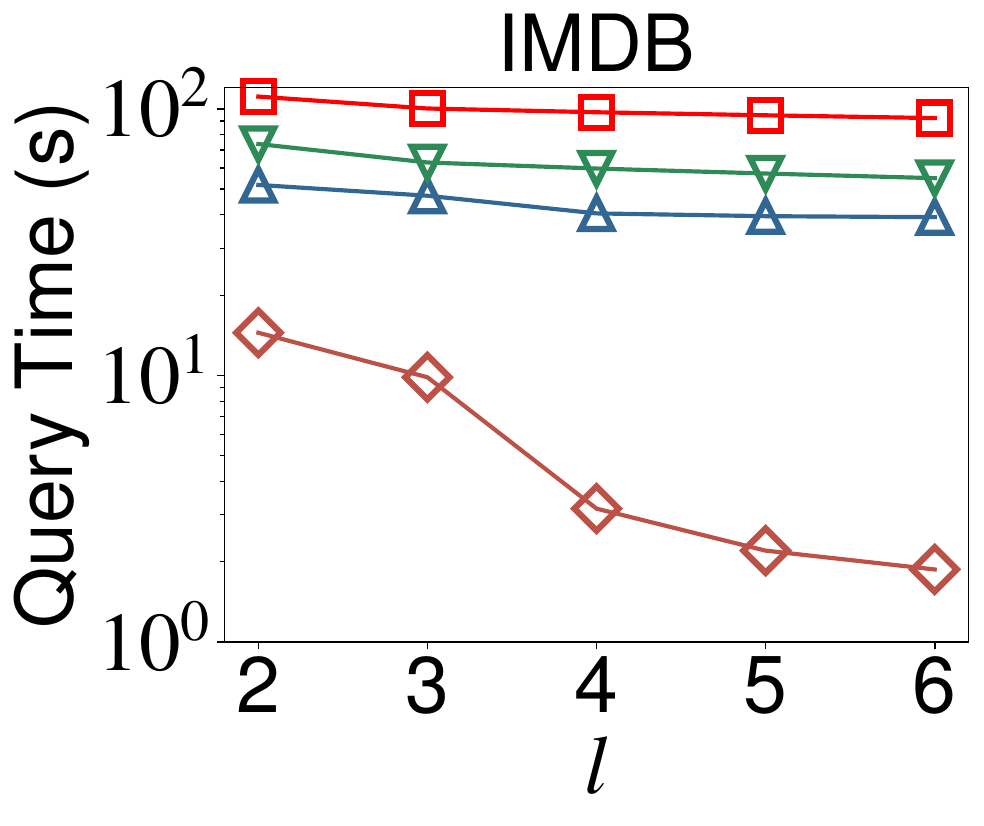}
    \includegraphics[width=0.192\textwidth]{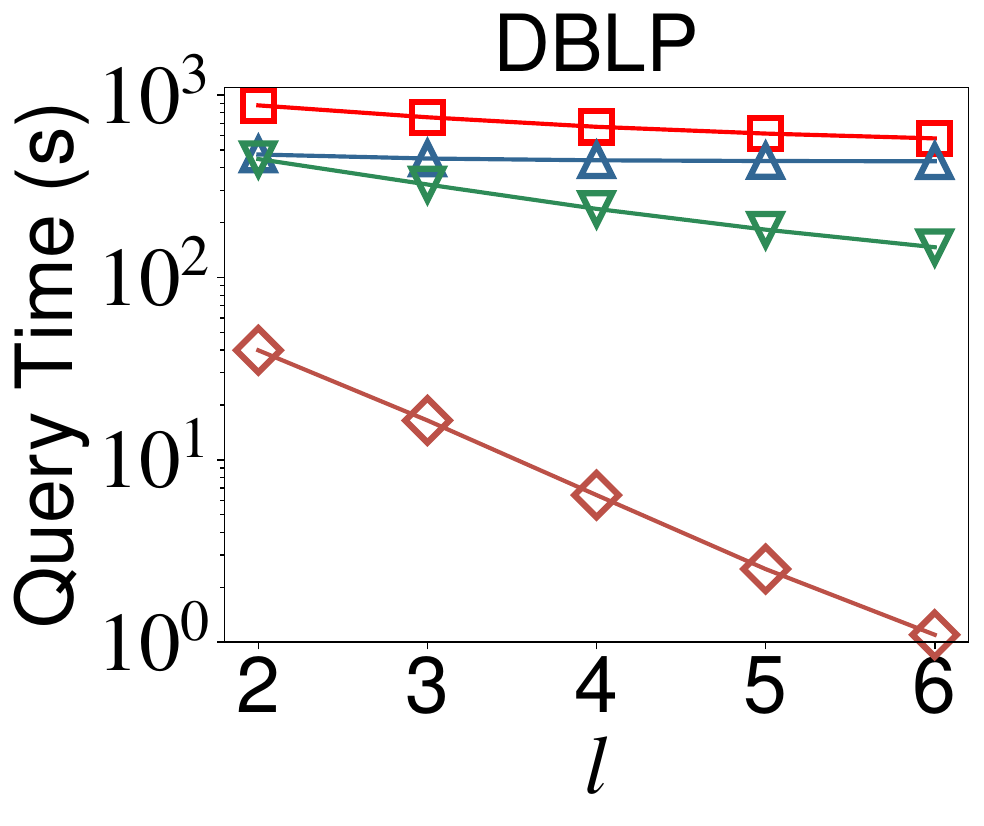}
    \includegraphics[width=0.192\textwidth]{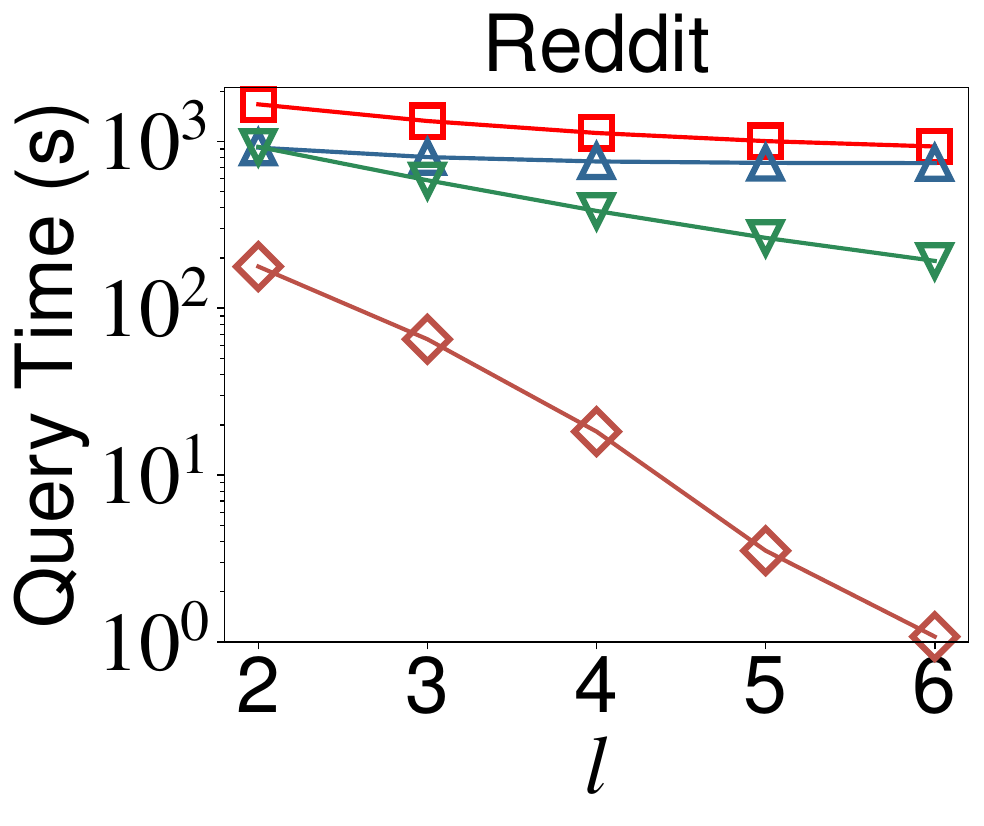}
    \includegraphics[width=0.192\textwidth]{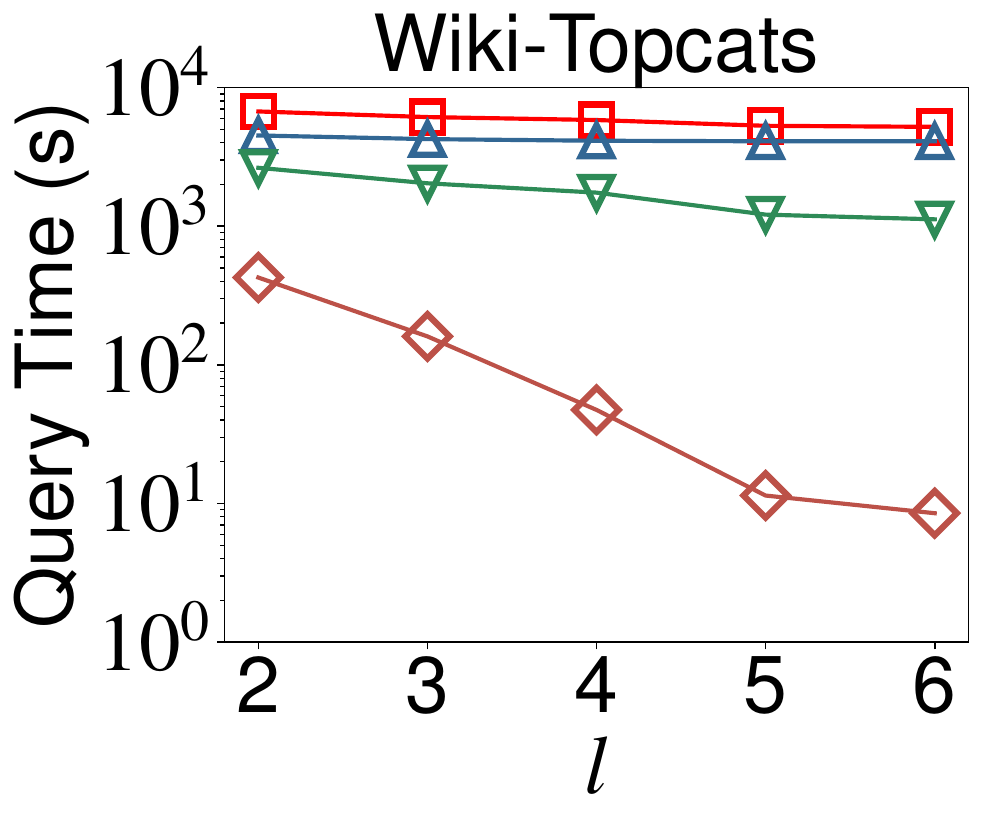}
    \\
    \vspace{1mm}
    \includegraphics[width=0.192\textwidth]{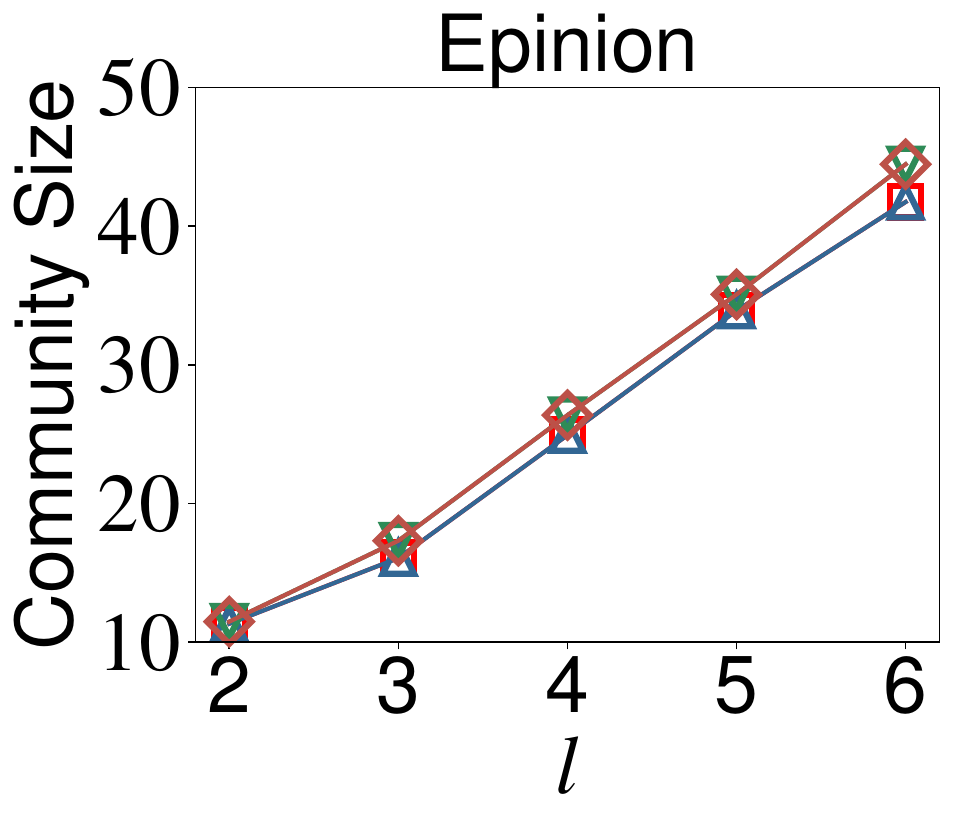}
    \includegraphics[width=0.192\textwidth]{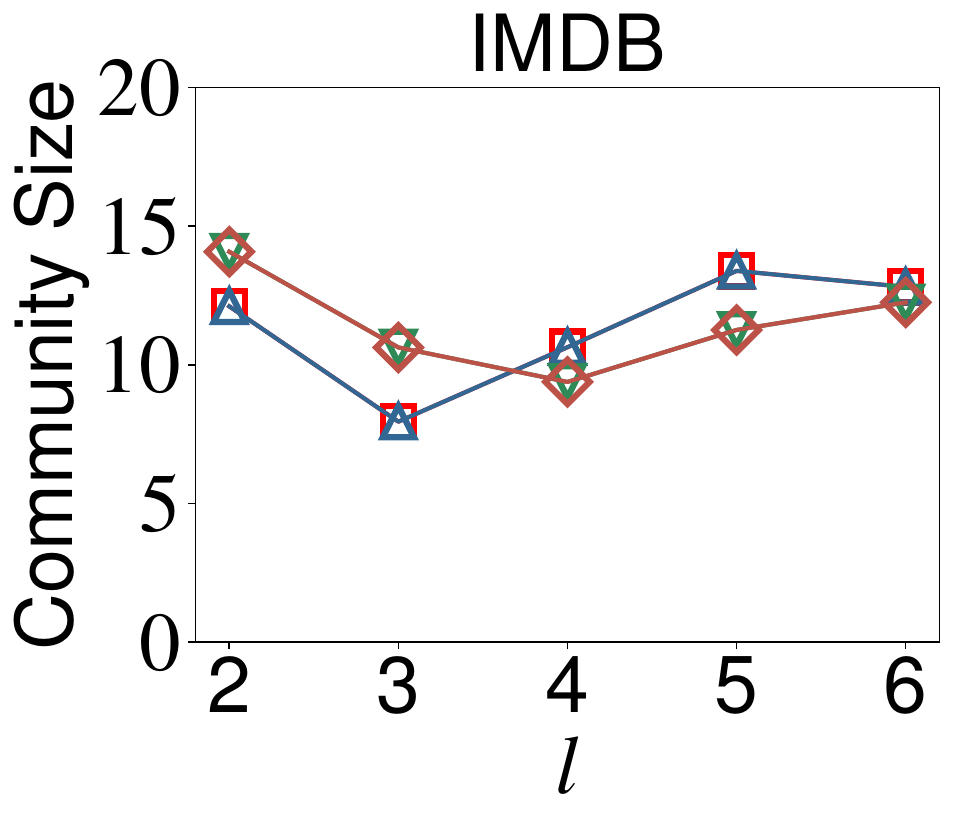}
    \includegraphics[width=0.192\textwidth]{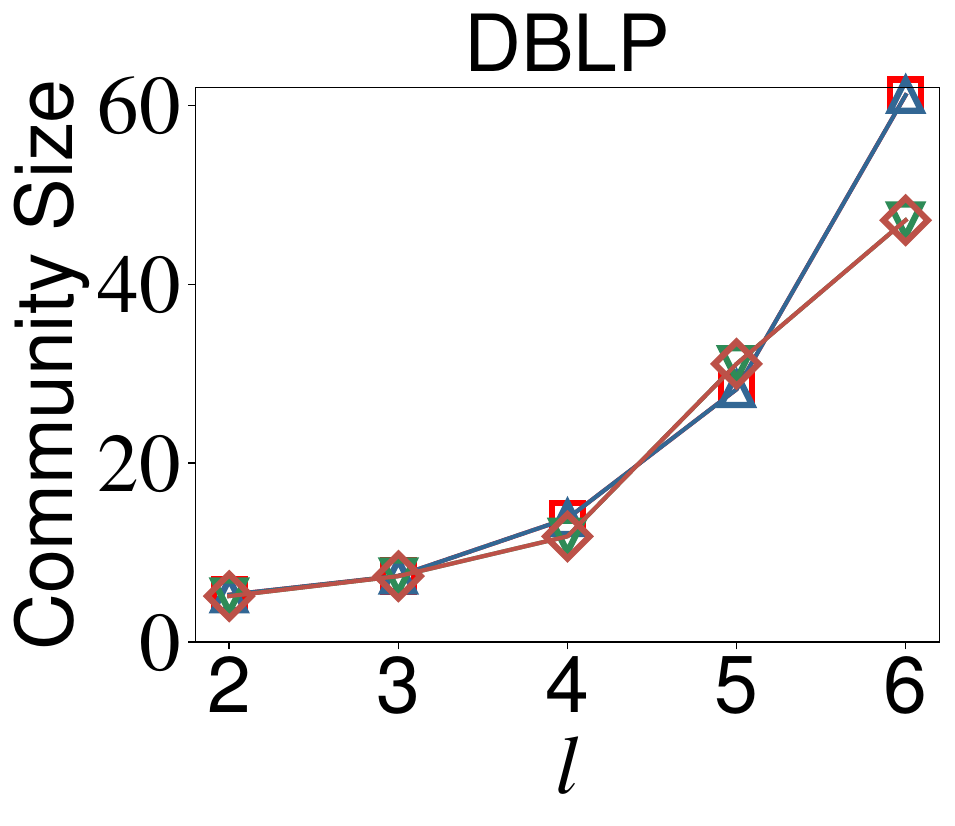}
    \includegraphics[width=0.192\textwidth]{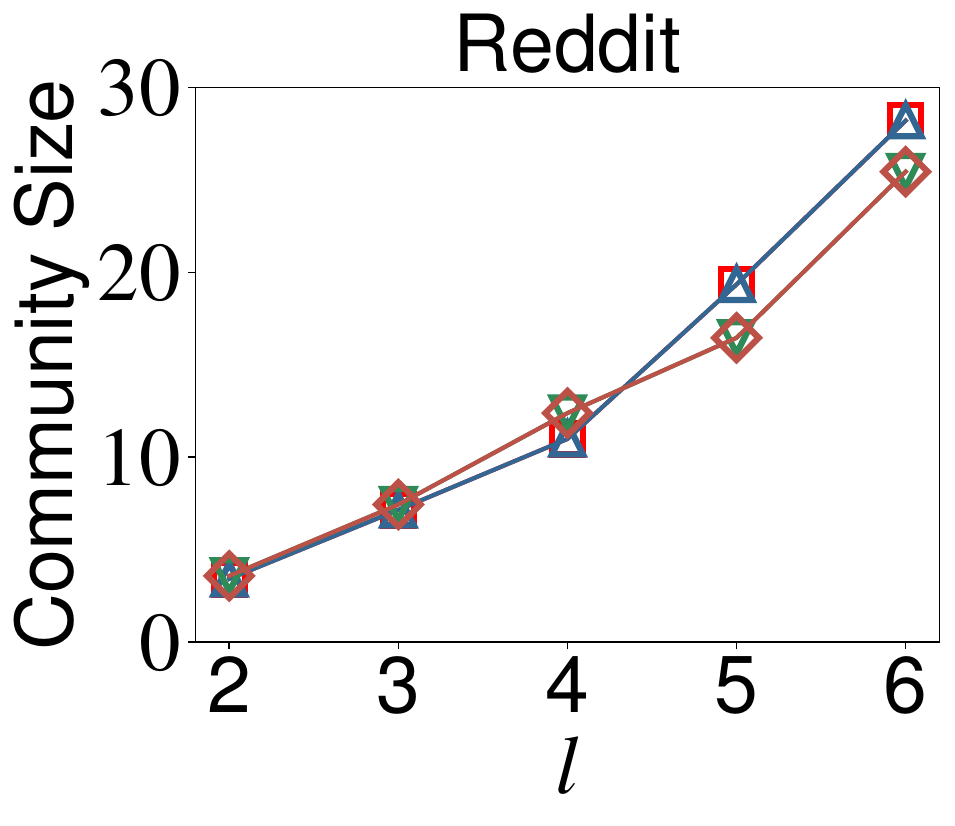}
    \includegraphics[width=0.192\textwidth]{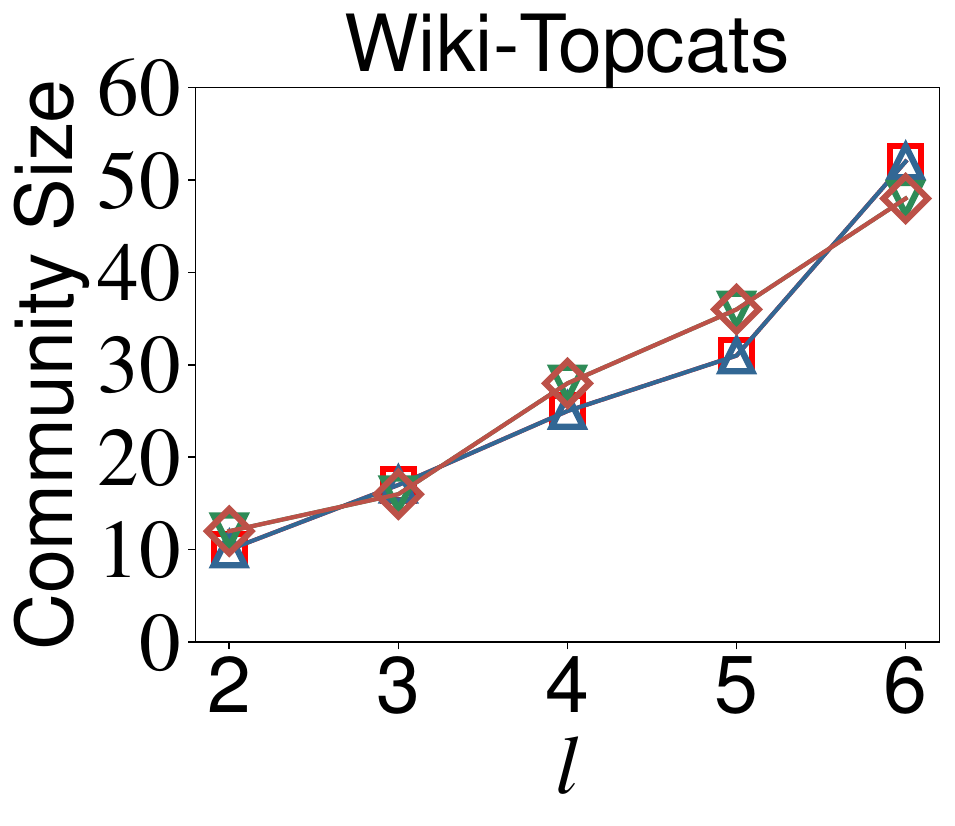}
    \\
    \vspace{1mm}
    \includegraphics[height=4.5mm]{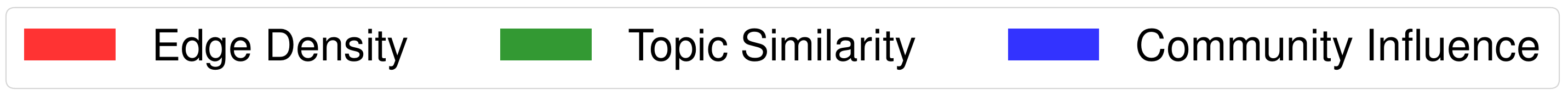}
    \\
    \vspace{1mm}
    \includegraphics[width=0.192\textwidth]{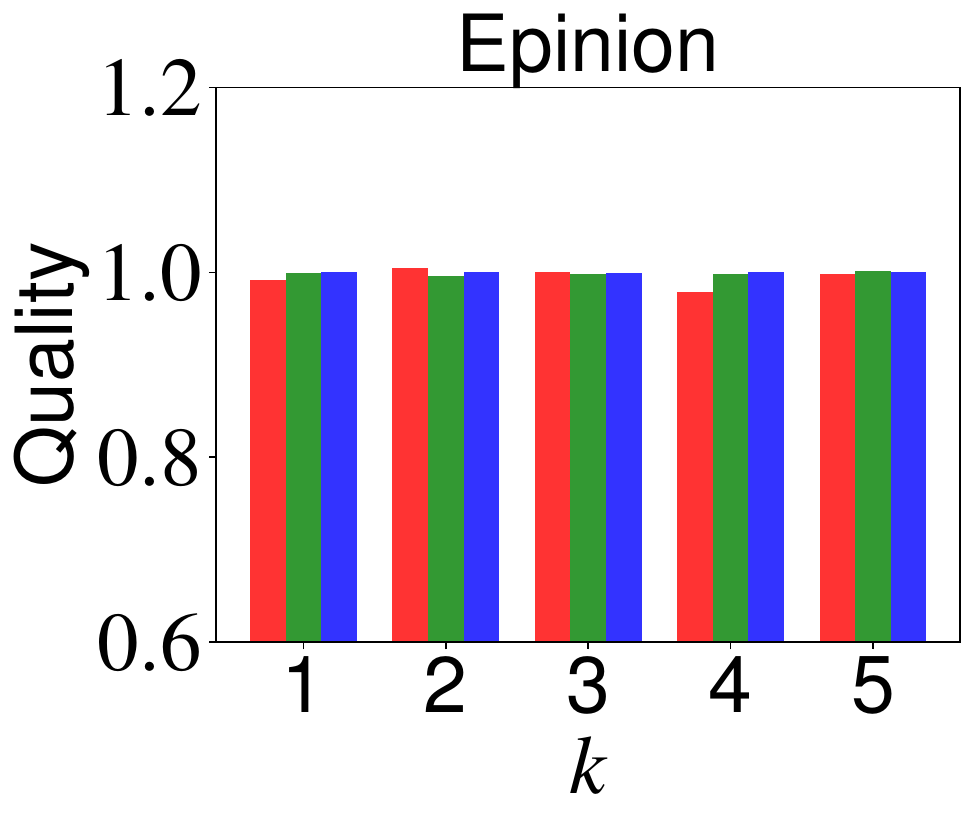}
    \includegraphics[width=0.192\textwidth]{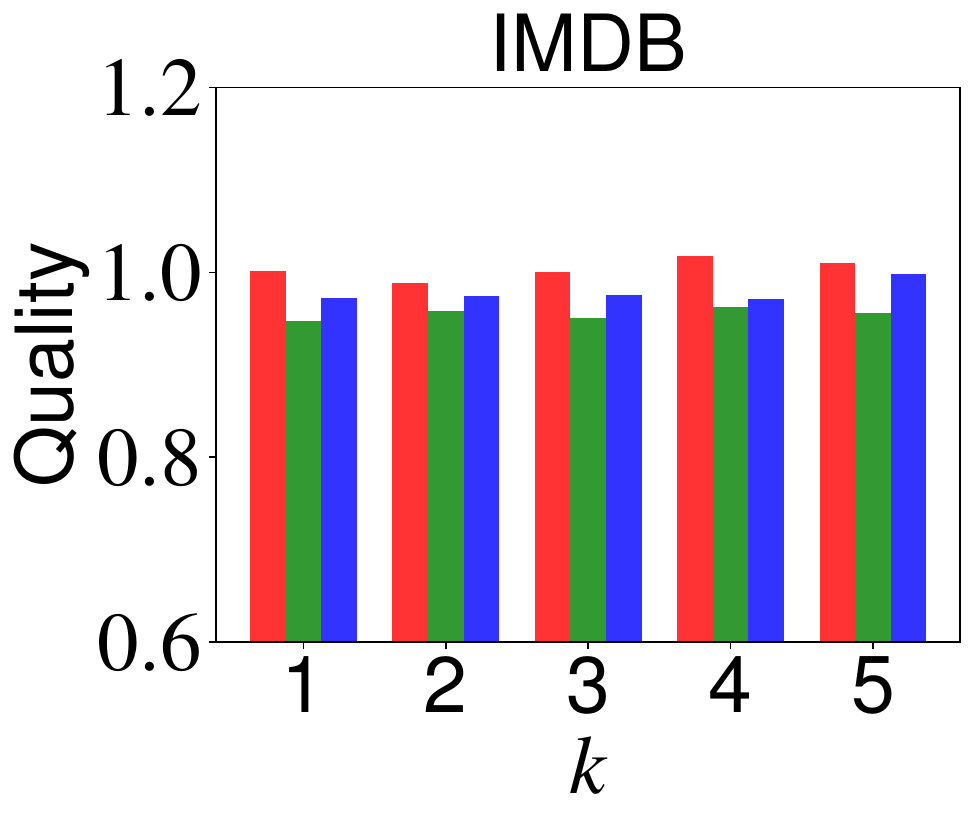}
    \includegraphics[width=0.192\textwidth]{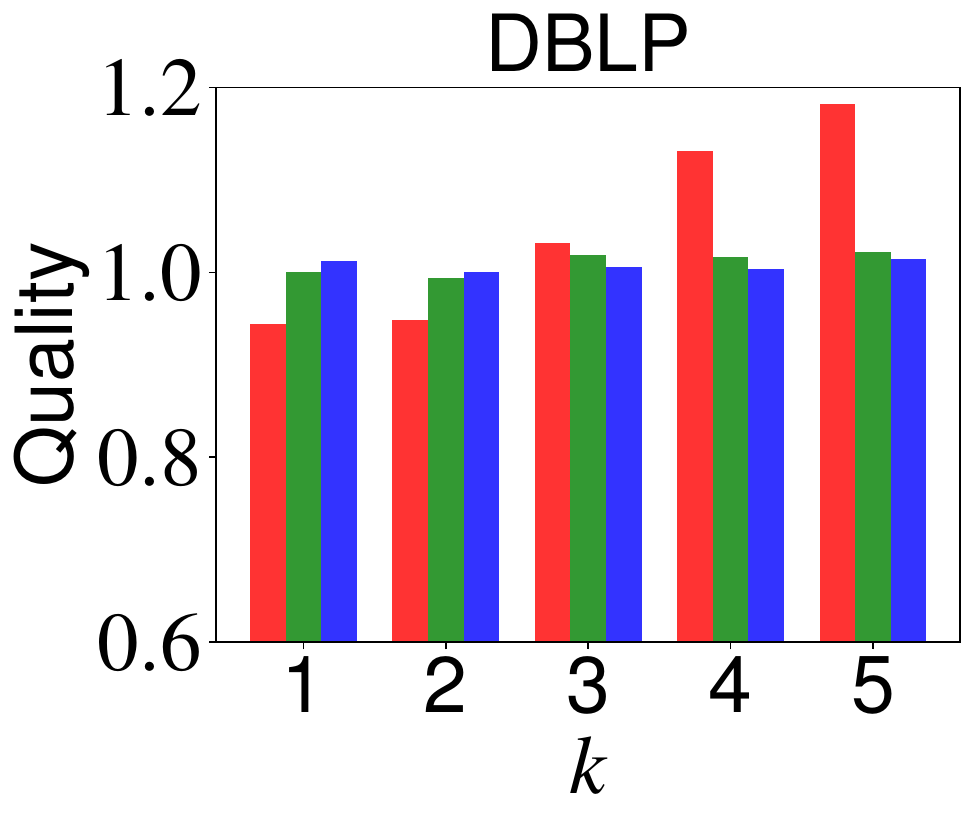}
    \includegraphics[width=0.192\textwidth]{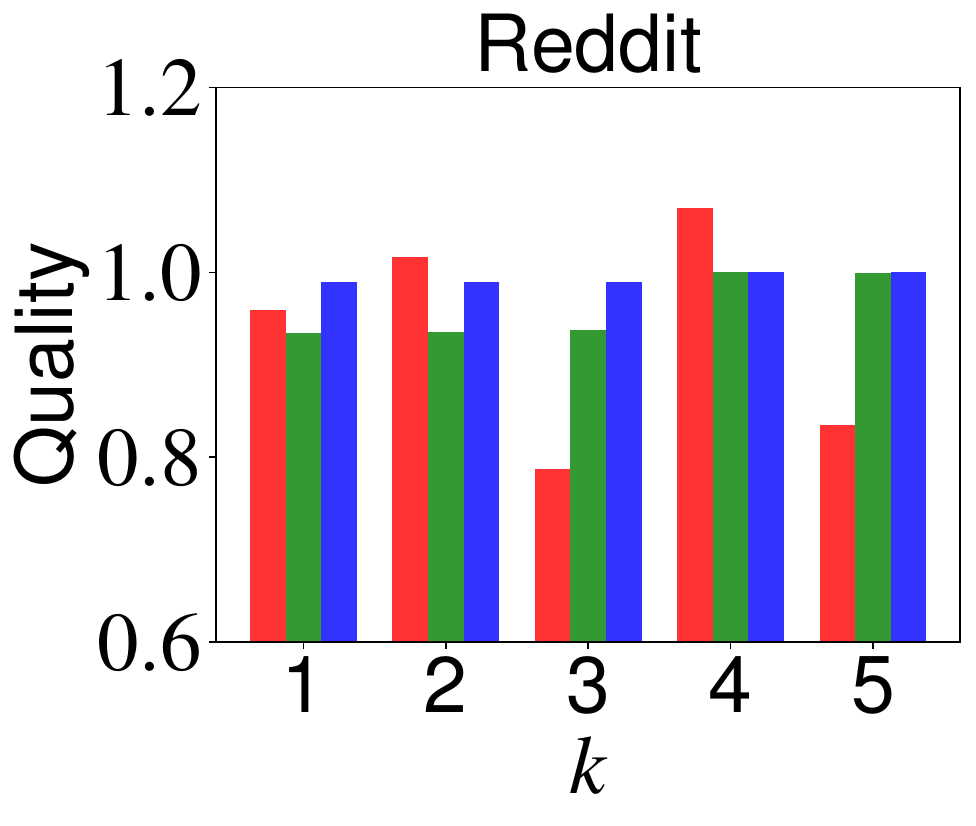}
    \includegraphics[width=0.192\textwidth]{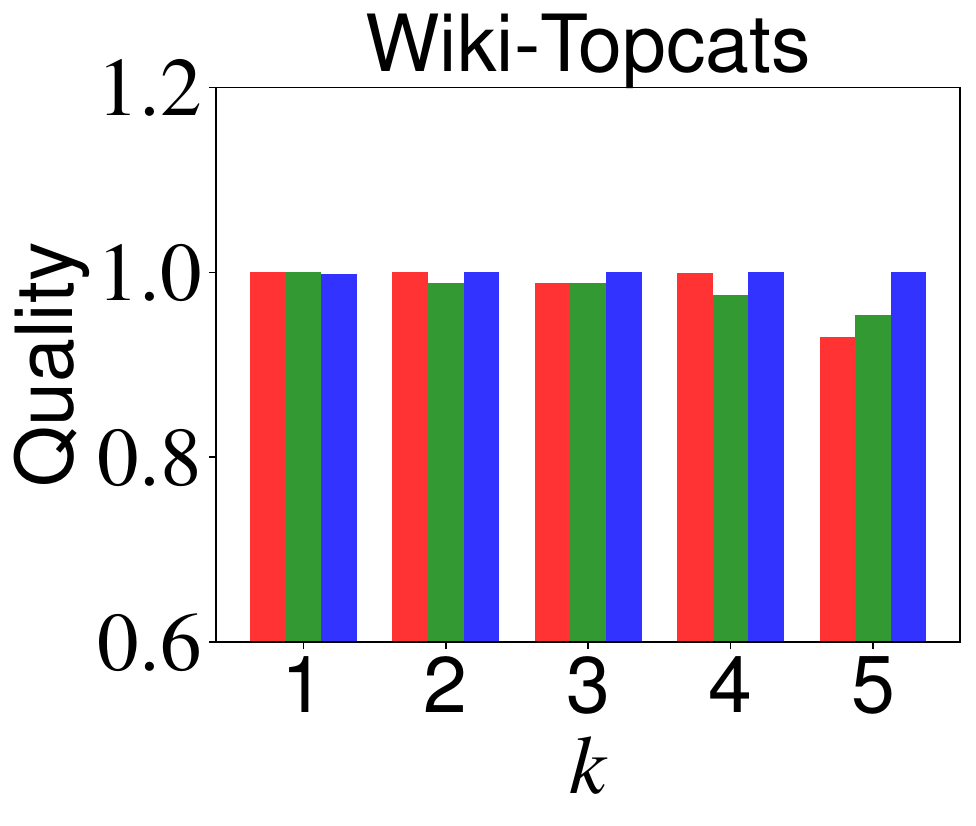}
    \\
    \vspace{1mm}
    \includegraphics[width=0.192\textwidth]{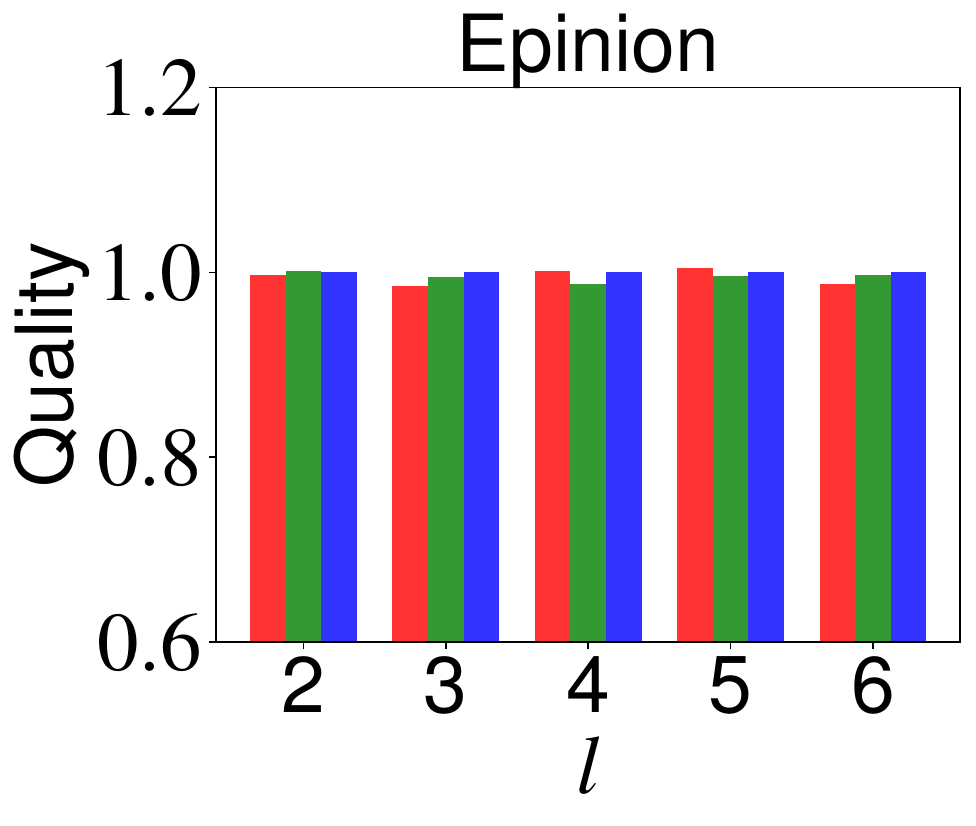}
    \includegraphics[width=0.192\textwidth]{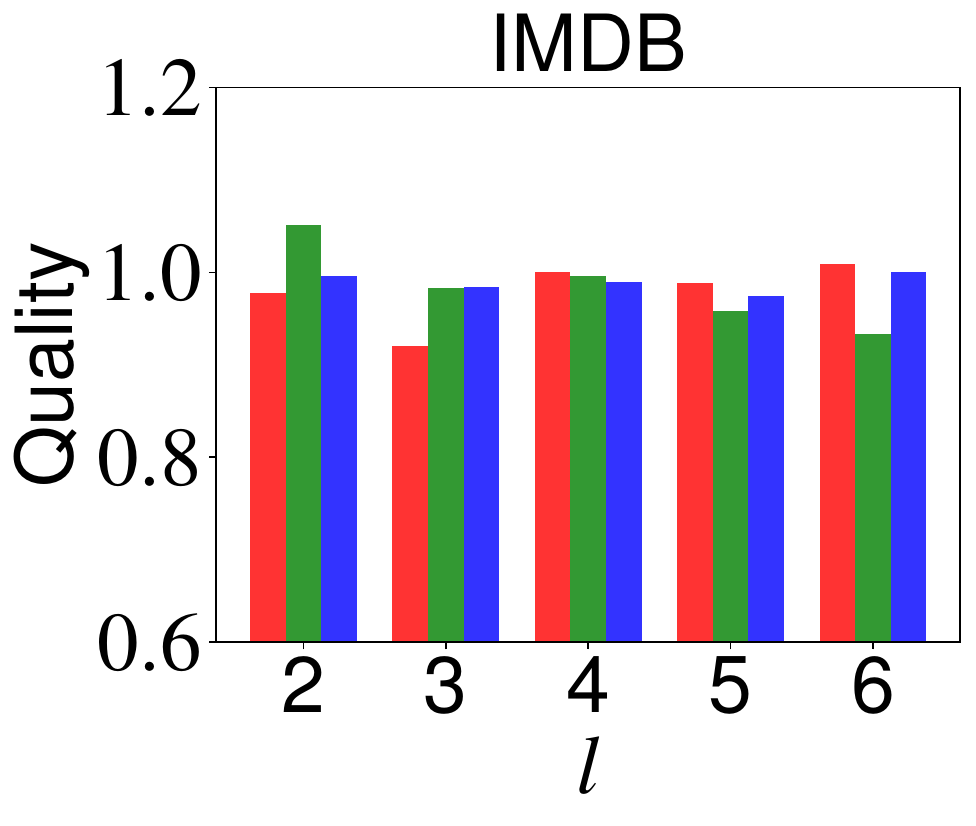}
    \includegraphics[width=0.192\textwidth]{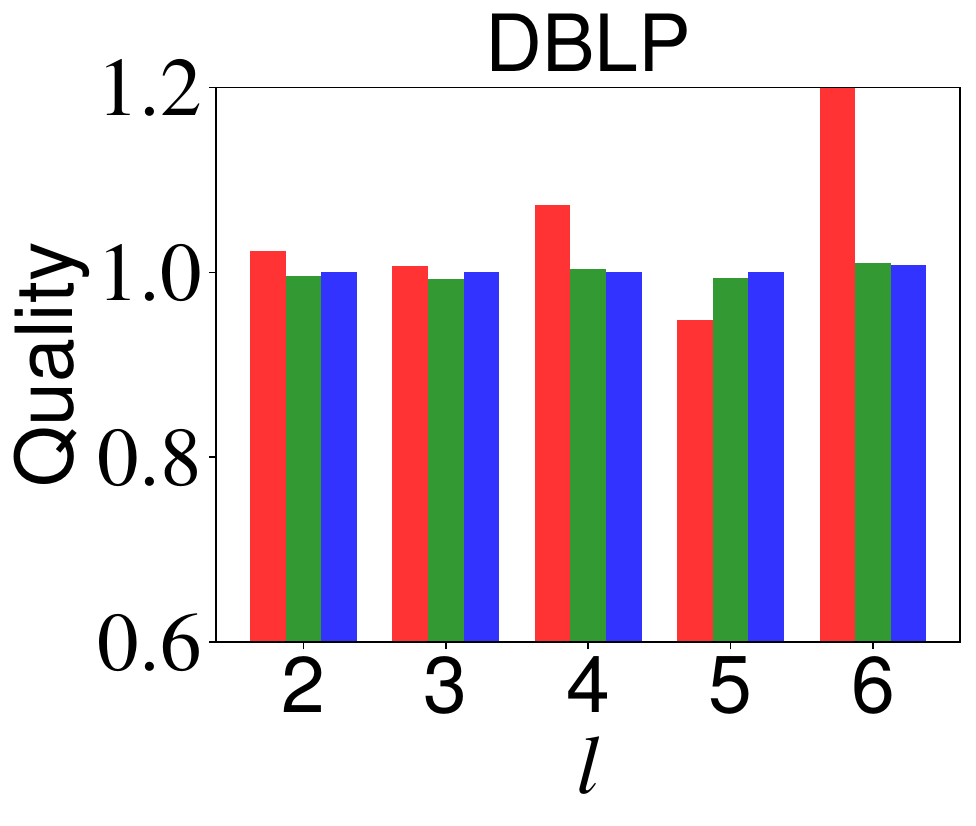}
    \includegraphics[width=0.192\textwidth]{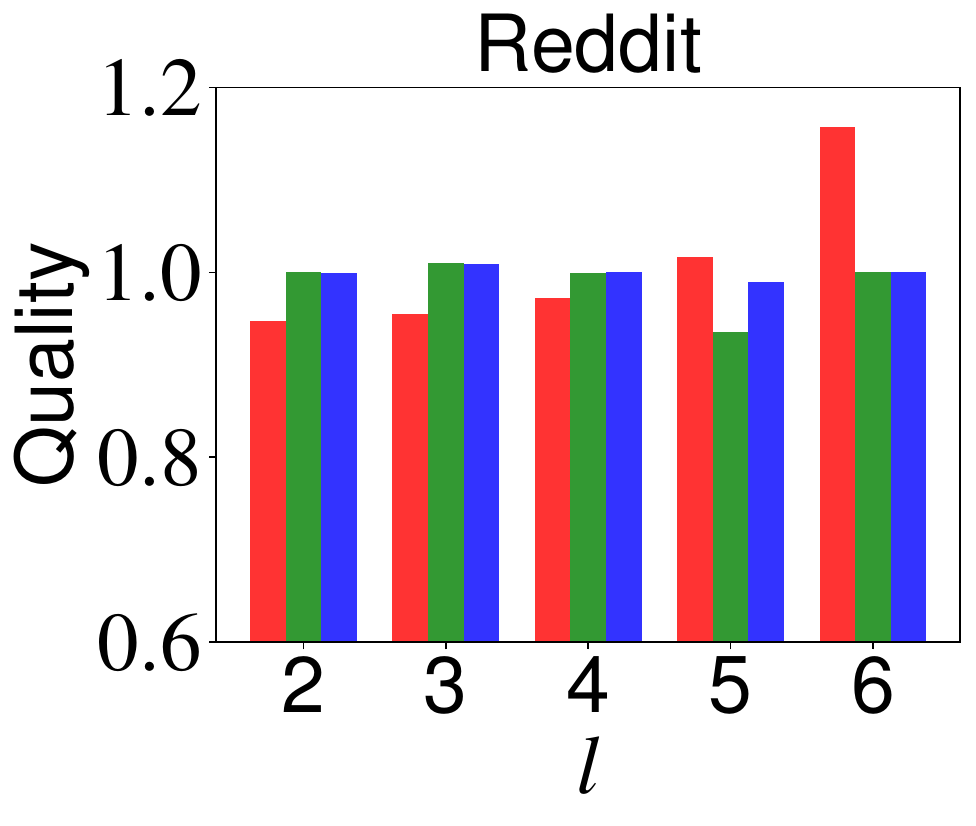}
    \includegraphics[width=0.192\textwidth]{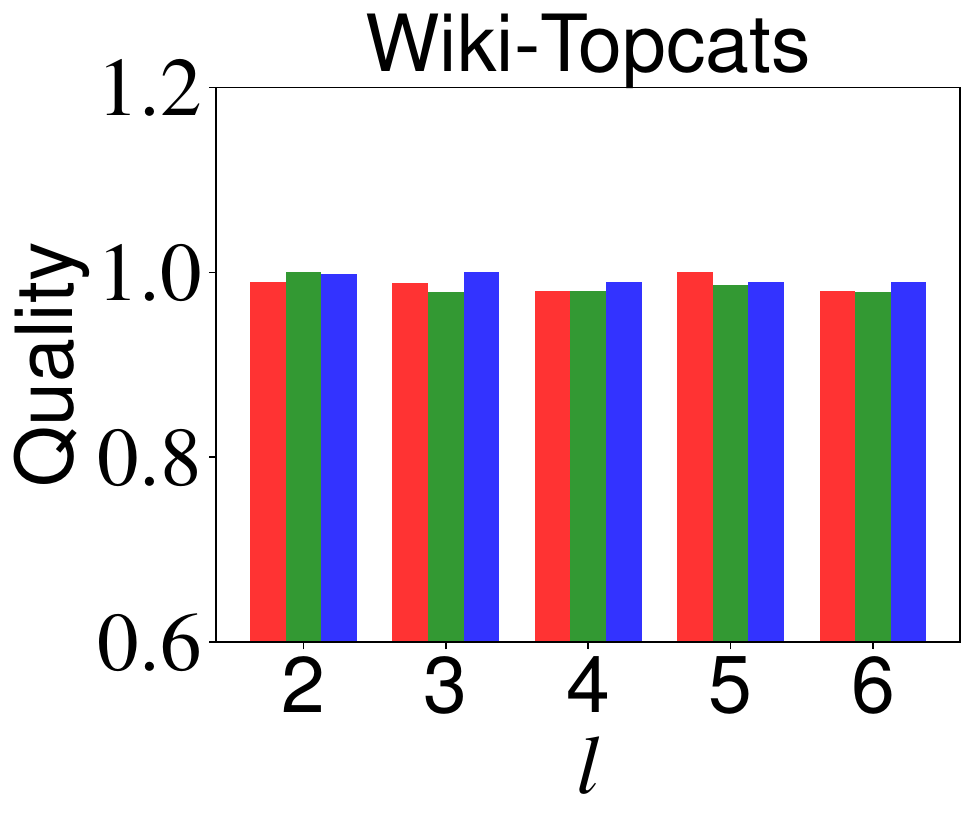}
    \\
    \caption{{Query time and community quality of online and index-based algorithms with varying $k = 1, 2, \dots, 5$ when $l = 5$ or varying $l = 2, 3, \dots, 6$ when $k = 2$ ($\eta = 0.2$).}}
    \label{fig_exp2_k}
\end{figure}

\begin{figure}[t]
    \centering
    \includegraphics[height=3.6mm]{figures/exp4_legend1.pdf}
    \\
    \vspace{1mm}
    \includegraphics[width=0.192\textwidth]{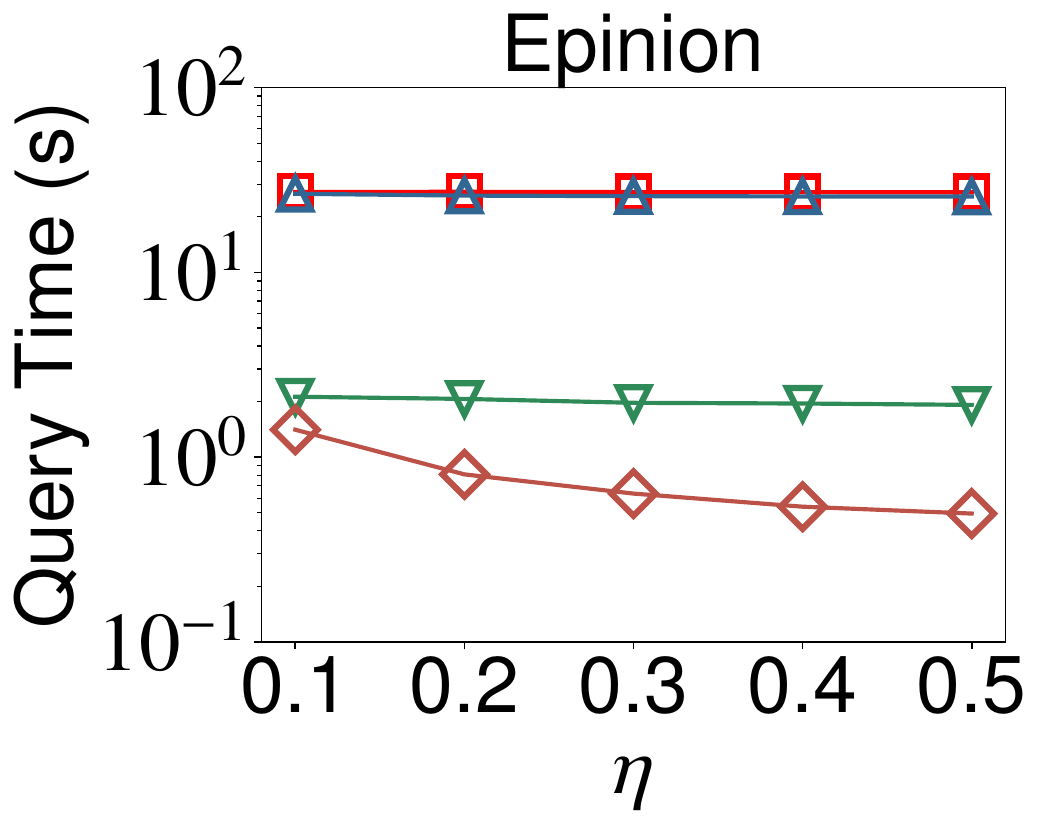}
    \includegraphics[width=0.192\textwidth]{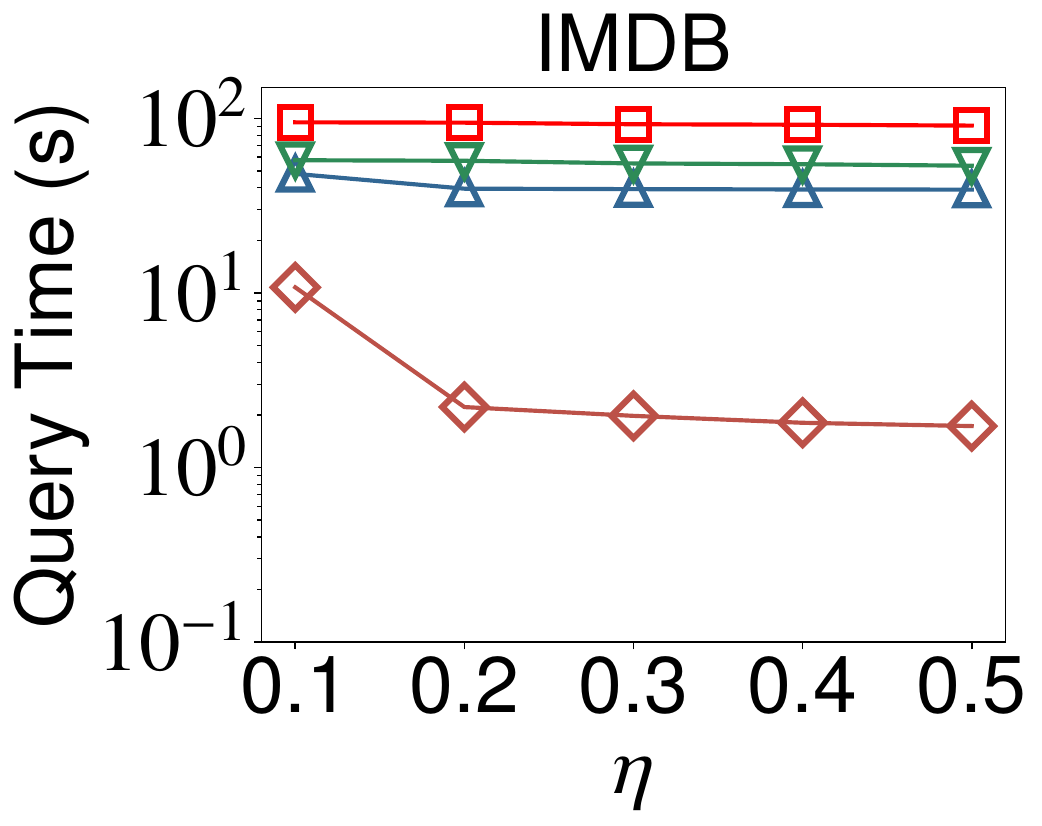}
    \includegraphics[width=0.192\textwidth]{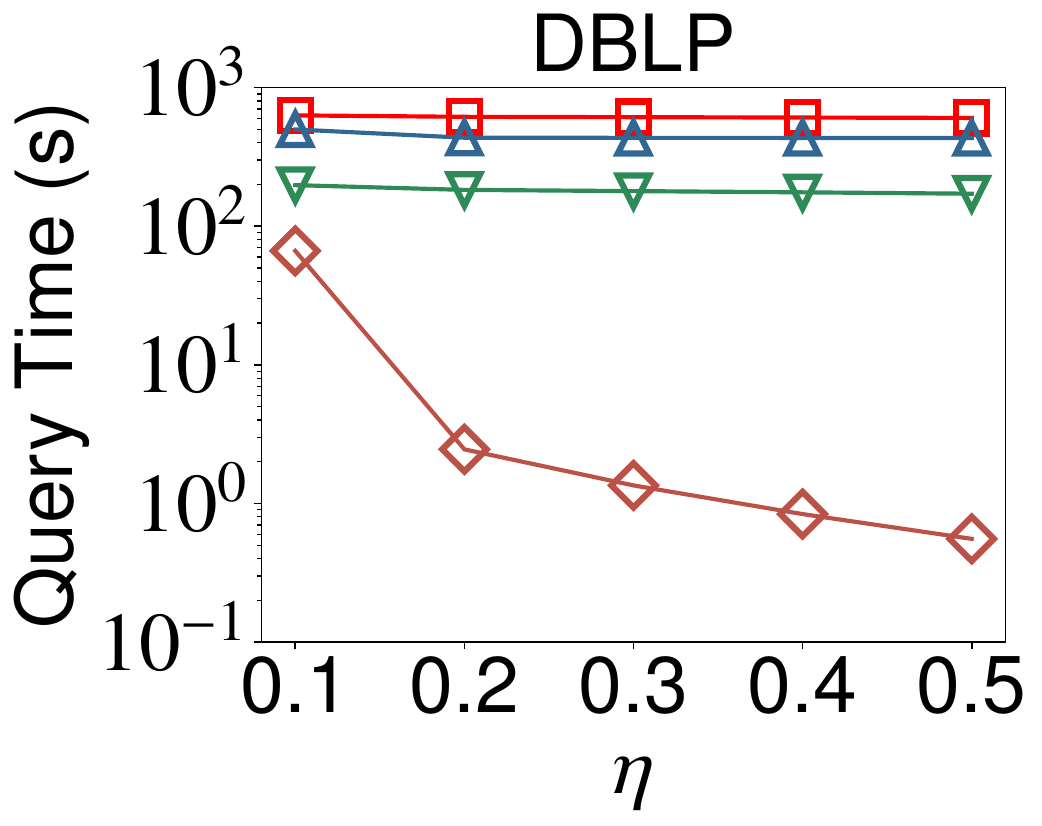}
    \includegraphics[width=0.192\textwidth]{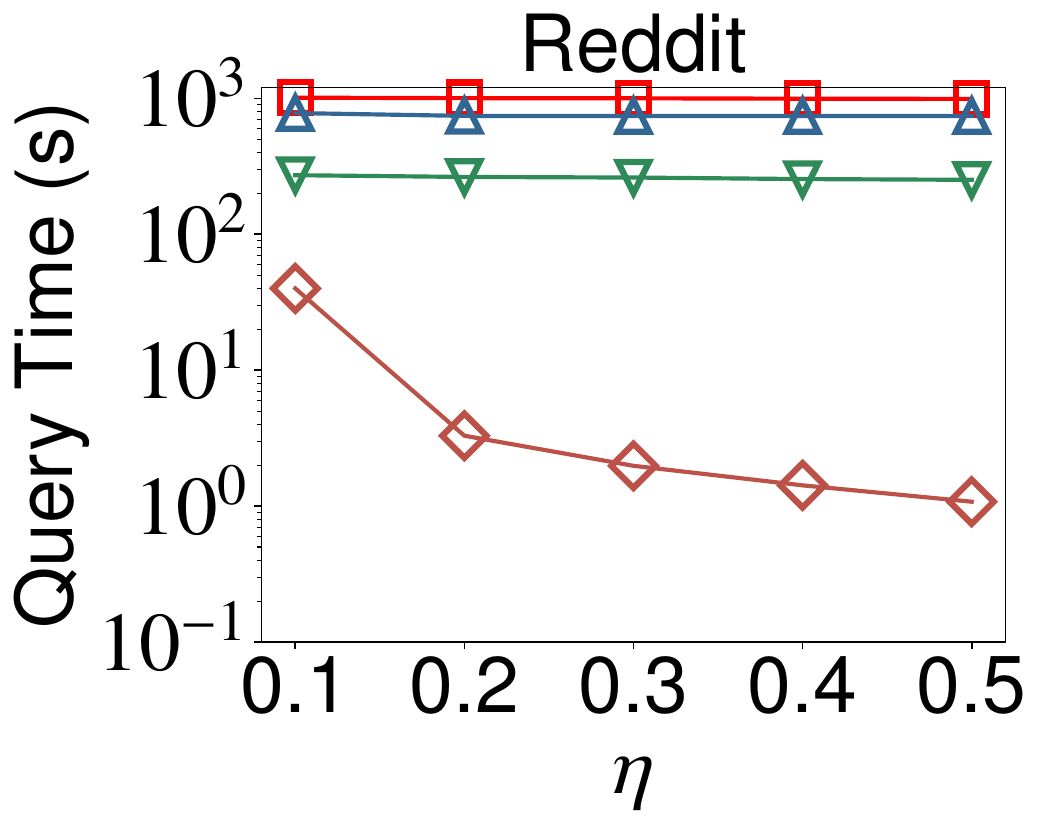}
    \includegraphics[width=0.192\textwidth]{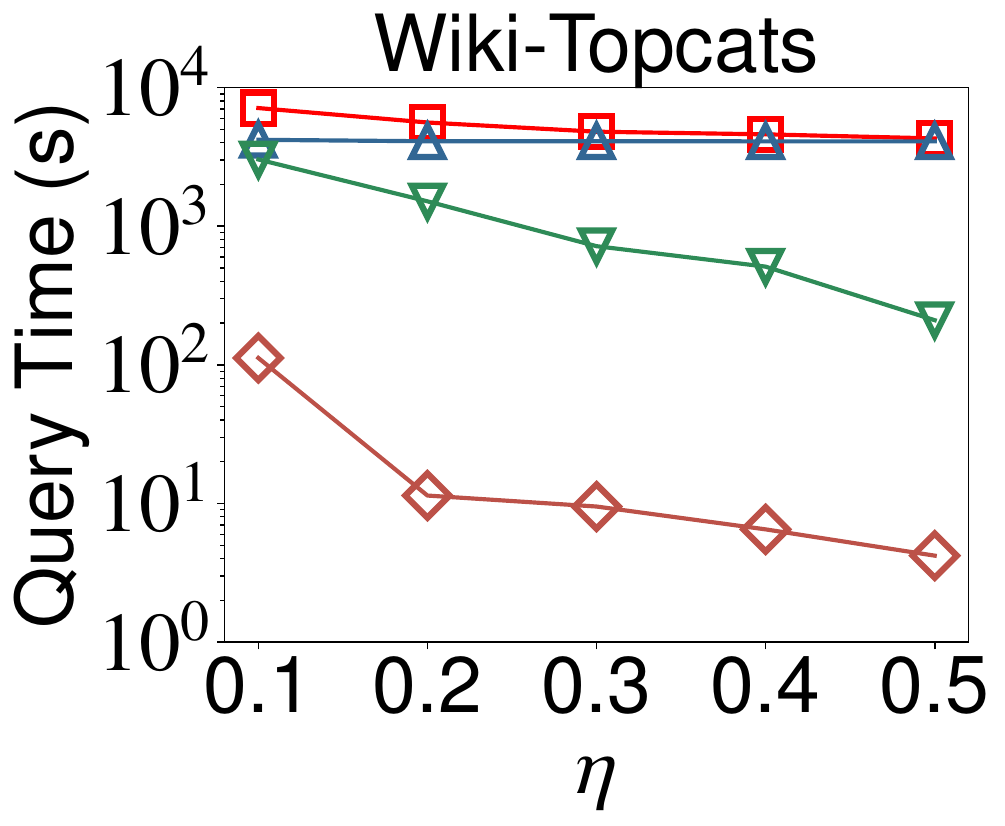}
    \\
    \vspace{1mm}
    \includegraphics[width=0.192\textwidth]{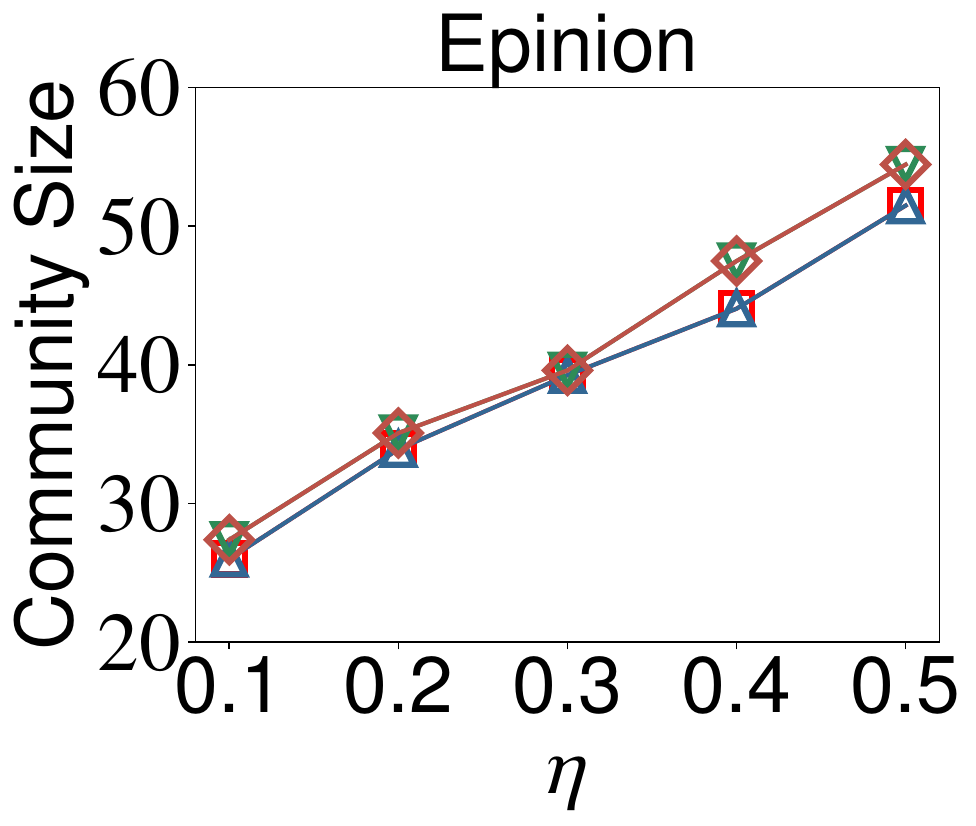}
    \includegraphics[width=0.192\textwidth]{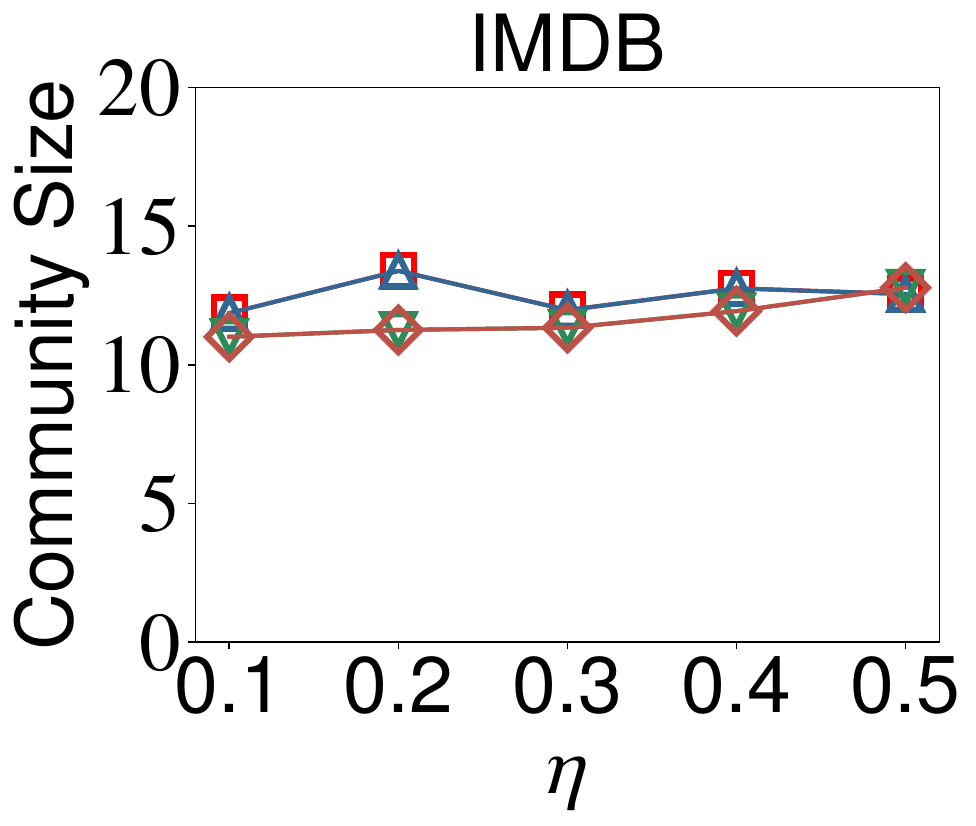}
    \includegraphics[width=0.192\textwidth]{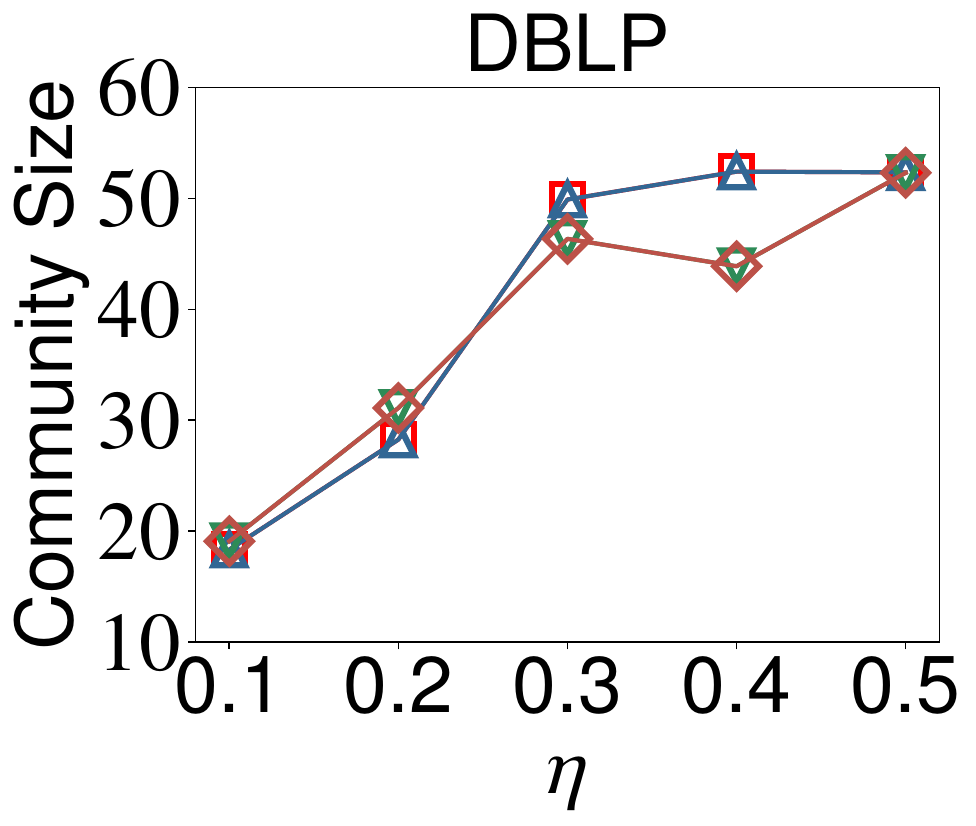}
    \includegraphics[width=0.192\textwidth]{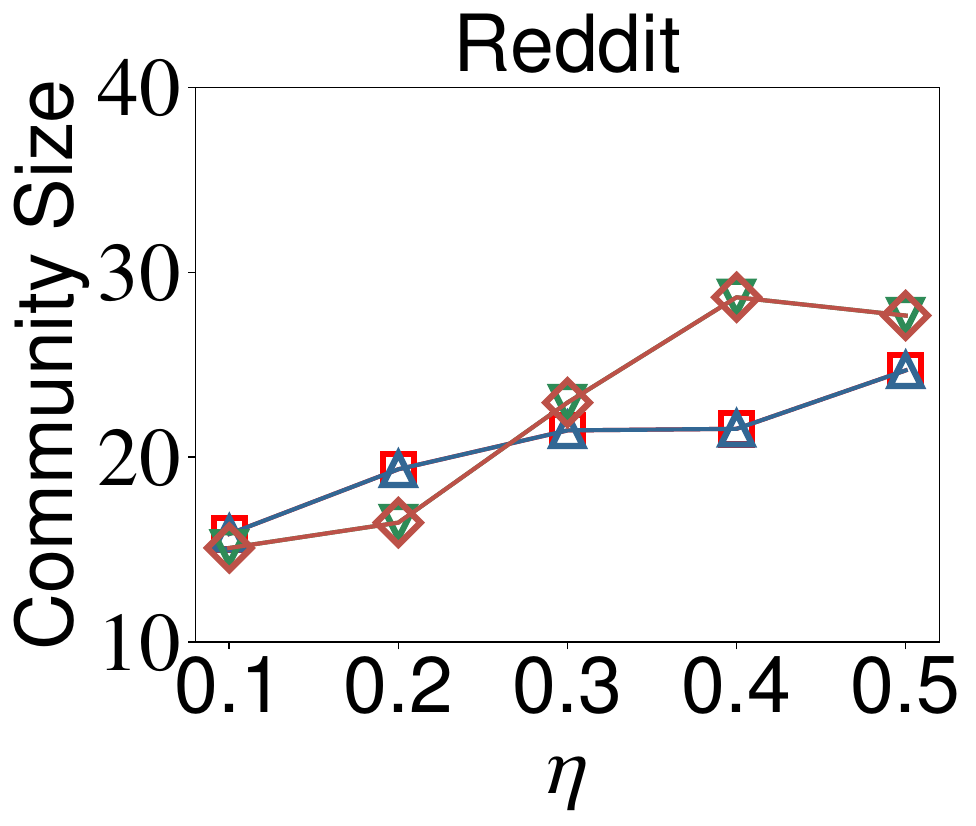}
    \includegraphics[width=0.192\textwidth]{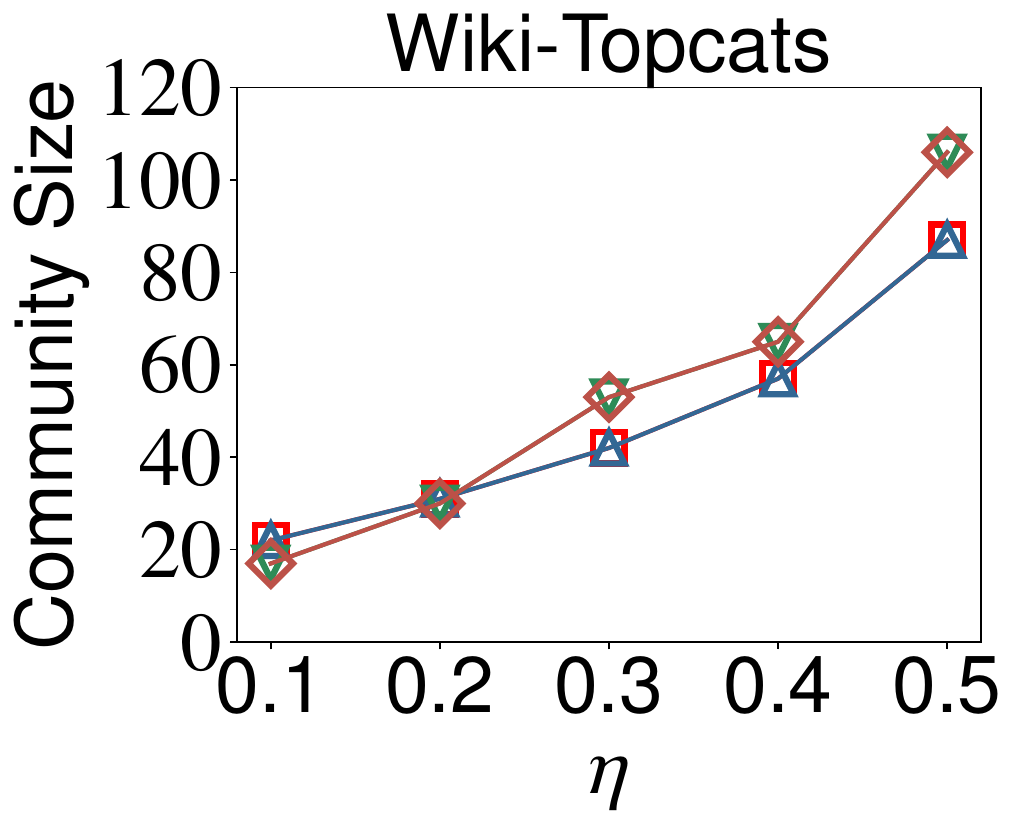}
    \\
    \vspace{1mm}
    \includegraphics[height=4.5mm]{figures/exp4_legend2.pdf}
    \\
    \vspace{1mm}
    \includegraphics[width=0.192\textwidth]{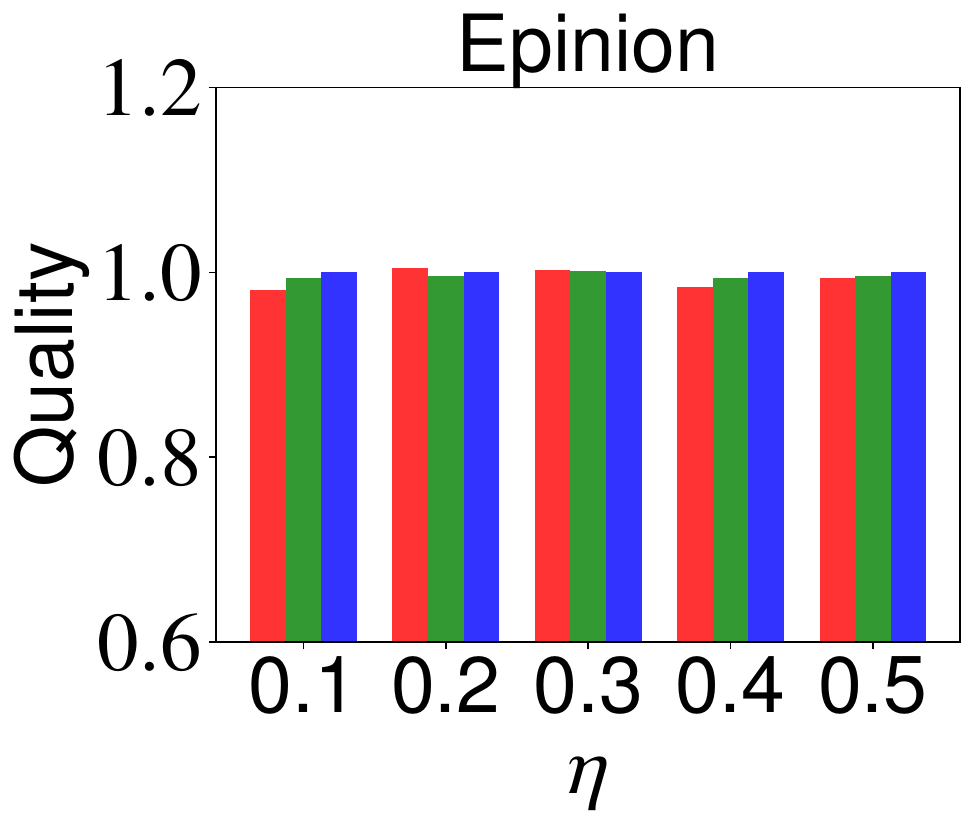}
    \includegraphics[width=0.192\textwidth]{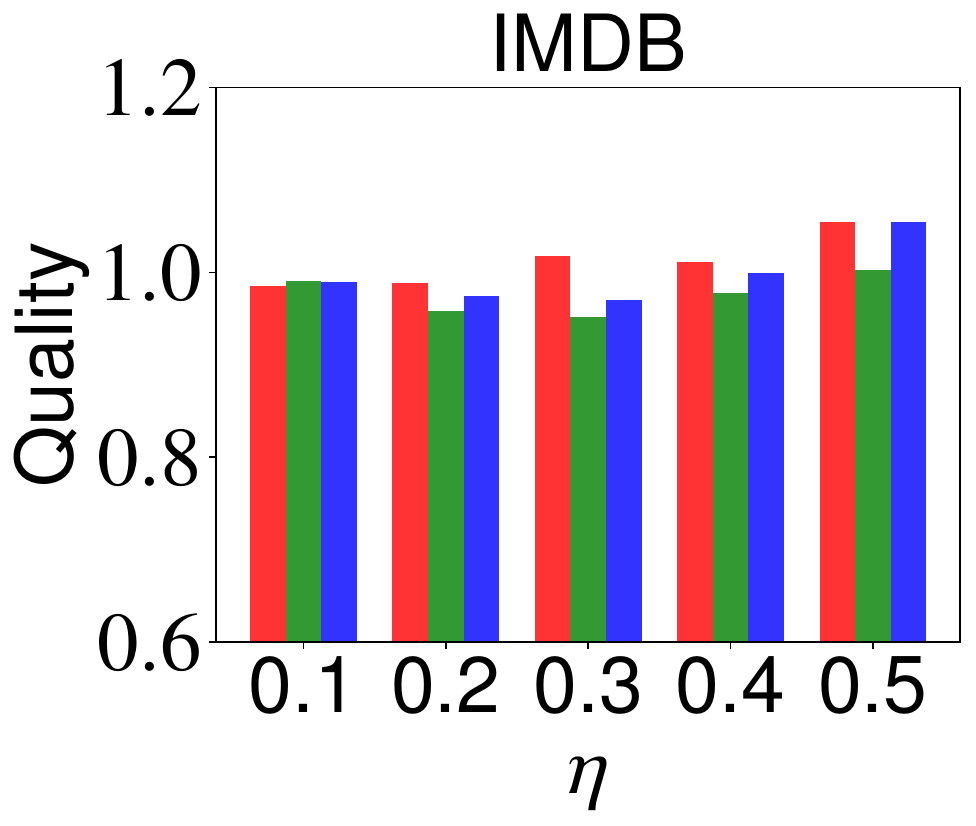}
    \includegraphics[width=0.192\textwidth]{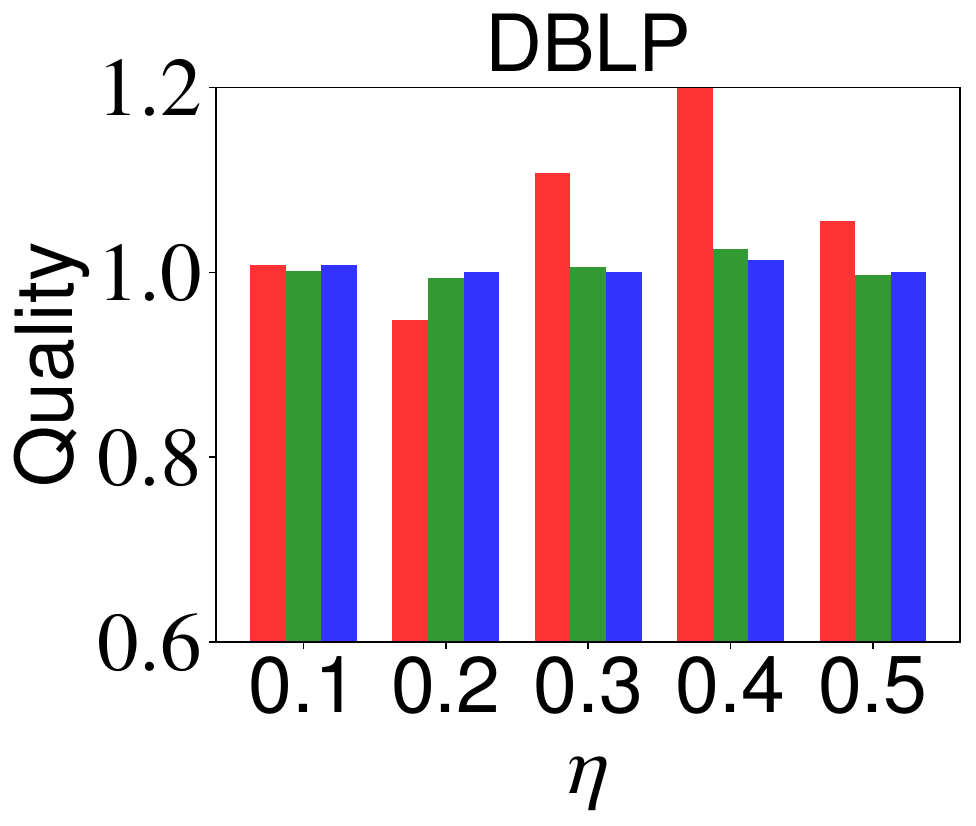}
    \includegraphics[width=0.192\textwidth]{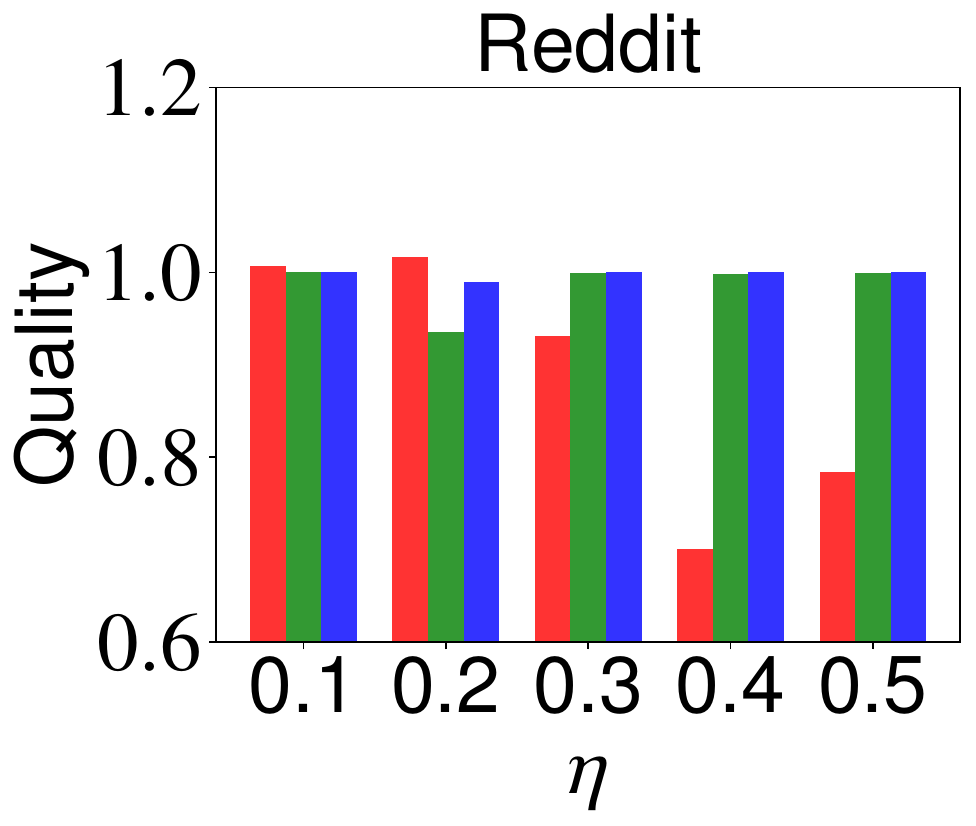}
    \includegraphics[width=0.192\textwidth]{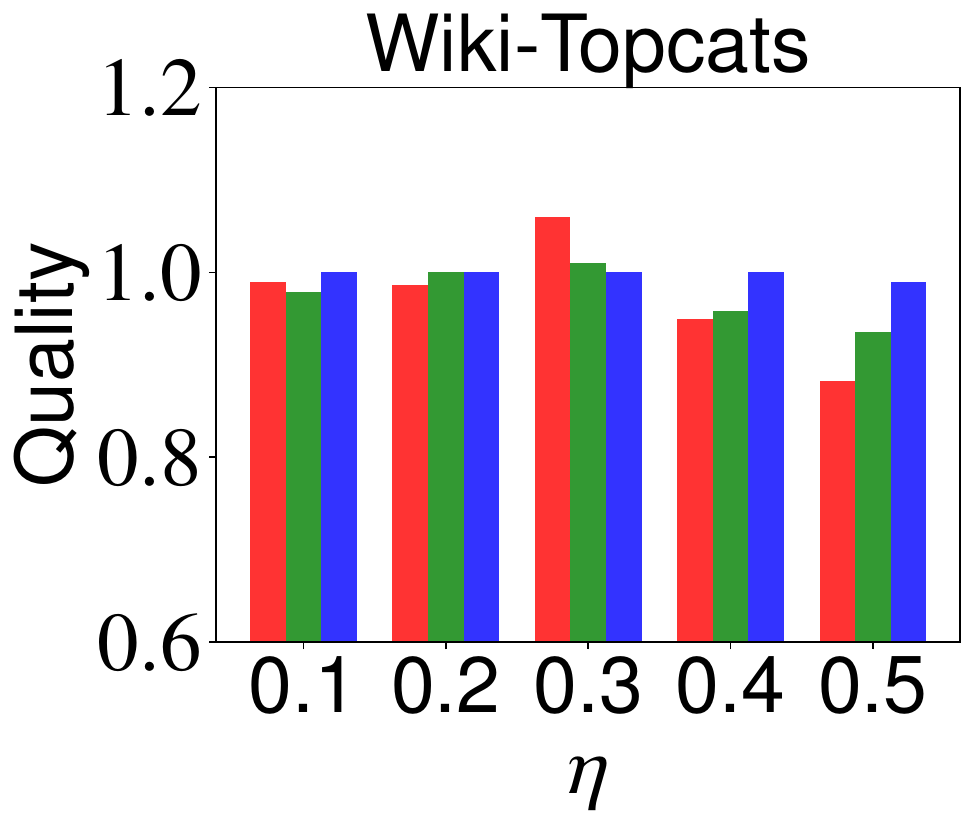}
    \\
    \caption{{Query time and community quality of online and index-based algorithms with varying $\eta = 0.1, 0.2, \dots, 0.5$ when $k = 2$ and $l = 5$.}}
    \label{fig_exp2_3}
\end{figure}

\paragraph{Exp-2: Effect of Parameters $k, l, \eta$}
We compare the time efficiency and community quality of online and index-based algorithms by varying the parameters $k$, $l$, and $\eta$.
The results are presented in Figs.~\ref{fig_exp2_k}--\ref{fig_exp2_3}.
{To analyze the impact of each structure on the index-based algorithm separately, we also present the results when using only a single indexing structure at a time (i.e., \textsf{Online} + \cidx and \textsf{Online} + \iidx).}

In terms of time efficiency, we observe that on all datasets, as the values of $k$, $l$, and $\eta$ increase, the online and index-based algorithms generally take shorter times for each \prob query.
This is because the sizes of maximal $(k, l, \eta)$-cores decrease as any of these values increases, and the time complexity of both algorithms is dominated by searching for maximal $(k, l, \eta)$-cores.
Furthermore, the index-based algorithm runs significantly faster than the online algorithm across all datasets.
The speed-up ratios of the index-based algorithm over the online algorithm can be up to three orders of magnitude in some cases.
{We also observe that \cidx and \iidx marginally reduce query time when either of them is used separately. Nevertheless, the acceleration ratios are much lower than when using them together. This confirms that both \cidx and \iidx are essential for \prob processing.}

In terms of community quality, we use the communities returned by the online algorithm as baselines and compare them with those of the index-based algorithm for the same three measures adopted in Exp-1: \emph{edge density}, \emph{topic similarity}, and \emph{influence}.
{Note that online search and \cidx provide the same results for $(k, l, \eta)$-core computation. The online algorithm and \textsf{Online} + \cidx provide the same communities, while the index-based algorithm and \textsf{Online} + \iidx also return the same communities. Therefore, \textsf{Online} + \cidx and \textsf{Online} + \iidx are omitted from the community quality comparison.}
We present the ratios of the communities of two algorithms for each measure.
As shown in Figs.~\ref{fig_exp2_k}--\ref{fig_exp2_3}, the ratios are greater than $0.9$ in most cases.
Meanwhile, their community sizes are also close to each other.
In some cases, especially on the Reddit dataset, the communities of the index-based algorithm show inferior quality compared to those of the online algorithm.
This is because the influence scores obtained from \iidx often have errors larger than those computed by RIS.
Due to errors in influence scores, the index-based algorithm may include or miss some vertices in its results.
In the former case, the edge density scores drop significantly.
In the latter case, the edge density scores increase and the ratios are greater than $1$, while the community sizes are smaller.
However, the ratios always remain at least $0.7$ in all cases.
In summary, the index-based algorithm still provides communities whose quality is comparable to that of the online algorithm in most cases.

\begin{table}[t]
    \footnotesize
    \centering
    \caption{Index size and construction time of \cidx and \iidx.}
    \label{tbl-Index}
    \newcolumntype{C}{>{\centering\arraybackslash}X}
    \begin{tabularx}{\textwidth}{CCCCC}
        \toprule
        \multirow{2}{*}{\textbf{Dataset}} & \multicolumn{2}{c}{\textbf{Index Size (MB)}} & \multicolumn{2}{c}{\textbf{Construction Time (s)}} \\
        \cmidrule(lr){2-3} \cmidrule(lr){4-5}
        & \cidx & \iidx  & \cidx  & \iidx \\
        \midrule
        Epinion & 0.65 & 89.02 & 12.39 & 846.5 \\
        IMDB & 6.92 & 303.9 & 430.2 & 2468.8 \\
        DBLP & 20.57 & 1453.4 & 3001.9 & 14156.6 \\
        Reddit & 11.34 & 5488.3 & 1652.5 & 16622.7 \\
        Wiki-Topcats & 82.51  & 7175.7 & 13742.1 & 31565.5 \\
        \bottomrule
    \end{tabularx}
\end{table}

\vspace{1mm}
\paragraph{Exp-3: Index Overhead}
Table~\ref{tbl-Index} illustrates the index size and construction time of the \cidx and \iidx on each dataset.
We can see that the \iidx takes more time and space than the \cidx, mostly due to running RIS on $h$ topic vectors and storing the estimated influence scores of all vertices for each of them.
For example, on the Reddit dataset, the \iidx occupies approximately $5$GB space and takes about $3$ hours to construct, while the \cidx uses only around $11$MB space and $27$ minutes. 
The size and construction time of the \iidx generally increase with the size of the dataset.
This confirms the result of Lemma~\ref{lm-inf-est}, where the value of $\theta$ increases linearly with the number of vertices $n$.
Unlike the \iidx, the size and construction time of the \cidx do not always increase with $n$.
On sparser datasets with fewer $(k, l, \eta)$-cores, the overhead for constructing the \cidx can decrease even when $n$ is larger.

\begin{figure}[t]
    \centering
    \includegraphics[width=0.6\textwidth]{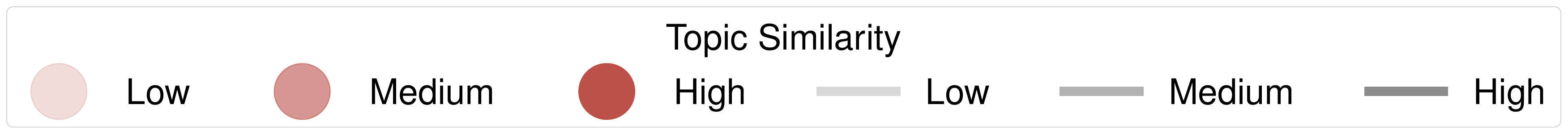}
    \hspace{2mm}
    \includegraphics[width=0.3\textwidth]{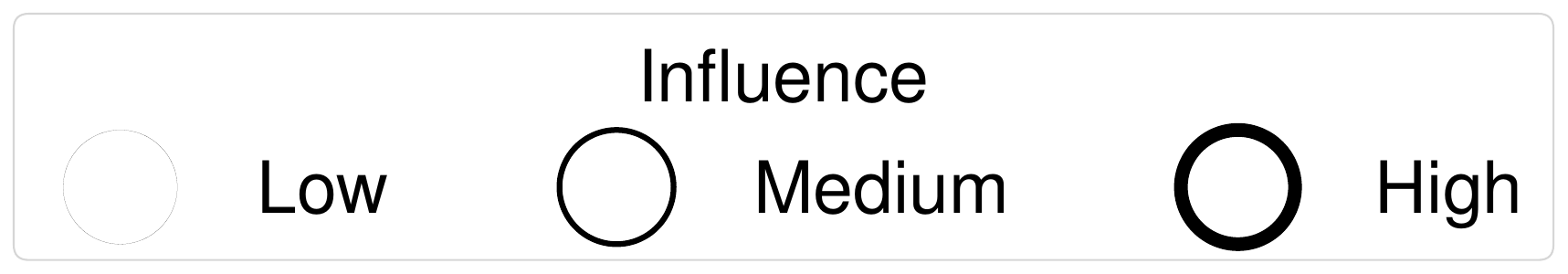}
    \\
    \vspace{2mm}
    \begin{subfigure}{0.24\textwidth}
        \includegraphics[width=\linewidth]{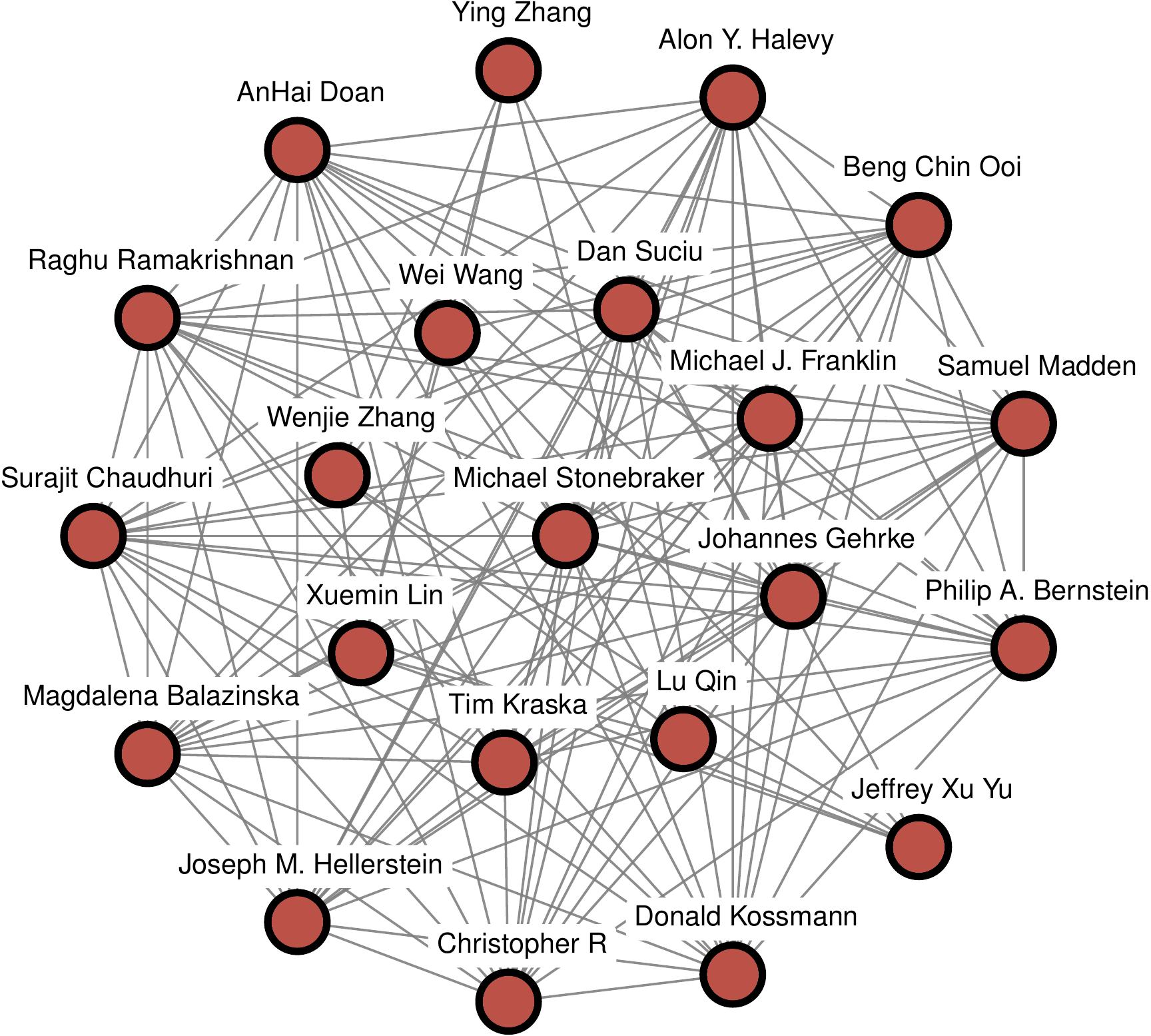}
        \caption{\prob}
    \end{subfigure}
    \hfill
    \begin{subfigure}{0.24\textwidth}
        \includegraphics[width=\linewidth]{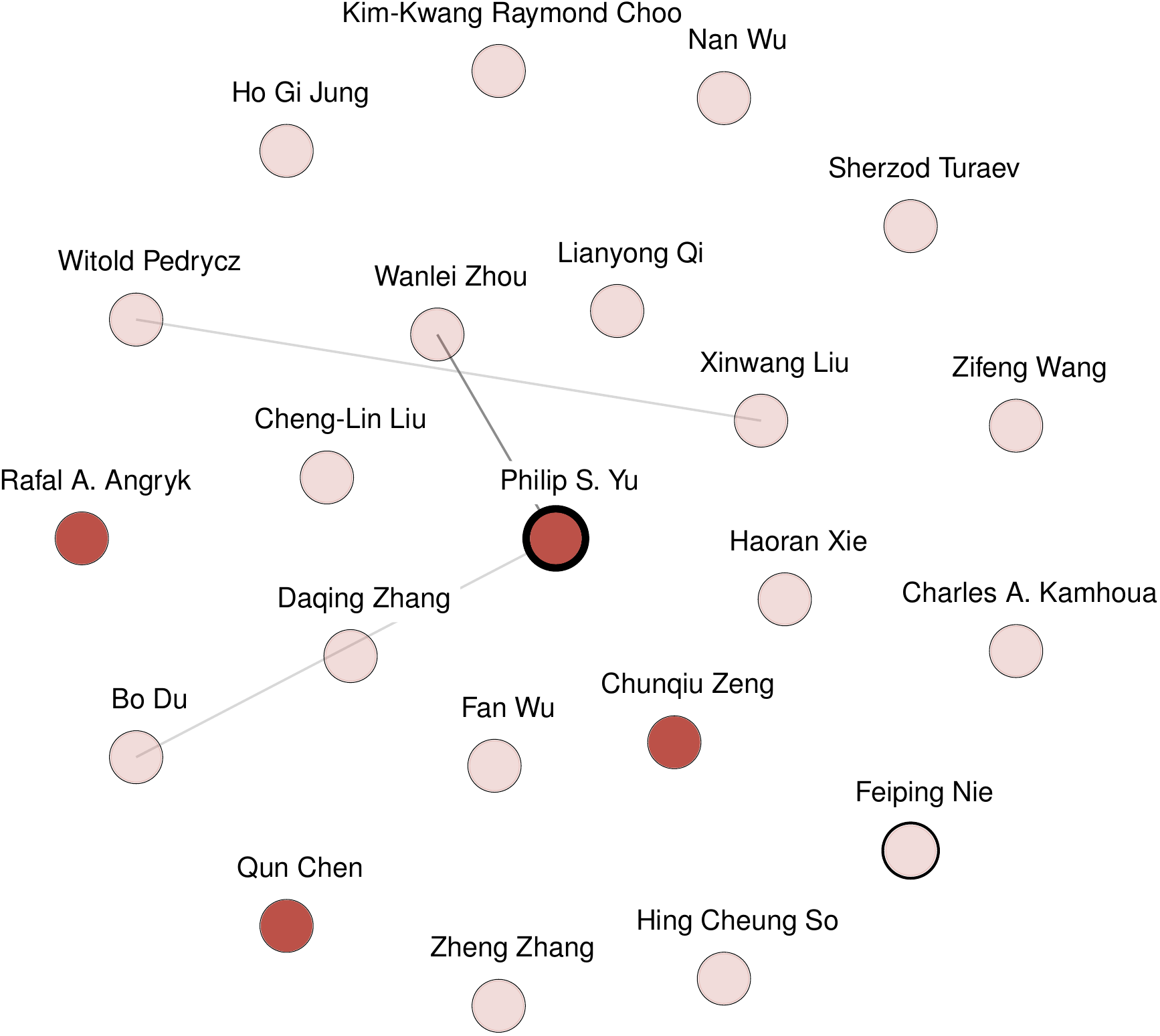}
        \caption{\textsf{TIM}}
    \end{subfigure}
    \hfill
    \begin{subfigure}{0.24\textwidth}
        \includegraphics[width=\linewidth]{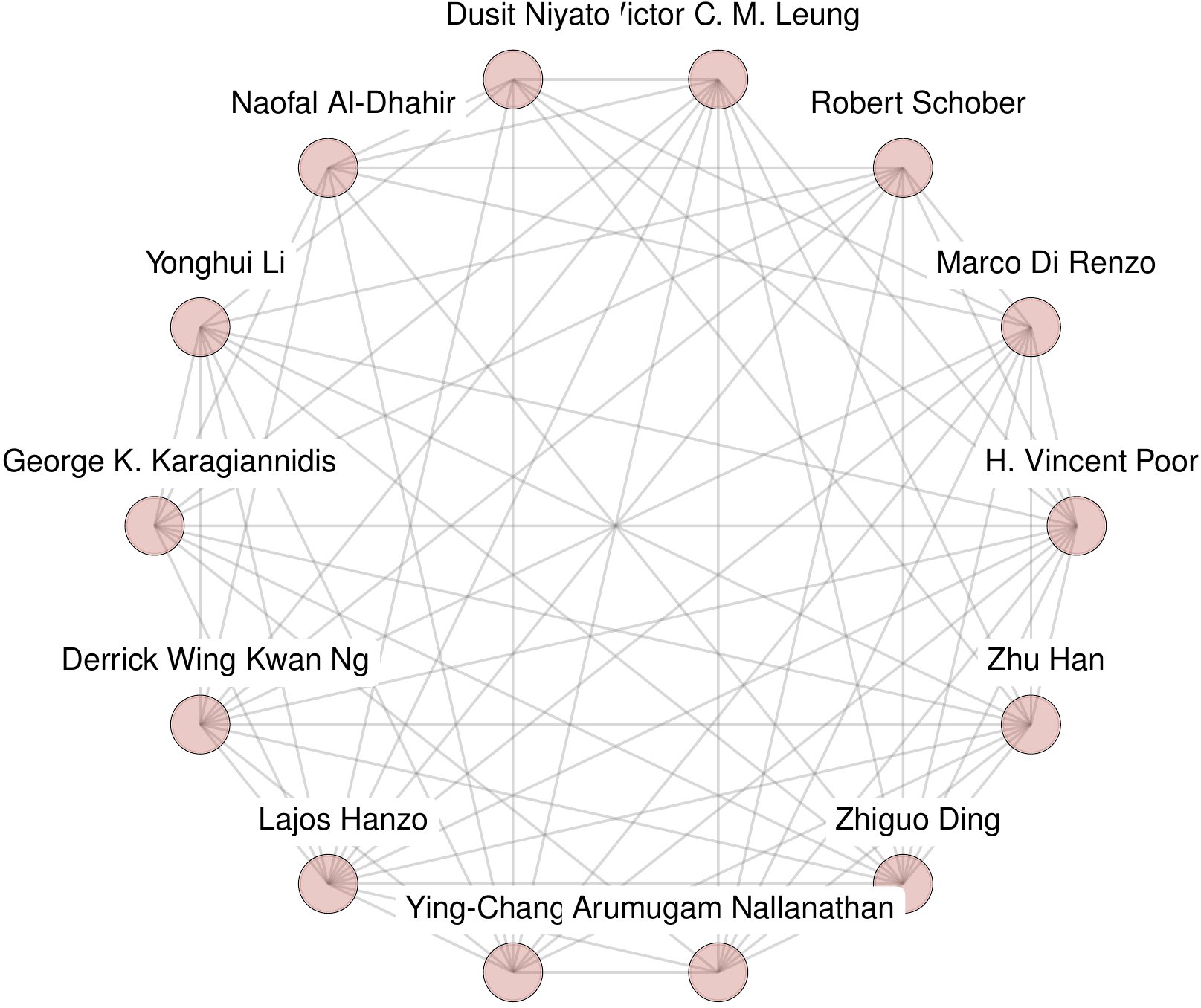}
        \caption{\textsf{UICS}}
    \end{subfigure}
    \hfill
    \begin{subfigure}{0.24\textwidth}
        \includegraphics[width=\linewidth]{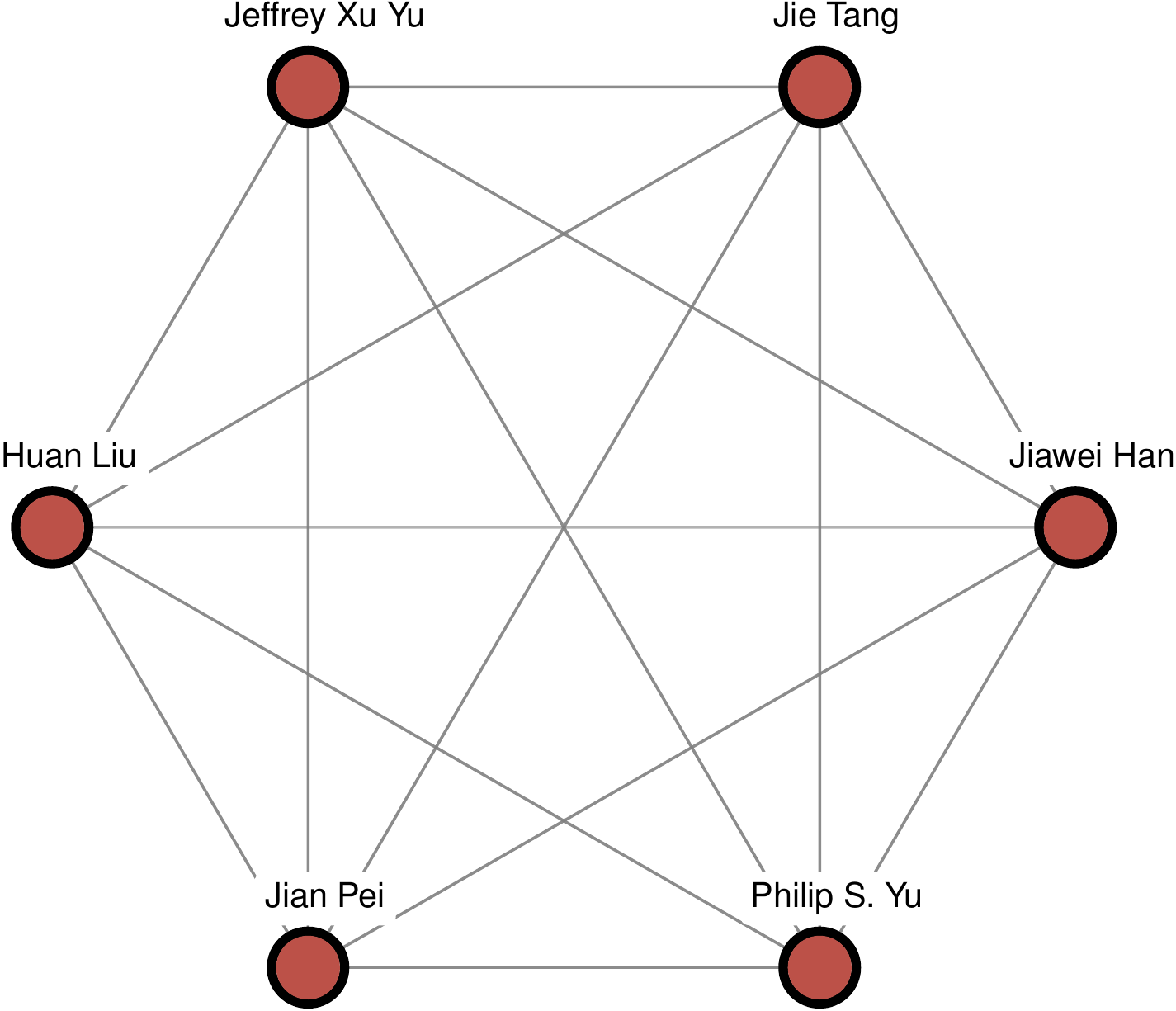}
        \caption{\textsf{KICQ}}
    \end{subfigure}
    \\
    \vspace{2mm}
    \begin{subfigure}{0.24\textwidth}
        \includegraphics[width=\linewidth]{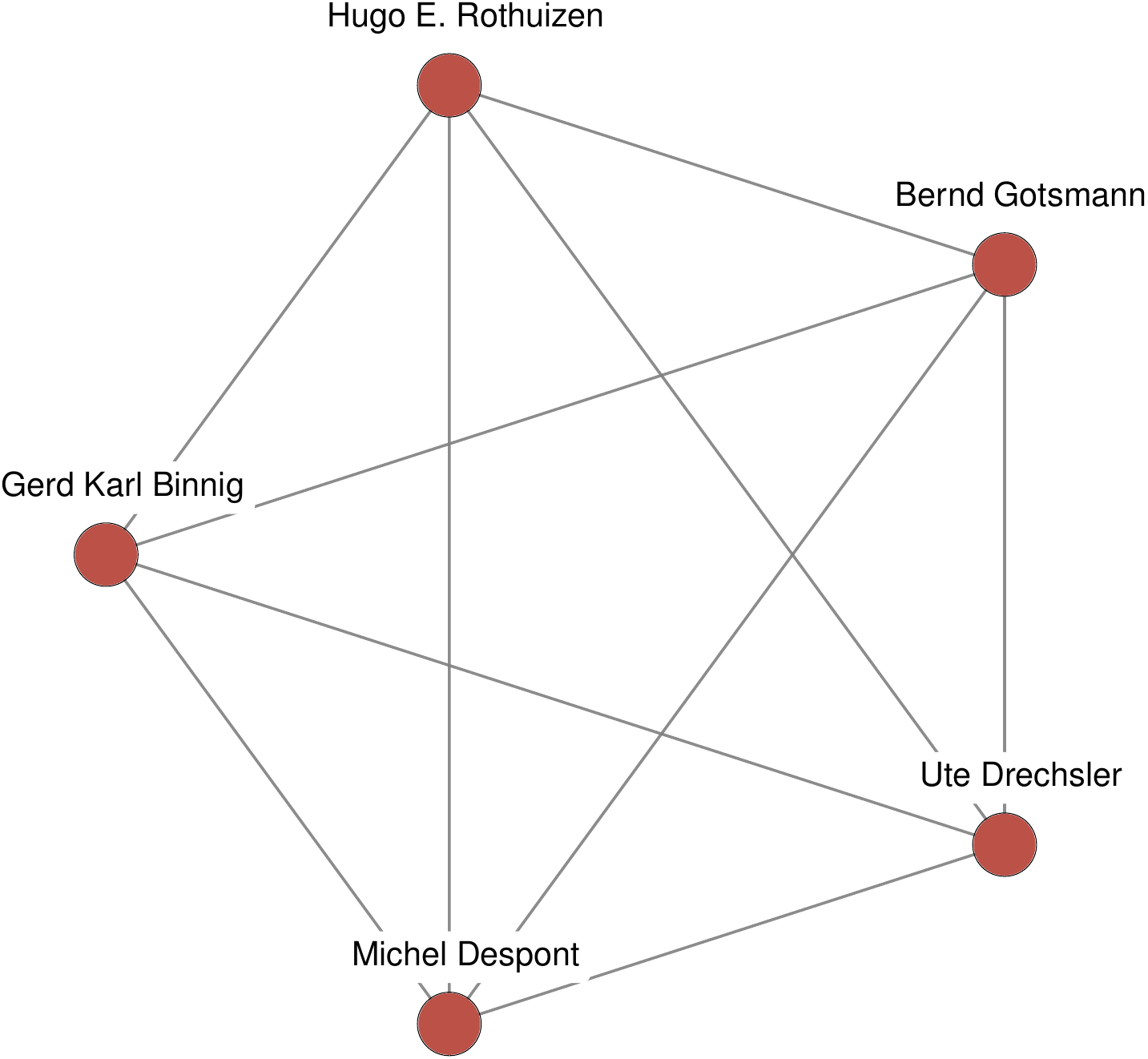}
        \caption{\textsf{VAC}}
    \end{subfigure}
    \hfill
    \begin{subfigure}{0.24\textwidth}
        \includegraphics[width=\linewidth]{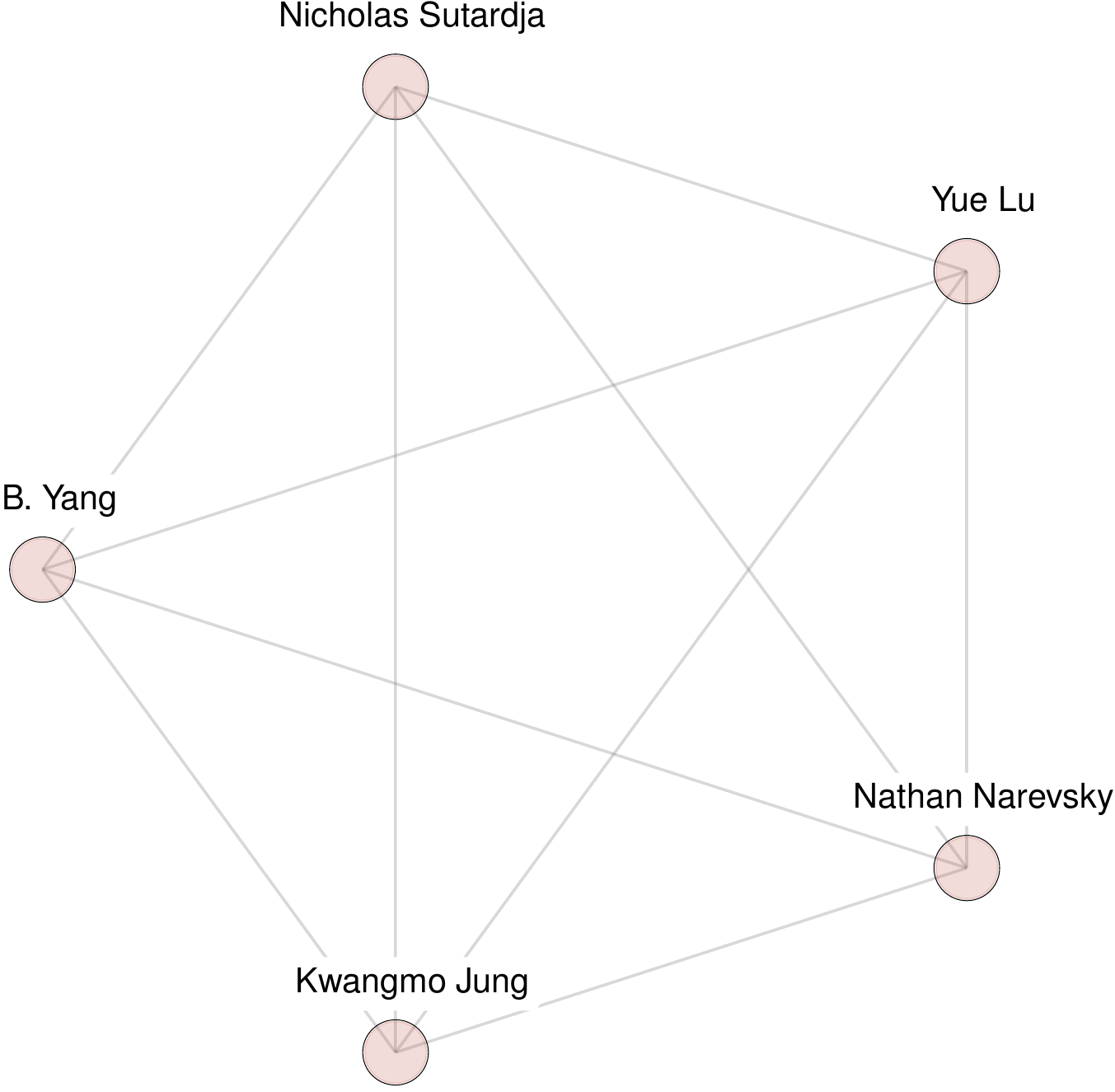}
        \caption{\textsf{EACS}}
    \end{subfigure}
    \hfill
    \begin{subfigure}{0.24\textwidth}
        \includegraphics[width=\linewidth]{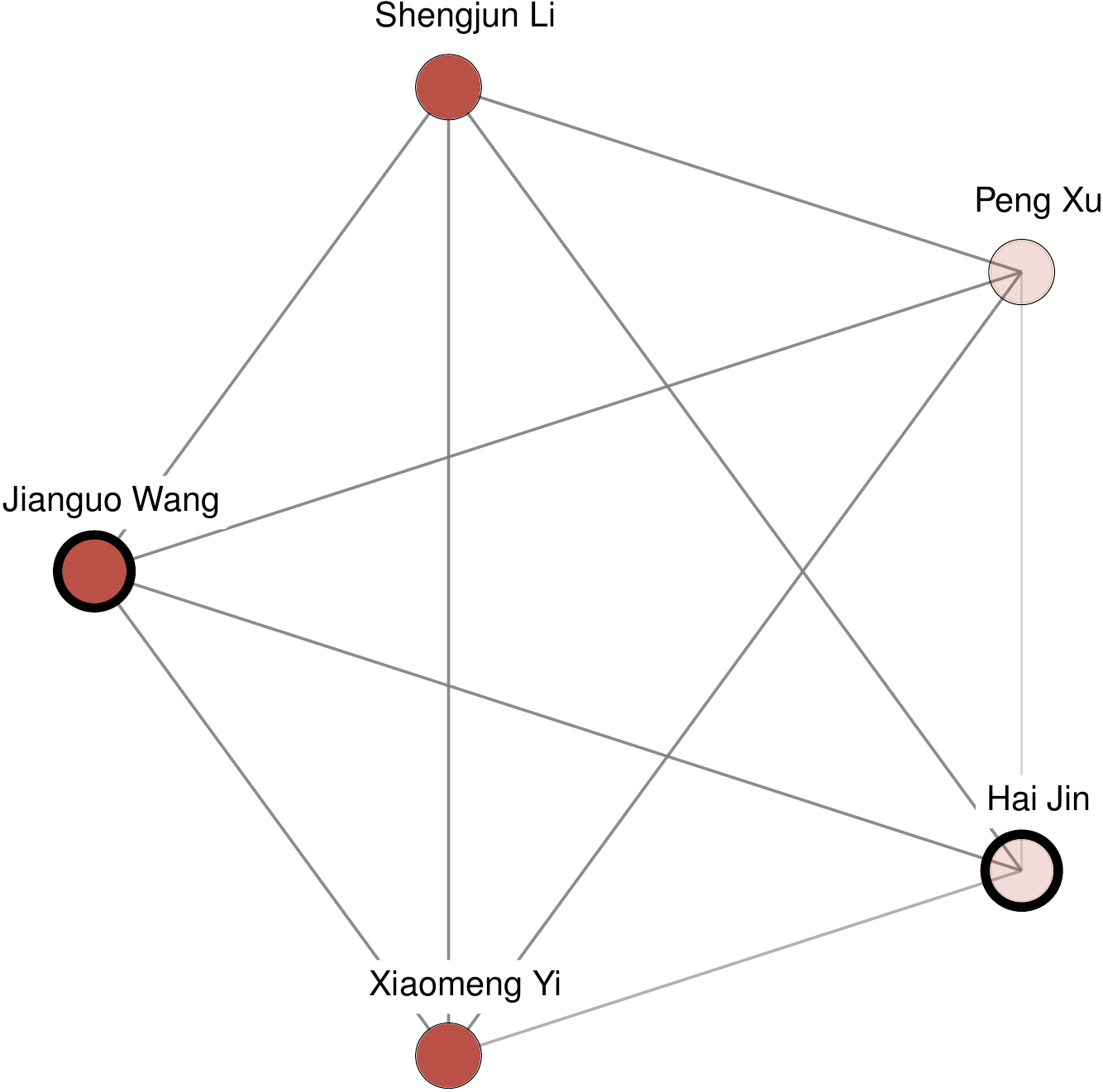}
        \caption{\textsf{TOPL-ICDE}}
    \end{subfigure}
    \hfill
    \begin{subfigure}{0.24\textwidth}
        \includegraphics[width=\linewidth]{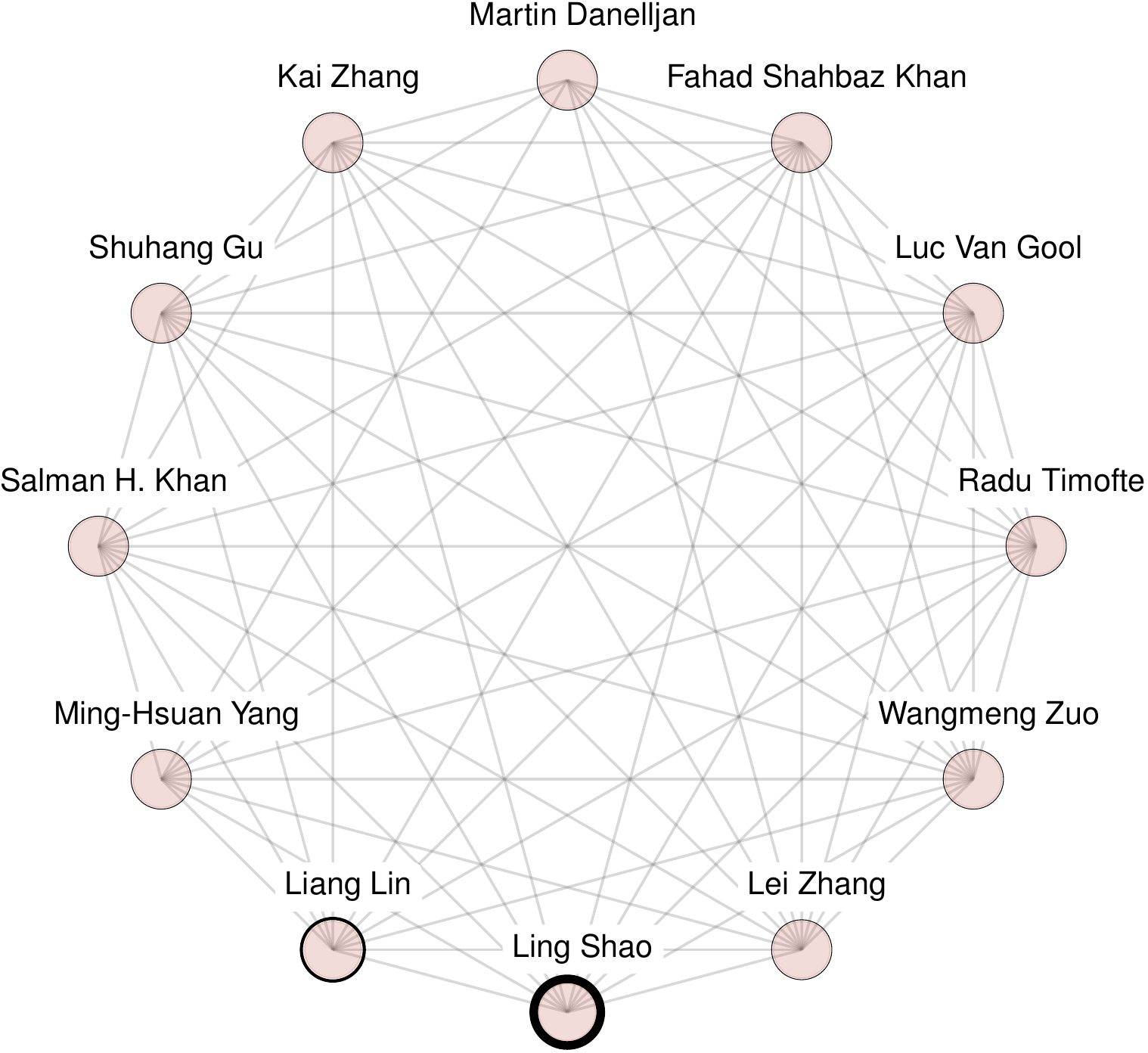}
        \caption{\textsf{MICS}}
    \end{subfigure}
    \\
    \caption{{Examples of communities returned by different CS methods on the DBLP dataset for subcategory ``Databases \& Information Systems.''}}
    \label{fig_CaseStudy_DBLP}
\end{figure}

\begin{figure}[t]
    \centering
    \includegraphics[width=0.6\textwidth]{figures/exp6_legend1.pdf}
    \hspace{2mm}
    \includegraphics[width=0.3\textwidth]{figures/exp6_legend3.pdf}
    \\
    \vspace{2mm}
    \begin{subfigure}{0.24\textwidth}
        \includegraphics[width=\linewidth]{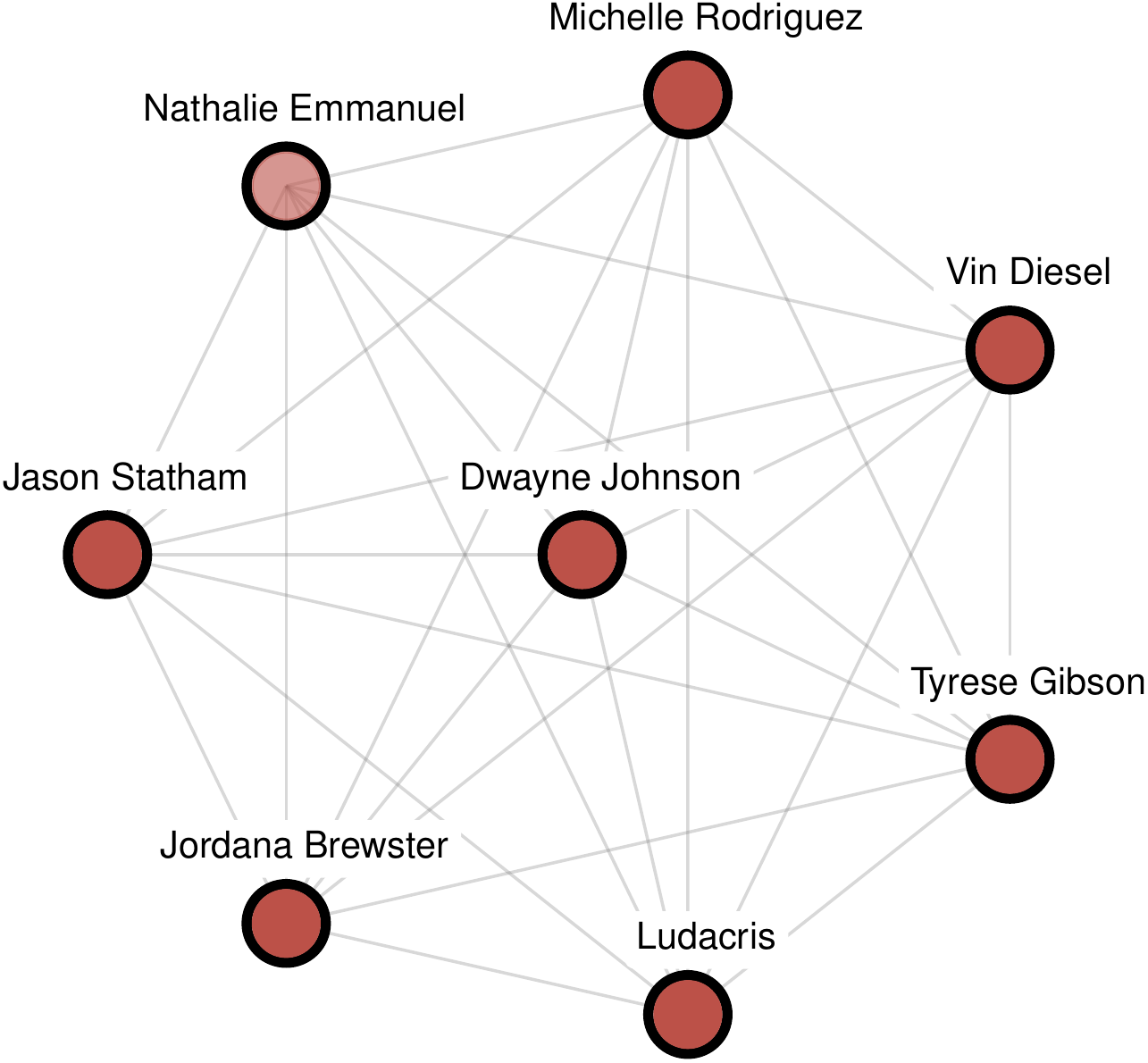}
        \caption{\prob}
    \end{subfigure}
    \hfill
    \begin{subfigure}{0.24\textwidth}
        \includegraphics[width=\linewidth]{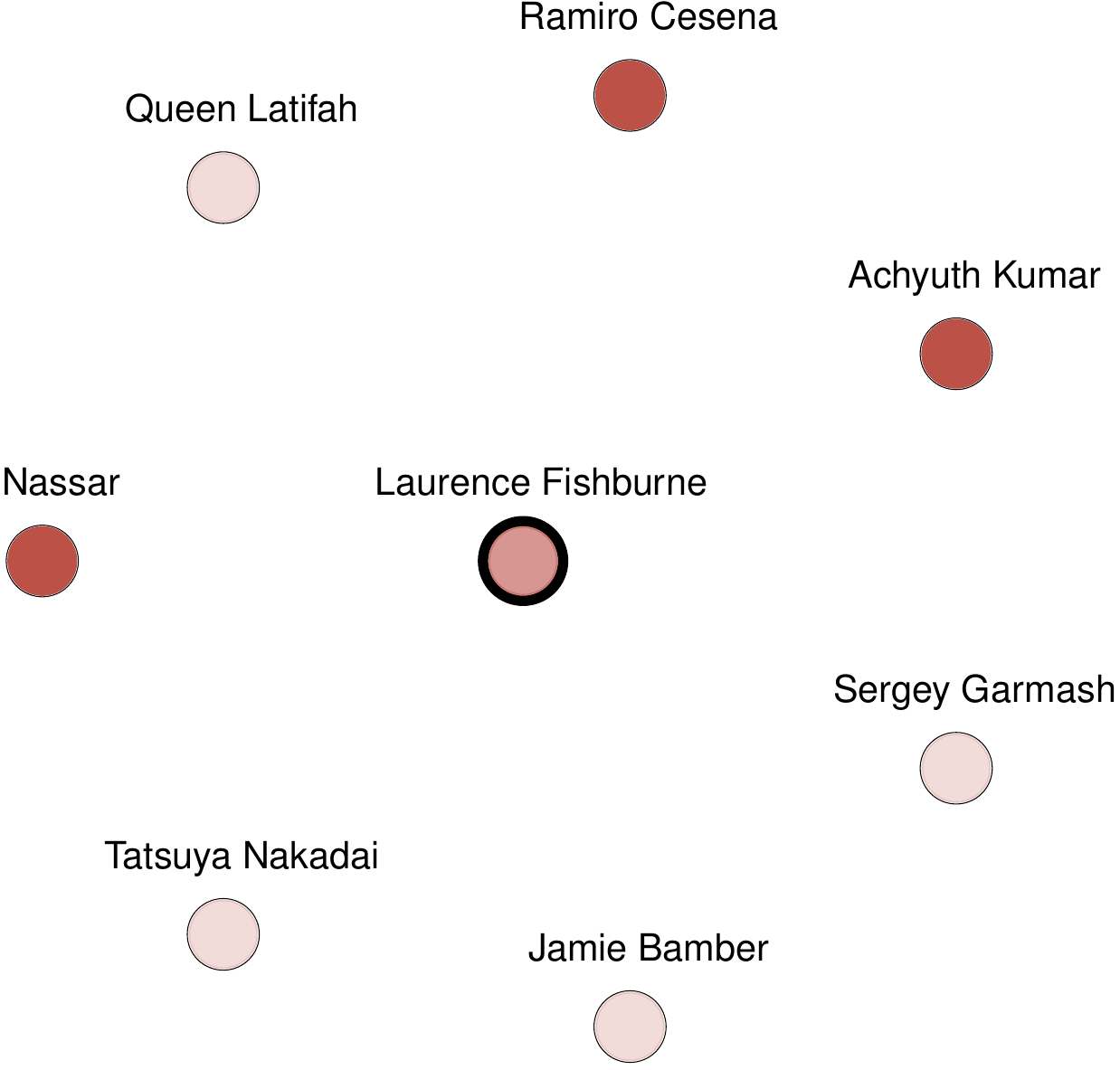}
        \caption{\textsf{TIM}}
    \end{subfigure}
    \hfill
    \begin{subfigure}{0.24\textwidth}
        \includegraphics[width=\linewidth]{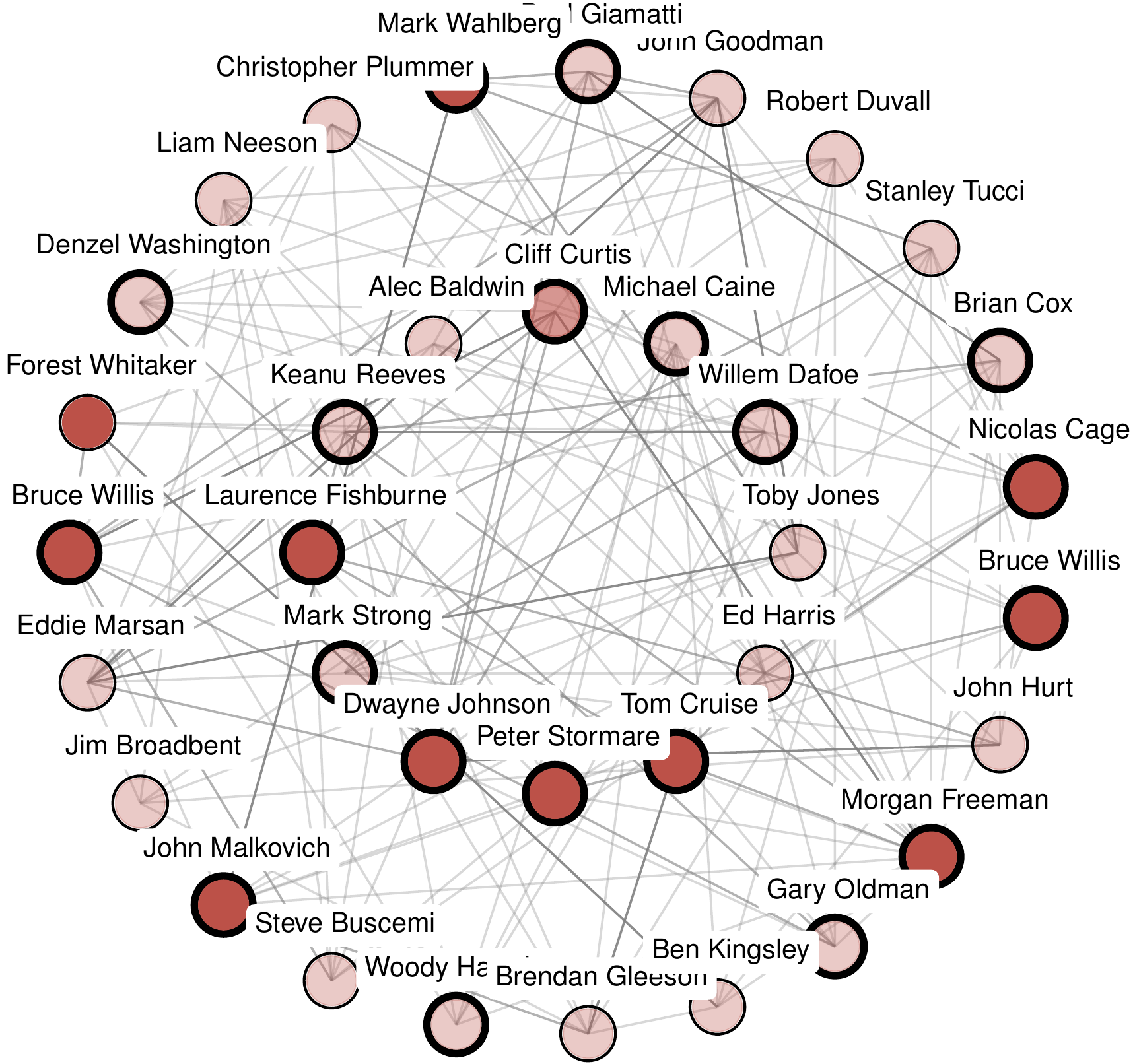}
        \caption{\textsf{UICS}}
    \end{subfigure}
    \hfill
    \begin{subfigure}{0.24\textwidth}
        \includegraphics[width=\linewidth]{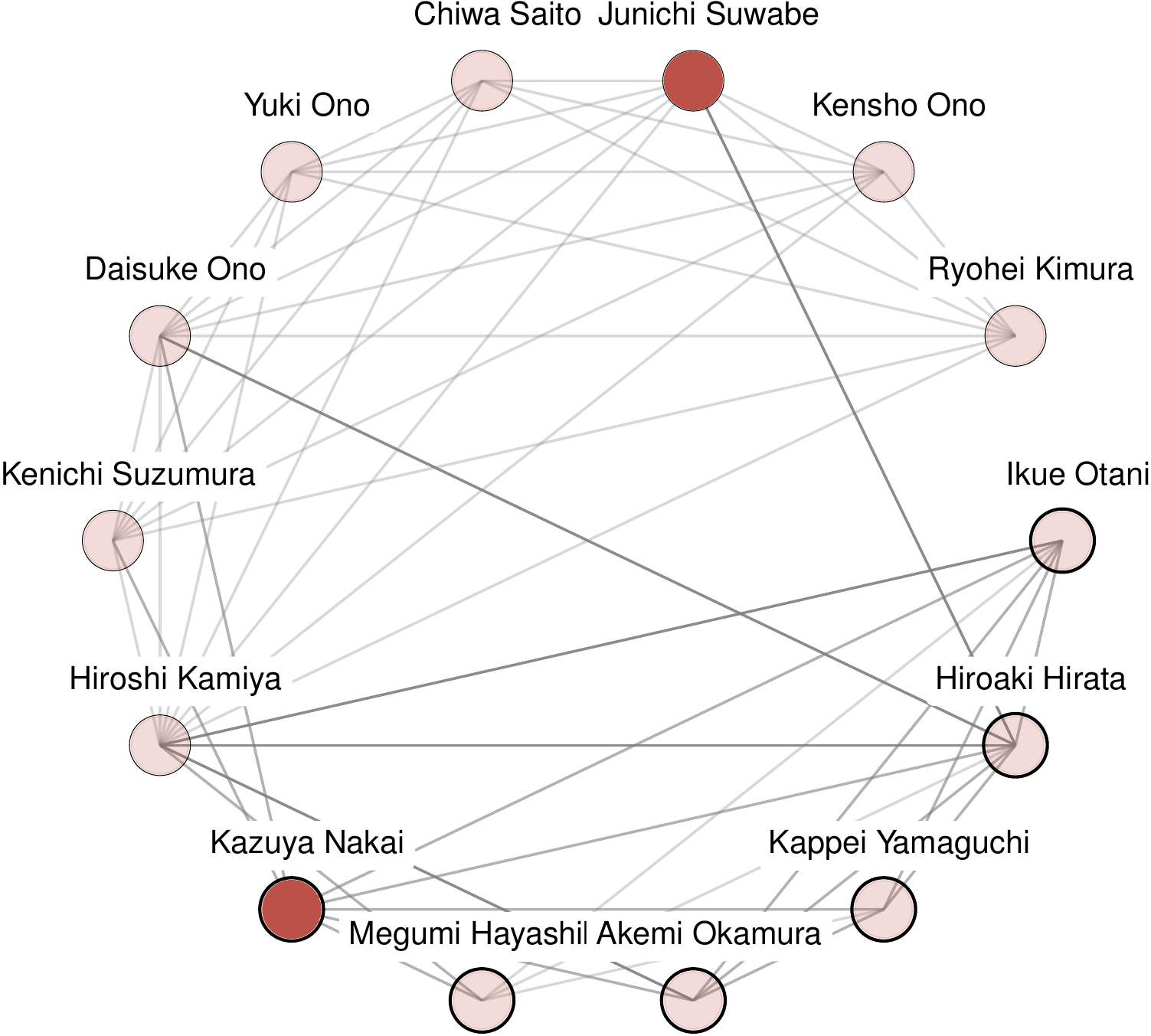}
        \caption{\textsf{KICQ}}
    \end{subfigure}
    \\
    \vspace{2mm}
    \begin{subfigure}{0.24\textwidth}
        \includegraphics[width=\linewidth]{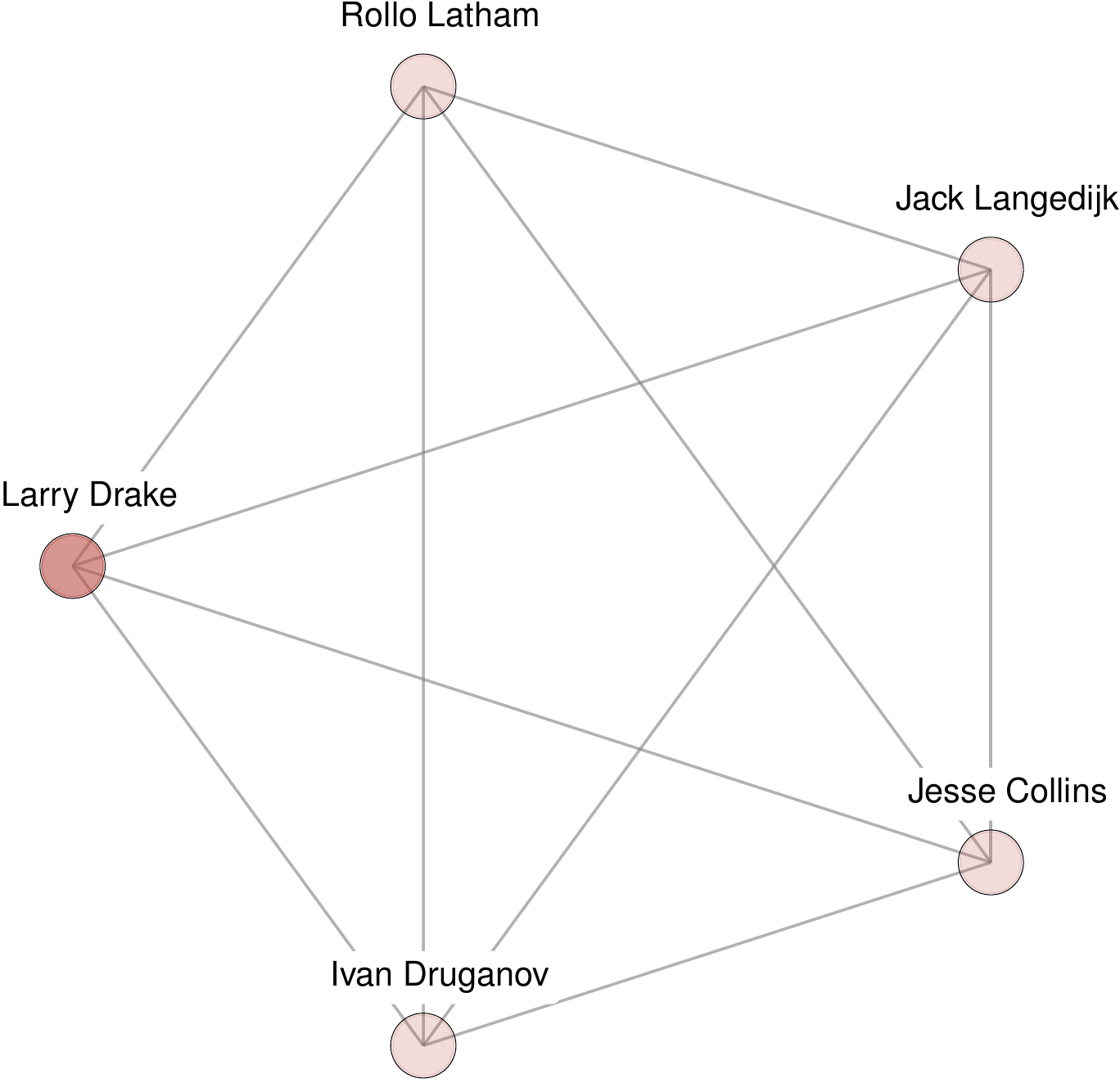}
        \caption{\textsf{VAC}}
    \end{subfigure}
    \hfill
    \begin{subfigure}{0.24\textwidth}
        \includegraphics[width=\linewidth]{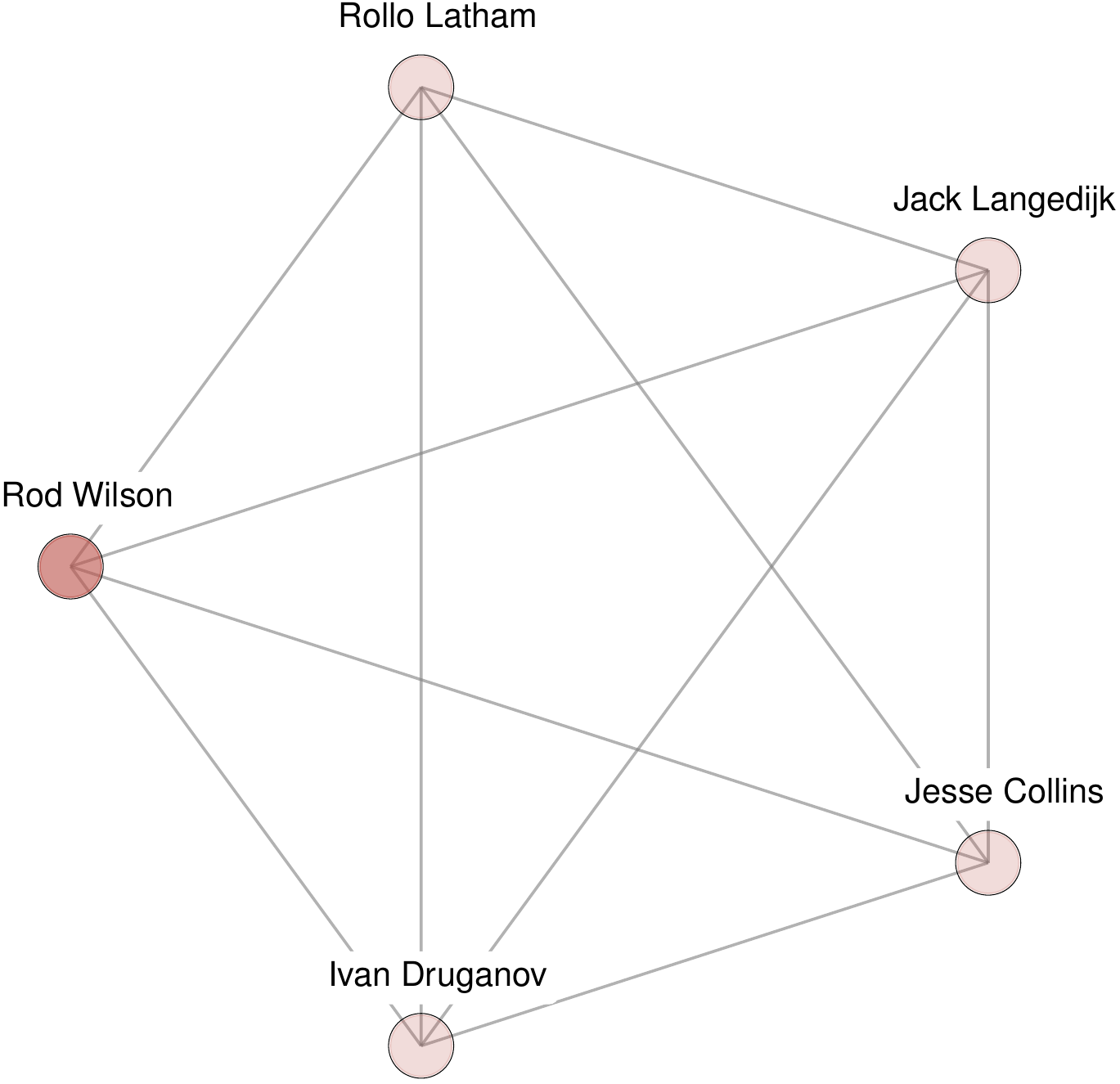}
        \caption{\textsf{EACS}}
    \end{subfigure}
    \hfill
    \begin{subfigure}{0.24\textwidth}
        \includegraphics[width=\linewidth]{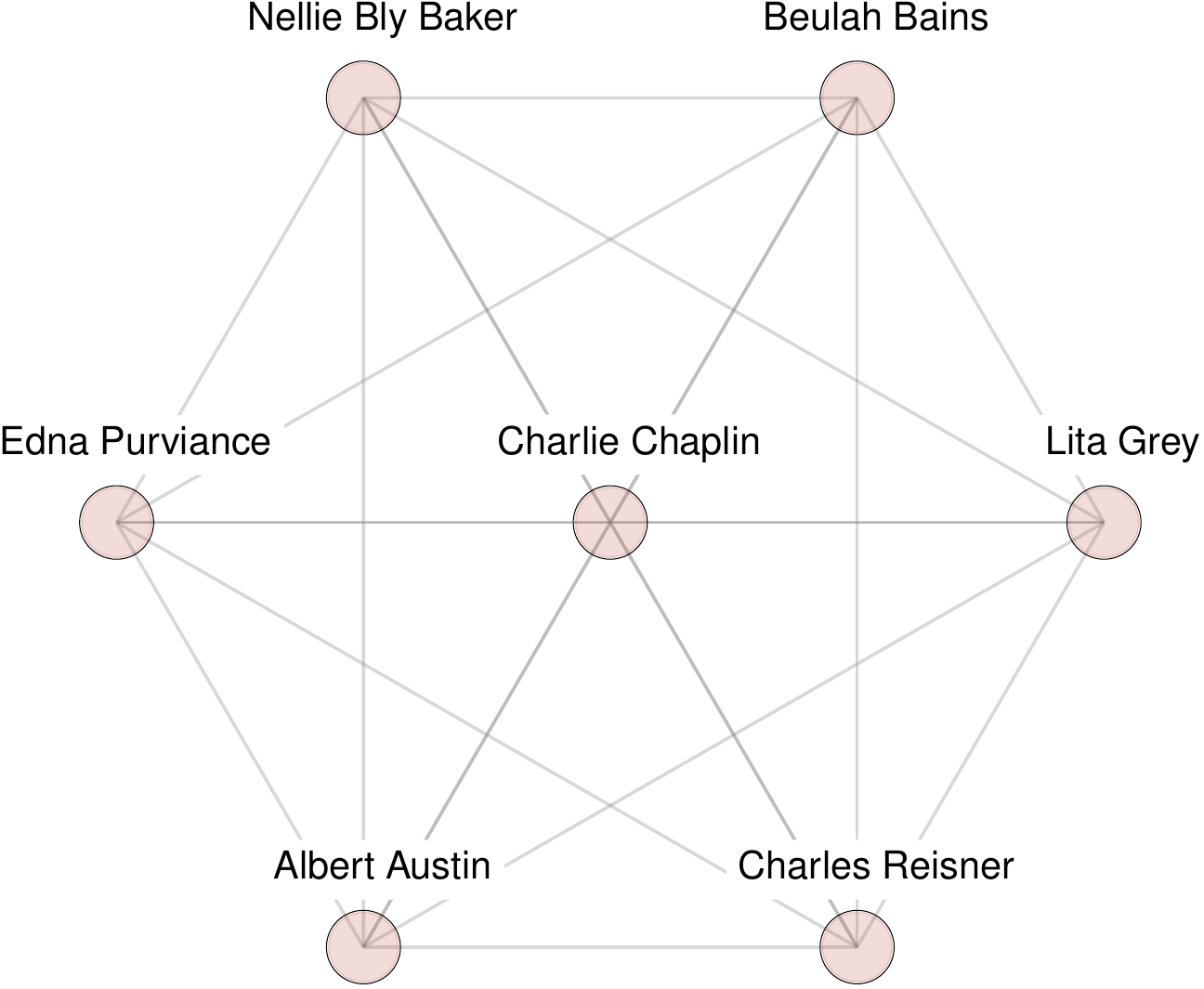}
        \caption{\textsf{MICS}}
    \end{subfigure}
    \hfill
    \begin{subfigure}{0.24\textwidth}
        \includegraphics[width=\linewidth]{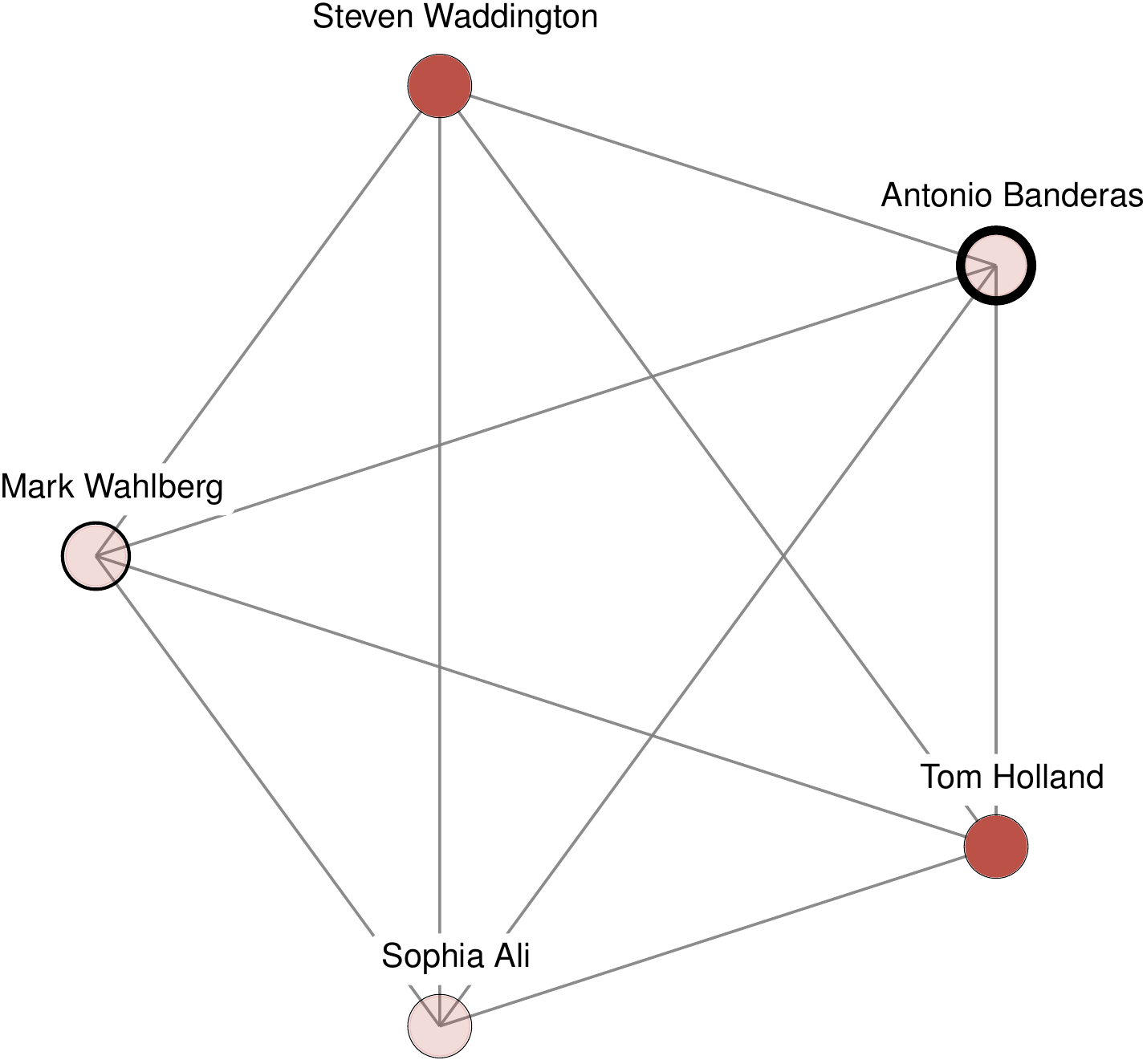}
        \caption{\textsf{TOPL-ICDE}}
    \end{subfigure}
    \\
    \caption{{Examples of communities returned by different CS methods on the IMDB dataset for genre ``Action.''}}
    \label{fig_CaseStudy_IMDB}
\end{figure}

\vspace{1mm}
\paragraph{Exp-4: Case Study}
We conduct case studies on the DBLP and IMDB datasets to illustrate the communities returned by different CS methods.
The results are presented in Figs.~\ref{fig_CaseStudy_DBLP} and~\ref{fig_CaseStudy_IMDB}.
We use a darker color to indicate that the vertex or edge is more relevant to the query topic (High: $>0.7$; Medium: $[0.4, 0.7]$; Low: $<0.4$).
We use a thicker black outline to indicate that the vertex has a higher influence score (High: top-$100$; Medium: top-$1,000$; Low: not in top-$1,000$).
The results further confirm that \prob provides high-quality communities.
On the DBLP dataset, the community provided by \prob consists of reputational scholars with close collaborations in the field of ``Databases \& Information Systems.''
On the IMDB dataset, the community provided by \prob includes many famous actors who co-starred with each other in action movies.
However, other CS methods cannot provide satisfactory communities.
The vertices returned by \textsf{TIM} have only a few connections with each other, which does not satisfy the concept of communities.
{\textsf{UICS} and \textsf{MICS}} retrieves vertices that are largely unrelated to the query topic due to its topic unawareness.
\textsf{VAC} and \textsf{EACS} do not take social influence into account and thus can only guarantee that the returned vertices are highly coherent but may have low influence.
Due to the limitation of keyword matching in representing semantic information about topics, {\textsf{VAC}, \textsf{EACS}, and \textsf{TOPL-ICDE}} also include less relevant vertices containing some of the query keywords.
Finally, by considering keywords and social influence collectively, \textsf{KICQ} also performs well on the DBLP dataset.
However, \textsf{KICQ} does not show good performance on the IMDB dataset.
We observe that \textsf{KICQ} returns a set of voice actresses and actors for Japanese animations, many of which also contain ``Action'' as a genre.
This is because query keywords do not align well with query topics.

%% file: tables/dataset.tex
\begin{table}[t]
    \footnotesize
    \centering
    \caption{Statistics of graphs in the experiments}
    \label{tab_data set}
    \begin{tabular}{cccccc}
        \toprule
        \textbf{Dataset} & $n$ & $m$ & $\Delta^{-}_{max}$ & $\Delta^{+}_{max}$ & $z$\\
        \midrule
        Epinion & 22,166    & 355,589   & 1,552 & 2,024 & 20\\
        IMDB    & 75,805    & 1,835,574 & 1,237 & 1,237 & 20\\
        DBLP    & 362,828   & 4,292,802 & 1,012 & 1,012 & 10\\
        Reddit  & 1,370,155 & 3,652,493 & 463   & 804   & 19\\
        Wiki-Topcats & 1,791,489 & 28,511,807 & 3,907 & 238,040 & 5\\
        \bottomrule
    \end{tabular}
\end{table}

%% file: sections-springer/7-conclusion.tex
\section{Conclusion}
\label{sec_conclusion}

In this paper, we study a novel problem of topic-aware most influential community search (\prob) on social networks.
The \prob problem is based on a novel community model in which the uncertain $(k,l,\eta)$-core is used for cohesiveness definition and the topic-aware independent cascade model is adopted for influence calculation.
We design an online algorithm and analyze its theoretical bound for \prob.
Furthermore, we devise an index-based heuristic algorithm to improve the efficiency of \prob processing.
{Finally, we conduct extensive experiments and case studies on real-world datasets to demonstrate that the communities of \prob have higher relevance and social influence w.r.t.~the query topics as well as structural cohesiveness than those of several state-of-the-art topic-aware and influential CS methods. Furthermore, the index-based algorithm achieves speed-ups of up to three orders of magnitude over the online algorithm while having a little impact on community quality.}

This paper still has several limitations to address in future work. First, the \prob model and the index structures do not fully consider the dynamic changes in user behavior. When user interest changes over time, they should be rebuilt from scratch. Therefore, how to introduce dynamic topic models and handle index updates in \prob would be a promising question to explore. Second, this work does not discuss the robustness of \prob against abnormal and extreme cases, such as misinformation or malicious users who intentionally inflate social influence. The robustness of community search would also be an interesting problem to investigate.